\documentclass[12pt,showpacs,preprintnumbers,amsmath,amssymb]{revtex4}
\usepackage{psfrag}

\newcommand{\be}{\begin{equation}}
\newcommand{\ee}{\end{equation}}

\newcommand{\bal}{\begin{align}}
\newcommand{\eal}{\end{align}}

\newcommand{\bea}{\begin{eqnarray}}
\newcommand{\eea}{\end{eqnarray}}
\newcommand{\n}{n}

\newcommand{\q}{q}
\newcommand{\sll}{\it}
\newcommand{\itt}{\sl}

\newcommand{\Z}{v}
\newcommand{\W}{v}
\newcommand{\f}{f_{1}}
\newcommand{\p}{f_{2}}
\newcommand{\Y}{u}
\newcommand{\Om}{u}
\newcommand{\om}{\sigma}

\newcommand{\mz}{m_{\mbox{\tiny Z}}}
\newcommand{\mh}{m_{\mbox{\tiny H}}}
\newcommand{\mw}{m_{\mbox{\tiny W}}}
\newcommand{\thetaw}{\theta_{\mbox{\tiny W}}}

\newcommand{\YY}{u}
\newcommand{\OOmega}{u}
\newcommand{\WW}{{\rm W}}

\newcommand{\F}{{B}}
\newcommand{\A}{{B}}

\renewcommand{\theequation}{\arabic{section}.\arabic{equation}}
\begin{document}

\

\vspace{3 cm}

%\title{Superconducting and Charged Electroweak Vortices}

\title{Superconducting non-Abelian vortices in Weinberg-Salam \\ 
theory --
electroweak thunderbolts}
%\title{Superconducting and Charged Electroweak Cosmic Strings}

\author{Julien Garaud and Mikhail S. Volkov}

 \affiliation{ {Laboratoire de Math\'{e}matiques et Physique Th\'{e}orique
CNRS-UMR 6083, \\ Universit\'{e} de Tours,
Parc de Grandmont, 37200 Tours, FRANCE}
}

\begin{abstract}

We present a detailed analysis of 
classical solutions in the bosonic sector of the 
electroweak theory which describe vortices carrying a 
constant electric current ${\cal I}$. These vortices  exist 
for any value of the Higgs boson mass and  
for any weak mixing angle, and  in 
the zero current limit they reduce to  
Z strings.
Their current is produced by 
the condensate of vector W bosons 
and typically it can attain billions of Amperes. 
For large ${\cal I}$ the vortices show a compact 
condensate core of size $\sim 1/{\cal I}$,
embedded into a region of size $\sim{\cal I}$ 
where the electroweak gauge symmetry is 
completely restored, followed by a transition zone   
where the Higgs field interpolates between the symmetric and broken phases. 
Outside this zone the fields are the same as for the ordinary electric wire. 
An asymptotic approximation of the large ${\cal I}$ solutions 
suggests that the current can be {arbitrarily} large, due to
 the scale invariance of the  vector boson condensate.  
 Finite vortex segments whose length grows with ${\cal I}$ 
seem to be perturbatively stable. This suggests that they can 
transfer electric charge between different regions of space, 
similarly to thunderbolts. It is also possible that they 
can form loops stabilized by the centrifugal force -- electroweak vortons.

\end{abstract}


\maketitle

\newpage

\tableofcontents

\section{Introduction}
\setcounter{equation}{0}

More than 20 years ago Witten proposed a field theory model that admits 
classical  solutions describing vortices carrying a constant
current 
-- `superconducting strings' \cite{Witten}. 
This model has a local U(1)$\times$U(1) invariance and 
 consists of two copies of the Abelian Higgs model for fields 
$(A^{(1)}_\mu,\phi_{1})$ and $(A^{(2)}_\mu,\phi_{2})$. 
It also includes  an interaction between 
the two complex scalars chosen such that
in vacuum one has $\phi_{1}\neq 0$ but $\phi_{2}=0$ so that the vector 
field $A^{(1)}_\mu$ 
is massive while $A^{(2)}_\mu$ is massless and can be identified with the 
electromagnetic field.  

The model admits as a solution the Abrikosov-Nielsen-Olesen (ANO) vortex
\cite{ANO} made of the `vortex fields'  $(A^{(1)}_\mu,\phi_{1})$,
with vanishing `condensate fields' $(A^{(2)}_\mu,\phi_{2})$. 
This embedded vortex is however unstable,   
but being topological it does not unwind into vacuum and relaxes to 
a `dressed vortex'  which contains 
a condensate of charged scalar bosons in the core and has $\phi_2\neq 0$. 
Giving then a non-trivial phase
to the condensate field 
$\phi_2$ produces a current and promotes the `dressed' vortex to the superconducting 
string supporting  the long-range 
Biot-Savart field represented by  $A^{(2)}_\mu\neq 0$.

The Witten string superconductivity  has been much studied 
\cite{SS}, \cite{Carter},
mainly in the cosmological context \cite{Vilenkin-Shellard}, \cite{Hindmarsh-Kibble},  
since Witten's model can be viewed as sector of some high energy 
Grand Unification Theory (GUT) \cite{Witten} that could perhaps be relevant 
at the early stages of the cosmological evolutions. 
Using the typical values of the GUT parameters for estimates gives 
for the string current enormous values of order $10^{20}$ Amperes, which 
suggests interesting applications \cite{Vilenkin-Shellard}, \cite{Hindmarsh-Kibble}.   
String superconductivity in the GUT-related non-Abelian models has also been studied
\cite{Everett},
in which case the string current is produced by a condensate of charged vector bosons.

Although the GUT physics could be important, one may 
wonder whether a similar 
string superconductivity could exist also in a less exotic context, 
at lower energies,  as for example in the electroweak sector of Standard Model.
In fact, the U(1)$\times$U(1) symmetry of Witten's model is contained  
in the SU(2)$\times$U(1) electroweak gauge symmetry. In addition, 
the electroweak theory contains 
 a pair of complex Higgs scalars,
one of which could well be responsible for the formation of the vortex while
the other one -- for the condensate. 
Since the ANO vortices can be embedded into the electroweak 
theory in the form of  Z strings \cite{Z}, one might expect that 
 current-carrying generalizations of the latter  
could exist. These would be electroweak 
analogues of Witten's superconducting strings.

However, no attempts to construct such solutions 
have ever been undertaken. This can probably be explained by 
the following reasons. The current-carrying Witten strings 
are usually viewed as excitations over the 
`dressed' currentless vortex  obtained by minimizing the 
energy of the `bare'  embedded ANO vortex.   
This explains 
 their essential properties,
as for example the value of the critical current \cite{Vilenkin-Shellard}. 
Now, the `bare' electroweak  Z strings 
are also unstable \cite{GH}, and it was conjectured \cite{per,Olesen} that 
they could similarly relax to `W-dressed Z strings'. However, a systematic search 
for such solutions gave no result \cite{perk}. This can  
probably be explained by the fact that      
Z strings are non-topological and can unwind into vacuum \cite{KO},
so that there is 
little chance that they could be stabilized by the condensate. 
But if one does not find currentless `dressed' Z strings, it is  natural to think that 
current-carrying strings in the electroweak theory do not exist either. 
This 
presumably explains why there has been
almost no activity on electroweak vortices during the last 10-15 years.

At the same time, it is difficult to believe that Z strings are the only possible vortex
solutions in the electroweak theory. 
The theory admits rather non-trivial solutions,  such as sphalerons \cite{sphalerons},
periodic BPS solutions \cite{period}, spinning dumbbells 
\cite{Nambu}, \cite{dumb},
oscillons \cite{oscillon}, spinning sphalerons \cite{RV}, and others 
(see \cite{achuc} for a review). 
In addition, in the semilocal limit where the SU(2) field decouples  the theory admits 
current-carrying vortices \cite{SL}, so that it is plausible that they could exist 
also for generic values of the weak mixing angle. 

We have therefore analysed the problem and found that    
superconducting  vortices indeed exist in the electroweak theory, 
despite the non-existence of the `W-dressed Z strings'. 
In other words, these two types of solutions are not necessarily related. 
To construct the solutions, we essentially reverse the standard 
`engineering' procedure used
in Witten's model.  There one starts from the 
`dressed' currentless vortex and {\it increases}
it `winding number density' that determines the phase of the condensate;
in what follows we shall call this parameter `twist'. 
This produces a  current that first increases with the twist but then starts to quench 
 and finally vanishes when the solution reduces to the `bare' ANO vortex 
\cite{Vilenkin-Shellard}. The superconducting strings thus comprise a one-parameter
family that interpolates between the `dressed' vortex and the `bare' ANO vortex.  

We construct  this family in the opposite direction, by starting from the `bare' 
vortex and then {\it decreasing} 
its twist. Within Witten's model this gives of course 
the same solutions but in the reversed order
 -- their current first increases then starts to quench 
and vanishes for zero twist when the solution reduces to the `dressed' string.  
However, the advantage of our method is that it can be applied also within the 
Weinberg-Salam theory, where there are no `dressed' currentless vortices but only   
 the `bare' ones -- Z strings.
 Their  twist is determined by the eigenvalue of the second variation of the energy
functional.  
Decreasing the twist gives us solutions with a non-zero current. 
Further decreasing it shows that the current always grows
and  tends to infinity when the twist approaches zero. As a result, 
contrary to what one would normally expect, and presumably because of the 
vector character of their current-carriers,  the electroweak  
 vortices do not (generically) exhibit the current quenching, so that  they 
do not need to admit the `dressed' currentless limit.  
%The absence of current
%quenching in this case can be related to the fact that the current carriers
%are {\it vector} and not scalar bosons.  

The stability analysis of these vortices  shows that
their finite  segments  are perturbatively stable. More precisely,
so far this property has been demonstrated only in the semilocal limit \cite{stab}, 
but it is very likely that the result extends for generic values of the weak 
mixing angle. The length of the stable segments increases with the current and 
can in principle attain any value, since  
there is no upper bound for the current.  

This may have interesting consequences. First of all, this suggests 
that  loops made of the stable  segments and balanced by the centrifugal force 
could perhaps be stable as well. Although studying this goes 
far beyond the scope of the present paper, the possibility 
to have stable solitonic objects in the Standard Model could be very important,
since if such  
 {\it electroweak vortons}  exist, they could perhaps 
contribute to the dark matter. 

Another, perhaps a more direct consequence, is related to the well-know fact that,
 since Z strings are non-topological, they can have finite length. 
%The magnetic flux through
%their two extremities then looking like a monopole-antimonopole pair.   
This suggests
that their current-carrying generalizations  could also exist in the form of finite
(and stable) 
segments connecting oppositely charged regions of space.  
They would then be similar to thunderbolts. For the solutions we could 
explicitly construct the current can be as large as $10^{10}$ Amperes, which exceeds
by several orders of magnitudes the power of the strongest thunderbolts in 
the Earth atmosphere. This suggests that superconducting vortices 
could be important for the dynamics of the Standard Model, although analysing   
this issue in detail would again lead us too far away from our main subject. 

The present paper is devoted to  
 a systematic analysis of the string/vortex-type solutions in the electroweak theory
(terms string and vortex are assumed to be synonyms).    
In what follows we shall show that every Z string 
admits a three-parameter  family of non-Abelian, current carrying generalizations. 
In
the gauge where the radial components of the gauge fields vanish, 
the upper and lower components of the Higgs field doublet
behave quite analogously to the vortex field and condensate field in Witten's model. 
The new vortices  
exist for (almost) any value of the weak mixing angle and for any 
Higgs boson mass. They show a compact, regular core 
containing the W-condensate that carries a constant electric current, 
producing the Biot-Savart electromagnetic field outside 
the vortex. In the comoving reference frame the vortex is characterized by the
current ${\cal I}$ and by the 
electromagnetic and Z fluxes through its cross section.
After performing a Lorentz boost, it develops also a non-zero electric charge density,
as well as momentum and angular momentum directed along the vortex.

It seems that 
 the vortex current can be as large as one wants, at least we could not find 
an upper bound for it. We have also managed to construct a 
simple approximation in the  large current limit which 
suggests that the current can be {\it arbitrarily} large, due to
 the fact that it is carried by {\it vectors} 
whose condensate exhibits the scale invariance. 
To the best of our knowledge,
a similar effect has never  been reported before.
For large currents the charged W boson 
condensate is contained in the vortex core
of size  $\sim 1/{\cal I}$, which is  
surrounded by a large  region of size $\sim {\cal I}$ 
where the Higgs field is driven to zero by the string magnetic field so that 
the electroweak gauge symmetry is completely restored. 
However, this does not destroy the vector boson superconductivity, 
since the scalar Higgs field is not the relevant order parameter in this case.
Outside the symmetric phase region
there is a transition zone where the Higgs field  relaxes to the broken phase 
and the theory reduces to the ordinary electromagnetism. This reminds 
somewhat of the electroweak vacuum polarization scenario discussed by 
Ambjorn and Olesen \cite{period}, in which the system stays  in the Higgs 
vacuum if only the magnetic field is not too strong, while a very strong 
field restores the full gauge symmetry.

The rest of the paper is made maximally self-contained.
In Sec.II the essential elements  of the Weinberg-Salam theory are 
introduced.  Sec.III presents the reduction to the stationary sector, 
the derivation of the conserved quantities, and the discussion of the 
Bogomol'nyi-Prasad-Sommerfield (BPS) limit. Sec.IV contains the further reduction to the 
axially symmetric sector, the basic field ansatz \eqref{003}, and the field equations 
\eqref{ee1}--\eqref{CONS}. In Sec.V the boundary conditions 
are discussed, while 
Sec.VI and Sec.VII contain the derivation of the charges and a brief description
of the known solutions. The new solutions are first considered 
in  Sec.VIII in the small current limit, when 
they can be treated 
as small deformations of Z strings. In Sec.IX they are 
presented at the full non perturbative level for generic values of the current.  
Sec.X describes the large current limit.   
Solutions for special parameter values are considered
in Sec.XI, while Sec.XII contains concluding remarks.

There are also three appendices.  Appendices A and B contain the derivation
of the local solutions at the symmetry axis and at infinity. 
Appendix C describes the  superconducting strings in Witten's model.
This Appendix is made self-consistent, but it uses the same methods and 
notation as in the main text, so that it illustrates our procedure 
on a simple example.

A very brief and preliminary summary of our results 
has been announced in Ref.\cite{V07}. 

Recently there has been an intense  activity on the non-Abelian strings 
with gauge group SU(N)$\times$U(1) \cite{QCD}. 
These strings have nothing to do with ours,
since they are obtained within the context of a 
supersymmetric theory that 
is different from the electroweak theory even for 
N=2. They are currentless and are supposed 
to be relevant in the QCD context, as models of gluon  tubes. 
There has also been  a report on current-carrying strings \cite{Verbin}, 
but again in a model 
with a different Higgs sector in which solutions 
taking values in the Cartan subalgebra of SU(N)$\times$U(1) 
are possible.

\section{Weinberg-Salam theory}
\setcounter{equation}{0}

The bosonic part of the Weinberg-Salam (WS) theory is 
described by the action 
\be          \label{0}
{\bf S}=\frac{1}{{\bf c}}\int\left(
-\frac{1}{4}\,{\bf W}^a_{\mu\nu}{\bf W}^{a\mu\nu}
-\frac{1}{4}\,{\bf \F}_{\mu\nu}{\bf \F}^{\mu\nu}
+({\bf D}_\mu \mbox{\boldmath $\Phi$})^\dagger {\bf D}^\mu 
\mbox{\boldmath $\Phi$}
-\mbox{\boldmath $\lambda$}\left(
\mbox{\boldmath $\Phi$}^\dagger\mbox{\boldmath $\Phi$} 
-\mbox{\boldmath $\Phi$}_0^2
%-\Phi_0^2
\right)^2
\right)d^4{\bf x}.
\ee
Here the complex Higgs field \mbox{\boldmath $\Phi$} is in the fundamental 
representation of SU(2) and 
\begin{align}          \label{bold}
{\bf W}^a_{\mu\nu}&={\boldmath \partial}_\mu{\bf W}^a_\nu
-\partial_\nu {\bf W}^a_\mu
+{\bf g}\epsilon_{abc}{\bf W}^b_\mu{\bf W}^c_\nu\, ,\nonumber \\
{\bf \F}_{\mu\nu}&=\partial_\mu{\bf \A}_\nu
-\partial_\nu{\bf \A}_\mu\, ,
\nonumber  \\
{\bf D}_\mu\mbox{\boldmath $\Phi$}
&=\left(\partial_\mu-\frac{i{\bf g}'}{2}\,{\bf \A}_\mu
-\frac{i{\bf g}}{2}\,\tau^a {\bf W}^a_\mu\right)\mbox{\boldmath $\Phi$},
\end{align}
where $\tau^a$ are the Pauli matrices and ${\bf c}$ in \eqref{0} is the 
speed of light. 
The SU(2) gauge field ${\bf W}^a_\nu$ and the U(1) field ${\bf \A}_\nu$
are sometimes called isospin and hypercharge 
fields, respectively. Here and below we denote dimensionfull quantities
by boldfaced symbols. In order to pass to dimensionless variables, 
let us first introduce the dimensionless coupling constants 
\begin{align}
g=\frac{\bf g}{{\bf g}_0}\equiv \cos\thetaw,~~~~~
g^\prime=\frac{{\bf g}^\prime}{{\bf g}_0}\equiv \sin\thetaw,  \label{scale0}
\end{align}
where the physical value of the Weinberg angle is 
\be
g^{\prime 2}=\sin^2\thetaw=0.23\ldots\, 
\ee
The parameter 
\be                \label{g0}
{\bf g}_0=\sqrt{{\bf g}^2+{\bf g}^{\prime 2}}
\ee
is related to the electron charge via 
$
{\bf e}=gg^\prime{\bf \hbar c}\, {\bf g}_0
$
so that its dimension is $[{\bf g}_0]=[1/{\bf e}]$ and the value 
\be
\frac{\bf e^2}{4\pi\hbar \bf c}=\frac{\hbar {\bf c}}{4\pi}\,{(gg^\prime{\bf g}_0)^2}
=\frac{1}{137}.
\ee
The Higgs field vacuum expectation value 
$\mbox{\boldmath $\Phi$}_0$
has the dimension $[\mbox{\boldmath $\Phi$}_0]=[{\bf e/L}]$ and the value 
\be                             \label{volts}
\mbox{\boldmath $\Phi$}_0=54.26\times 10^9~{\rm Volts}.
\ee
%(in units where ${\bf \hbar=c}=1$ the ratio $\mbox{\boldmath $\Phi$}_0/{\bf e}$ 
%corresponds to $246/\sqrt{2}$ GeV). 
Introducing the length scale
$
{\bf L}={1}/{{\bf g}_0\mbox{\boldmath $\Phi$}_0}=1.52\times 10^{-18}
$~metres ($\sqrt{2}{\bf L}$ is the Z boson Compton length)
one can define the dimensionless 
variables 
\begin{align}                      \label{scale}
x^\mu&={\bf g}_0\mbox{\boldmath $\Phi$}_0\,{\bf x}^\mu,~~~~
{\A}_\mu={g}^\prime\,\frac{{\bf \A}_\mu}{\mbox{\boldmath $\Phi$}_0}\,,~~~~
\WW^a_\mu=g\,\frac{{\bf W}^a_\mu}{\mbox{\boldmath $\Phi$}_0}\,,~~~~
\Phi=\frac{\mbox{\boldmath $\Phi$}}{\mbox{\boldmath $\Phi$}_0}\equiv
\left(\begin{array}{c}
\phi_1 \\
\phi_2
\end{array}\right),~~~~\beta=\frac{8\mbox{\boldmath $\lambda$}}{{\bf g}_0^2},
\end{align}
such that the action \eqref{0} becomes 
\be                                     \label{00}
{\bf S}=\frac{1}{{\bf c\,g}_0^2} \int {\cal L}\, d^4 x. 
\ee
Here
\be                             \label{L}
{\cal L}=
-\frac{1}{4g^2}\,\WW^a_{\mu\nu}\WW^{a\mu\nu}
-\frac{1}{4g^{\prime 2}}\,{\F}_{\mu\nu}{\F}^{\mu\nu}
+(D_\mu\Phi)^\dagger D^\mu\Phi
-\frac{\beta}{8}\left(\Phi^\dagger\Phi-1\right)^2,
\ee
and  
\begin{align}
\WW^a_{\mu\nu}&=\partial_\mu\WW^a_\nu
-\partial_\nu \WW^a_\mu
+\epsilon_{abc}\WW^b_\mu\WW^c_\nu\, ,\nonumber \\
{\F}_{\mu\nu}&=\partial_\mu{\A}_\nu
-\partial_\nu{\A}_\mu\, ,
\nonumber  \\
D_\mu\Phi
&=\left(\partial_\mu-\frac{i}{2}\,{\A}_\mu
-\frac{i}{2}\,\tau^a \WW^a_\mu\right)\Phi\,. \label{unbold} 
\end{align}
The Lagrangian \eqref{L} is invariant under the  
SU(2)$\times$U(1) gauge transformations
\be                               \label{gauge}
\Phi\to {\rm U}\Phi,~~~~~~~~
{\cal W}\to {\rm U}{\cal W}{\rm U}^{-1}
+2i{\rm U}\partial_\mu {\rm U}^{-1}dx^\mu\,,
\ee
with 
\be                            \label{U}
{\rm U}=\exp\left(\frac{i}{2}\,\Theta+\frac{i}{2}\,\tau^a\theta^a\right)
\ee
where $\Theta$ and $\theta^a$ are functions of $x^\mu$
and 
\be                          \label{alg}
{\cal W}=
(B_\mu+\tau^a\WW^a_\mu)dx^\mu 
\ee
is the SU(2)$\times$U(1) Lie-algebra valued gauge field. 
Varying the action with respect to the fields 
gives the field equations,
\begin{align}
\partial^\mu {B}_{\mu\nu}&=g^{\prime 2}\,\frac{i}{2}\,
((D_\nu\Phi)^\dagger\Phi
-
\Phi^\dagger D_\nu\Phi
)\equiv g^{\prime 2}J^{0}_\nu, \label{P0}\\
D^\mu \WW^a_{\mu\nu}
&=g^{2}\,\frac{i}{2}\,
(
(D_\nu\Phi)^\dagger\tau^a\Phi
-\Phi^\dagger\tau^a D_\nu\Phi
)
\equiv g^{2} J^{a}_{\nu}, \label{P1}\\
D_\mu D^\mu\Phi&+\frac{\beta}{4}\,(\Phi^\dagger\Phi-1)\Phi=0,      \label{P2}
\end{align}
with $D_\mu\WW^a_{\alpha\beta}=\partial_\mu \WW^a_{\alpha\beta}
+\epsilon_{abc}\WW^b_\mu\WW^c_{\alpha\beta}$. 
Varying the action with respect to the spacetime metric gives the energy momentum tensor, 
\be                                    \label{T}
T^\mu_{\phantom\mu\nu}=2g^{\mu\sigma}\frac{\partial\mathcal{L}}
{\partial g^{\sigma\nu}}
-\delta^\mu_{\phantom\mu\nu}\mathcal{L}\,, 
\ee
which 
 evaluates to 
\be                      \label{TT}
T^\mu_{\phantom\mu\nu}=
-\frac{1}{g^2}\,\WW^{a\mu\sigma}\WW^a_{\nu\sigma}
-\frac{1}{g^{\prime\,2}}B^{\mu\sigma}B_{\nu\sigma}
+(D^\mu\Phi)^\dagger D_\nu\Phi
+(D_\nu\Phi)^\dagger D^\mu\Phi-\delta^\mu_{\phantom\mu\nu}\mathcal{L}\,.
\ee
Let us define {\it vacuum} as 
\be
\WW^a_\mu=\A_\mu=0,~~~~~ 
\Phi=\left(\begin{array}{c}
1 \\
0
\end{array}\right).
\ee
Any field with $T^\mu_{\phantom\mu\nu}=0$ is a gauge transformation of the vacuum.
Let us consider small fluctuations around the vacuum 
in the unitary gauge, assuming that $\WW^a_\mu$ and $\A_\mu$
are small and that $\Phi=\left(\begin{array}{c}
1+\phi \\
0
\end{array}\right)$ with $\phi$ real and small.
Linearising the field equations \eqref{P0}--\eqref{P2} with respect to the 
perturbations 
gives 
\begin{align}
\partial_\mu {F}^{\mu\nu}&=0,~~~~~~\nonumber \\
\partial_\mu Z^{\mu\nu}&+\mz^2\, Z^\nu=0,   \nonumber \\
\partial_\mu \WW_a^{\mu\nu}&+\mw^2\,\WW_a^\nu=0~~(a=1,2), \nonumber \\
\partial_\mu\partial^\mu\phi&+\mh^2\,\,\phi=0.
\end{align}
Here 
the electromagnetic and Z fields are defined as
\be                                  \label{az}
F_{\mu\nu}=\frac{g}{g^\prime}\, \A_{\mu\nu}-\frac{g^{\prime}}{g}\,\WW^3_{\mu\nu},~~~~~~
{Z}_{\mu\nu}=\A_{\mu\nu}+\WW^3_{\mu\nu},
\ee 
with the potentials 
\be                                  \label{az1}
A_{\mu}=\frac{g}{g^\prime}\, \A_{\mu}-\frac{g^{\prime}}{g}\,\WW^3_{\mu},~~~~~~
{Z}_{\mu}=\A_{\mu}+\WW^3_{\mu},
\ee 
while  the field masses 
\be                                   \label{masses}
\mz=\frac{1}{\sqrt{2}},~~~~~~
\mw=g\mz,~~~~~~
\mh=\sqrt{\beta}\,\mz.
\ee
 Multiplying these  by 
${\bf\hbar c\, g}_0\mbox{\boldmath $\Phi$}_0=
{\bf e}\mbox{\boldmath $\Phi$}_0/(gg^\prime)$
gives the dimensionfull masses, 
for example 
\be
{\bf m}_{\mbox{\tiny Z}}{\bf c}^2=
{\bf e}\mbox{\boldmath $\Phi$}_0/(\sqrt{2}gg^\prime),
%=52.68/(\sqrt{2}gg^\prime)~ {\rm GeV}
\ee
which evaluates %to $88.5$ GeV. This differs slightly from the 
%true value of the Z boson mass including 	
%the loop corrections, 
$91.18$ GeV \cite{book}. 
%but since our analysis is purely classical,
%we shall ignore this difference and use the classical  values 
%in what follows. 
The value of the parameter $\beta$ defining the Higgs field mass 
is currently unknown, however, there are indications that it belongs
 to the 
interval $1.5\leq\beta\leq 3.5$ \cite{book}.

The electromagnetic and Z fields in \eqref{az} are defined under the 
assumption that the system is close to the vacuum.  
In what follows we shall need to define 
these   fields
also off the vacuum.  This can be done in a 
gauge invariant way, however there is no {\it unique}
definition in this case \cite{Coleman},
although all known 
definitions reduce to \eqref{az} when the fields approach the vacuum.   
Unless the otherwise is stated, we shall 
adopt   the definition 
of Nambu \cite{Nambu},
\be                                  \label{Nambu}
F_{\mu\nu}=\frac{g}{g^\prime}\,  
\A_{\mu\nu}-\frac{g^{\prime}}{g}\,n^a\WW^a_{\mu\nu}\,,~~~~~~
{Z}_{\mu\nu}=\A_{\mu\nu}+n^a\WW^a_{\mu\nu}\,,
\ee
where 
%$\xi=\Phi/\sqrt{\Phi^\dagger\Phi}$ one can define 
\be						\label{n}
n^a=\xi^\dagger\tau^a\xi,~~~~~~~~~~~
\xi=\Phi/\sqrt{\Phi^\dagger\Phi}.
\ee
It seems that this definition 
gives the most sensible results as compared to the 
other definitions. 
Its only disadvantage is that the 2-forms \eqref{Nambu} are 
not closed in general, so that there are no field potentials.  
However, this cannot be considered as a drawback, since there is no reason 
why the Maxwell equations should hold off the Higgs vacuum. 
It is worth noting that a similar 
preference in favour of the Nambu definition     
was made also in Ref. \cite{Hindmarsh-James}. 

Another well known definition of the fields is according to 't Hooft 
 \cite{hoft}, 
\be                                  \label{Hoft}
F^{\rm H}_{\mu\nu}=\frac{g}{g^\prime}\, \A_{\mu\nu}-\frac{g^{\prime}}{g}\,
{\cal V}_{\mu\nu}\, ,~~~~~~
{Z}^{\rm H}_{\mu\nu}=\A_{\mu\nu}+{\cal V}_{\mu\nu}\,
\ee
where
\be						\label{calV}
{\cal V}_{\mu\nu}=n^a\WW^a_{\mu\nu}-\epsilon_{abc}n^a D_\mu n^b D_\nu n^c
=\,\partial_\mu{\cal V}_\nu-\partial_\nu{\cal V}_\mu\,
\ee
with
\be						\label{Dn}
{\cal V}_\mu=n^a{\rm W}^a_\mu+2i\xi^\dagger\partial_\mu\xi\,,~~~~~
D_\mu n^a=\partial_\mu n^a+\epsilon_{abc}\WW^b_\mu n^c.
\ee 
The field strengths are closed in this case and the  
potentials are known explicitly.   
In the unitary gauge where $n^a=\delta^a_3$
the potentials
reduce simply to 
${A}_\mu=\frac{g}{g^\prime}\,  \A_\mu-\frac{g^{\prime}}{g}\,\WW^3_\mu$
and 
${Z}_\mu=\A_\mu+\WW^3_\mu$
in which form they are most often used (see, e.g., \cite{period}). 
However, as we shall see below, the t' Hooft definition gives sometimes 
singular results, while applying instead the Nambu 
definition makes things 
perfectly sensible.

\section{Symmetry reduction}

\setcounter{equation}{0}

In what follows we shall be considering stationary and translationally 
symmetric fields invariant under the action of two 
spacetime isometries generated by the Killing vectors 
\be                      \label{Kil}
K_{(0)}=\frac{\partial}{\partial x^0},~~~~~~~~~
K_{(3)}=\frac{\partial}{\partial x^3}.
\ee
%Such fields could describe vortices parallel to the $x^3$-axis. 
Since these
vectors commute and all the internal symmetries of the theory are gauged, 
there exists a gauge where the symmetric fields do not depend
on $x^0\equiv t$ and  $x^3\equiv z$ \cite{Forgacs}. 

Let  $\sigma_\alpha$ be a constant 
(co)vector in the $(x^0,x^3)$ plane and 
$\tilde{\sigma}_\alpha=\epsilon_{\alpha\beta}\sigma^\beta$ its dual.  
These two vectors are orthogonal and have the norm
\be
\sigma^2\equiv 
-\sigma^\alpha\sigma_\alpha=
+\tilde{\sigma}^\alpha\tilde{\sigma}_\alpha=
-(\sigma_0)^2+(\sigma_3)^2\,.
\ee
Using these two vectors, the fields respecting
the symmetries \eqref{Kil} can be written as
\begin{align}                           \label{03}
{\cal W}=&[\YY\sigma_\alpha +\tilde{\YY}\tilde{\sigma}_\alpha ]
dx^\alpha+\A_i dx^i              
+\tau^a
[\OOmega_a\sigma_\alpha +\tilde{\OOmega}_a\tilde{\sigma}_\alpha ]
dx^\alpha+ \tau^a\WW^a_i dx^i %\right\}
,~~~
\Phi=\left(\begin{array}{c}
\phi_1 \\
\phi_2
\end{array}\right),          
\end{align}
where $\alpha=0,3$ and $i,k=1,2$, while
$u,\tilde{u},u^a,\tilde{u}^a,B_i,\WW^a_i,\phi_1,\phi_2$
depend on $x^k$, so that there are in total 20 independent real functions.   
The ansatz (\ref{03}) keeps its form 
under gauge transformations (\ref{gauge}) with $\Theta,\theta^a$ 
depending only on $x^k$. 
%As a result, one can impose on 20 functions in (\ref{03}) 
%four gauge conditions by
%adjusting four  functions $\Theta,\theta^a$. 
%The conditions that we have use below are 
%\be                          \label{radial} 
%x_k A_k=0,~~~~x_k\WW^a_k=0\, . 
%\ee

Inserting this ansatz into \eqref{L} gives the reduced Lagrangian 
\bea                           \label{LL}
{\cal L}&=&-\left.\left.\frac{\sigma^2}{2g^{\prime\,2}}
\right( (\partial_k \YY)^2
-(\partial_k \tilde{\YY})^2\right)
-
\left.\left.\frac{\sigma^2}{2g^{2}}
\right((D_k\OOmega_a)^2
-(D_k \tilde{\OOmega}_a)^2 \right) \nonumber \\
&+&\left.\left.\frac{(\sigma^2)^2}{2g^{2}}
\right((\OOmega_a)^2(\tilde{\OOmega}_a)^2
-(\OOmega_a\tilde{\OOmega}_a)^2\right)
-\frac{1}{4 g^{\prime\,2}}\,(\F_{ik})^2
-\frac{1}{4 g^{2}}\,(\WW^a_{ik})^2 \nonumber \\
&-&|D_k\Phi|^2-
\left.\left.\frac{\sigma^2}{4}
\right(\Phi^\dagger(\YY+\OOmega_a\tau^a)^2\Phi
-
\Phi^\dagger(\tilde{\YY}+\tilde{\OOmega}_a\tau^a)^2\Phi
\right) 
-\frac{\beta}{8}\,(\Phi^\dagger\Phi-1)^2
\eea
where
$D_k\OOmega_a=\partial_k\OOmega_a+\epsilon_{abc}\WW^b_k\OOmega_c$
and $D_k\Phi=(\partial_k-\frac{i}{2}\,\A_k-\frac{i}{2}\,\tau^a\WW^a_k)\Phi$.
%the two-dimensional indices $i,k$ are raised and lowered 
%with the metric $g_{ik}=\delta_{ik}$ (if they refer to the Cartesian coordinates). 
Varying this gives 
the field equations, where, as is evident from the structure of 
the reduced Lagrangian, 
one can consistently set 
\be                         \label{YY}
\tilde{\YY}=\tilde{\OOmega}^a=0\,.
\ee
The remaining non-trivial equations then read 
\begin{subequations}   \label{eee}
\begin{align}
\partial_s\partial_s\, \YY&=
\frac{g^{\prime\,2}}{2}\,\Phi^\dagger(\YY+\OOmega_a\tau^a)\Phi\,,\label{e1}\\
D_s D_s\, \OOmega_a&=
\frac{g^{2}}{2}\,\Phi^\dagger(\OOmega_a+\YY\,\tau^a)\Phi\, ,\label{e2} \\
\partial_s \F_{sk}&=i\,\frac{g^{\prime\,2}}{2}
(\Phi^\dagger D_k\Phi-(D_k\Phi)^\dagger\Phi)\,,\label{e3} \\
D_s \WW^a_{sk}-\sigma^2\epsilon_{abc}\OOmega_b D_k\OOmega_c
&=i\,\frac{g^{2}}{2}
(\Phi^\dagger \tau^a D_k\Phi-(D_k\Phi)^\dagger\tau^a \Phi)\,,\label{e4}\\
D_s D_s\, \Phi-\frac{\sigma^2}{4}\,(\YY+\OOmega_a\tau^a)^2\Phi&=
\frac{\beta}{4}\,(\Phi^\dagger\Phi-1)\Phi \, . \label{e5}
\end{align}
\end{subequations}
Let us discuss the global   quantities associated 
to solutions of these equations -- 
the magnetic fluxes, charge and current,  
energy and momentum.

\subsection{Fluxes}
The electromagnetic and Z fields produce 
magnetic fluxes through the $x^1,x^2$ plane,  
\be                                   \label{fluxes}
\Psi_F=
\frac{1}{2} \int\epsilon_{ik}{F}_{ik}\, d^2{x}
\,,~~~~~~~~               
%\\
\Psi_{Z}
=\frac12 \int\epsilon_{ik}{Z}_{ik}\, d^2{x}\,.           % \label{fluxZ}
\ee
The dimensionfull values are 
$\mbox{\boldmath $\Psi$}_F
=\frac12 \int\epsilon^{ik}{\bf F}_{ik}\, d^2{\bf x}=\Psi_F/{{\bf g}_0}$
and $\mbox{\boldmath $\Psi$}_Z=\Psi_Z/{{\bf g}_0}$. 
%
%$\mbox{\boldmath $\Psi$}_Z
%=\frac12 \int\epsilon^{ik}{\bf Z}_{ik}\, d^2{\bf x}$
%are expressed as 
%$\mbox{\boldmath $\Psi$}_F=\Psi_F/{{\bf g}_0}$
%and $\mbox{\boldmath $\Psi$}_Z=\Psi_Z/{{\bf g}_0}$. 
The fluxes are gauge invariant, but their values depend 
on the definition of 
 $F_{\mu\nu}$ and $Z_{\mu\nu}$ chosen. 
If the fields  are defined according to 't Hooft, 
then they admit potentials, in which case one can use the Stokes theorem to 
express the fluxes in terms of integrals over the boundary at infinity.
%which implies that fluxes are {\it conserved}. 
With the Nambu definition one should calculate the surface integrals.

In general there is no reason for the fluxes to be quantized. Indeed,
let us briefly recall the standard argument in favour of the vortex flux quantization. 
Far away from the vortex the condition  
\be
(\partial_\mu-\frac{i}{2}\,\A_\mu-\frac{i}{2}\,\tau^a\WW^a_\mu)\,\Phi=0
\ee 
should be fulfilled, 
whose solution is 
\be
\Phi(x)={\cal P}\exp{\left(\frac{i}{2}\int_{{\cal C}}
(\A_\mu+\tau^a\WW^a_\mu)dx^\mu
\right)}\Phi(x_0)\,.
\ee
Here ${\cal P}$ stands for the path ordering along a contour
${\cal C}$
interpolating between $x_0$ and $x$.  Choosing ${\cal C}$ to be a large circle
around the vortex gives the holonomy condition
\be                                             \label{Holonom}
\Phi(x_0)={\cal P}\exp{\left(\frac{i}{2}\oint_{{\cal C}}
(\A_\mu+\tau^a\WW^a_\mu)d x^\mu
\right)}\Phi(x_0).
\ee
If $\WW^a_\mu=0$, as
for example in the semilocal limit $g\to 0$ (see Sec.\ref{sem} below), then the condition 
\eqref{Holonom} reduces to 
\be
\frac{1}{2}\int_{0}^{2\pi}\A_\varphi\, d\varphi =2\pi\n\,
\ee
and so the hypercharge field flux is quantized, $\Psi_B=4\pi\n$.

Let us now  
suppose that $\WW^1_\mu=\WW^2_\mu=0$ but 
$\WW^3_\mu\neq 0$ and let us denote $\Psi_{W}=
\int_{0}^{2\pi}\WW_\varphi\, d\varphi$. 
The condition \eqref{Holonom} then gives 
\be
\phi_1(x_0)=e^{\frac{i}{2}(\Psi_B+\Psi_W)}\phi_1(x_0),~~~~~
\phi_2(x_0)=e^{\frac{i}{2}(\Psi_B-\Psi_W)}\phi_2(x_0)\,,
\ee
and if, for example, $\phi_2(x_0)=0$, then one should have $\Psi_B+\Psi_W=4\pi\n$
 but the difference 
$\Psi_B-\Psi_W$ can be arbitrary.
The fluxes are therefore not quantized in general. 
%In addition, depending on their definition, the fields $F_{\mu\nu}$
%and $Z_{\mu\nu}$ may even have no potentials, in which case the 
%holonomy condition \eqref{Holonom} becomes completely irrelevant. 

\subsection{Current}

The conserved electromagnetic current is
\be					\label{cur1}
J_\nu=\partial^\mu F_{\mu\nu}.
\ee 
Since nothing depends on $x^\alpha=(x^0,x^3)$, one has 
$J_\alpha=-\partial_k F_{k\alpha}.$
This being a total derivative, 
integrating over the $x^1,x^2$ plane gives  
\be                         \label{cur2}
I_\alpha=\int J_\alpha\, d^2x=-\int \partial_k F_{k\alpha}\, d^2 x=
-\oint \epsilon_{ks}F_{k\alpha}\, dx^s\,.
\ee
Here $I_0$ is the total electric charge per unit length of the vortex 
and $I_3$ is the total electric current through the $x^1x^2$ plane.  
Since the integration in \eqref{cur2} is performed over the boundary 
at infinity, in the region where all  fields approach the vacuum and 
the electromagnetic field is uniquely defined, the value of $I_\alpha$
does not depend on the chosen definition of $F_{\mu\nu}$. 
Let us consider 
the dimensionfull version of \eqref{cur1},
\be					\label{cur3}
\frac{1}{\bf c}\,{\bf J}_\nu=\partial^\mu {\bf F}_{\mu\nu},
\ee 
where ${\bf F}_{\mu\nu}={\bf g}_0\mbox{\boldmath $\Phi$}_0^2 F_{\mu\nu}$ and 
the derivative is taken with respect to 
${\bf x}^\mu=x^\mu/{{\bf g}_0\mbox{\boldmath $\Phi$}_0}$.
One has 
\be                         \label{cur4}
\int {\bf J}_\alpha\, d^2{\bf x}=
-{\bf c}\int {\boldmath\partial}_k {\bf F}_{k\alpha}\, d^2 {\bf x}
=-{\bf c}\mbox{\boldmath $\Phi$}_0\int \partial_k F_{k\alpha}\, d^2 x=
{\bf c}\mbox{\boldmath $\Phi$}_0 I_\alpha,
\ee
from where it follows that $I_\alpha$ gives the current in units of
\be                             \label{Amperes}
{\bf c}\mbox{\boldmath $\Phi$}_0=
{\bf c}\times 54.26\times 10^9~{\rm Volts}=
1.8\times 10^9~{\rm Amperes}. 
\ee
This value is quite large. Below we shall present vortex solutions 
whose current is typically $I_3\sim 1-10$,  
which looks modest but corresponds in fact 
to billions of Amperes ! Very large currents are typical for superconducting
strings. In the GUT-related model of Witten the current can  be as large as 
$10^{20}$ Amperes \cite{Witten},\cite{SS} -- 
because the GUT Higgs field vacuum expectation value is much 
larger than $\mbox{\boldmath $\Phi$}_0$.

\subsection{Energy and momentum}
Contracting the Killing vectors with the energy-momentum tensor \eqref{TT} 
gives conserved Noether currents,
\be
j^\mu_{(K)}=T^\mu_{~\nu} K^\nu~~~~~~\Rightarrow~~~~~
\partial_\mu j^\mu_{(K)}=0.
\ee
Integrating over the $x^1,x^2$ plane
gives the conserved charges per unit vortex length -- the vortex 
energy and momentum densities,
\begin{align}
E&=\int T^0_{~\nu} K^\nu_{(0)}d^2x=\int T^0_{~0} d^2x,  \label{E} \\
P&=\int T^0_{~\nu} K^\nu_{(3)}d^2x=\int T^0_{~3} d^2x\,.   \label{P}
\end{align}
The dimensionfull values are ${\bf E}=\mbox{\boldmath $\Phi$}_0^2\, E$
and ${\bf P}=\mbox{\boldmath $\Phi$}_0^2\, P/{\bf c}$. 

Let us consider the momentum. According to Ref. \cite{VW},
in gauge field theories without global internal symmetries 
the density $j^0_{(K)}$ for 
a spacelike Killing vectors $K$ has the
total derivative structure. 
Calculating the energy-momentum tensor 
\eqref{TT} for the field \eqref{03} shows that $T^0_{~3}$ is indeed a total
derivative, 
\begin{align}                      \label{T03}
T^0_{~3}&=
-\frac{1}{g^2}\,\WW^{a 0 \nu}\WW^a_{3\nu}
-\frac{1}{g^{\prime\,2}}B^{0\nu}B_{3\nu}
+(D^0\Phi)^\dagger D_3\Phi
+(D_3\Phi)^\dagger D^0\Phi  \nonumber \\
&=\sigma_0\sigma_3\, \partial_k\Xi_k
\end{align}
with
\begin{align}
\partial_k\Xi_k&=\frac{1}{g^{\prime 2}}\,(\partial_k\YY)^2+
\frac{1}{g^{2}}\,(D_k\OOmega_a)^2+\frac12\,
\Phi^\dagger(\YY+\OOmega_a\tau^a)^2\Phi \nonumber \\
&=\partial_k\left(\frac{1}{g^{\prime 2}}\,\YY\partial_k\YY+
\frac{1}{g^{2}}\,\OOmega^a D_k\OOmega^a\right),  \label{Xi}
\end{align}
where the field equations \eqref{e1}--\eqref{e5} have been used 
to pass  to the 
second line. For globally regular fields the total momentum is therefore 
completely determined by the boundary term, 
\be
P=\sigma_0\sigma_3
\oint \epsilon_{ks}\Xi_k dx^s\,.
\ee

Let us now consider the energy. 
It is given by 
a sum of two manifestly positive terms which we shall call, respectively,
electric and magnetic energy,
\be
E=\int T^0_{~0}\, d^2x=\int({\cal E}_1+{\cal E}_2)\, d^2x\,
\equiv E_1+E_2 .
\ee
The structure of ${\cal E}_1$ is very similar to that of $T^0_3$ in \eqref{T03}, 
\be                                \label{EN1}
{\cal E}_1=\frac{\sigma_0^2+\sigma_3^2}{2}\,
\partial_k\Xi_k,
\ee
so that it is a total derivative. 
The magnetic energy is 
\be                                  \label{EN2}
{\cal E}_2=\frac{1}{4g^2}\,(\WW^a_{ik})^2
+\frac{1}{4g^{\prime\,2}}(\F_{ik})^2+
|D_s\Phi|^2+\frac{\beta}{8}\,(|\Phi|^2-1)^2.
\ee

\subsection{Bogomol'nyi bound and the electroweak condensate  \label{amjorn}}

It is instructive 
to mention the existence of the non-trivial lower bound for the 
magnetic energy, observed by Ambjorn and Olesen \cite{period}. 
Using the identity 
\be\label{bogo-id}
|D_s\Phi|^2=
\frac{1}{2}\,|D_i\Phi+ i\epsilon_{ij}D_j\Phi|^2
+\frac{1}{4}\,\epsilon_{ik}\Phi^\dagger(\F_{ik}
+\tau^a\WW^a_{ik})\Phi
-i\epsilon_{sk}\partial_s (\Phi^\dagger D_k\Phi)\, 
\end{equation}
the density ${\cal E}_2$ can be rewritten as
\bea
{\cal E}_2&=&\frac{1}{4g^2}
\left(\WW^a_{ik}+\frac{g^2}{2}\,\epsilon_{ik}\,\Phi^\dagger\tau^a\Phi
\right)^2
+
\frac{1}{4g^{\prime\,2}}
\left(\F_{ik}+\frac12\,\epsilon_{ik}
\left(g^{\prime\,2}\Phi^\dagger\Phi-1\right)\right)^2
\nonumber \\
&+&\frac{1}{2}\,|D_i\Phi+ i\epsilon_{ij}D_j\Phi|^2
+\frac{\beta-1}{2}\,(\Phi^\dagger\Phi-1)^2 \nonumber \\
&+&\frac{1}{4g^{\prime\,2}}\,\epsilon_{ik}\F_{ik}
-i\epsilon_{sk}\partial_s(\Phi^\dagger D_k\Phi)
-\frac{g^2}{8g^{\prime\,2}}\,.
\eea
From this it follows that for $\beta\geq 1$ one has 
\be						\label{bound}
E_2\geq 
\int(\frac{1}{4g^{\prime\,2}}\,\epsilon_{ik}\F_{ik}
-i\epsilon_{sk}\,\partial_s(\Phi^\dagger D_k\Phi)
-\frac{g^2}{8g^{\prime\,2}})d^2x,
\ee
which defines the lower energy bound. This bound is, however, somewhat 
special. Considering just one vortex and 
integrating over the whole plane, the first term in the integrand in
\eqref{bound}
gives
the magnetic flux, the second term gives zero, while the third one gives
the infinite area of the plane with a negative coefficient. The conclusion
is that $E_2$ is always 
larger than minus infinity, which is true but trivial.

A less trivial conclusion can be made if one considers  
an infinite number of parallel vortices 
forming  a periodic lattice  in the $x^1,x^2$ plane,
since the field $\F_{ik}$ then does not vanish at infinity and its 
contribution can overcome the effect of the area term.  
It is then sufficient to integrate just 
over the elementary periodicity cell with the area ${\cal A}$.
One uses the relation $B_{ik}=gg^\prime F_{ik}+g^{\prime 2}Z_{ik}$ where 
$Z_{ik}$ does not contribute to the flux  since its potentials  
(defined in \cite{period} according to t' Hooft)
are gauge invariant and therefore periodic if $Z_{ik}$ is periodic \cite{period}.  
The bound then reads 
\be                     \label{eB1}
E_2=\int_{{\cal A}}{\cal E}_2\, d^2 x\geq \frac{g}{2g^{\prime}}\,
\Psi_{F}
-\frac{g^2}{8g^{\prime\,2}}{\cal A}=
\frac{\mw^2}{e}(\Psi_F-\frac{\mw^2}{2e}{\cal A})
\, 
\ee
where $\Psi_{F}$ is the flux of $F_{ik}$ through ${\cal A}$. 
If
the system contains only a constant magnetic field $F_{12}>{\mw^2}/{e}$
then this relation gives 
\be
E_2=\frac12\,(F_{12})^2{\cal A}\geq 
\frac{\mw^2}{e}(F_{12}-\frac{\mw^2}{2e}){\cal A}>0.
\ee
This shows that for large $F_{12}$ the system becomes unstable, 
since its energy grows as $(F_{12})^2$ while the 
lower energy bound grows only as $F_{12}$. 
The bound can be achieved for $\beta=1$ if
the first order 
Bogomol'nyi equations are fulfilled,
\begin{subequations}    \label{bogom}
\begin{align}  \label{b1}
\WW^a_{ik}+\frac{g^2}{2}\,\epsilon_{ik}\,\Phi^\dagger\tau^a\Phi&=0\,, \\
\F_{ik}+\frac12\,\epsilon_{ik}
\left(g^{\prime\,2}\Phi^\dagger\Phi-1\right)&=0\,, \label{b2}\\
D_i\Phi+ i\epsilon_{ij}D_j\Phi&=0\,.                   \label{b3}
\end{align}
\end{subequations}
Periodic solutions of these equations describe a non-linear condensate of 
W,Z and Higgs fields forming an infinite number of non-Abelian vortices that
spontaneously appear in the very strong magnetic field \cite{period}.
It turns out that non-trivial solutions exist if only \cite{period}
\be                                        \label{F12}
\frac{\mw^2}{e}<\langle F_{12}\rangle <\frac{\mh^2}{e}
\ee
where $\langle F_{12}\rangle$ is the magnetic field averaged over ${\cal A}$. 
If the lower inequality here is violated, then the Higgs field falls to the
vacuum, while in the opposite limit it is driven to zero
\cite{period}. This can be interpreted as the electroweak symmetry 
restoration in the very strong magnetic field.

Since the magnetic fields satisfying the conditions 
\eqref{F12} are extremely strong, it was suggested to use the Witten 
superconducting 
strings of the GUT origin as their source \cite{period}. 
 At certain distances 
from the string core the magnetic field falls into the interval \eqref{F12},
so that an electroweak condensate can form within a cylindrical shell  
around the string.  
Such systems are sometimes called `W-dressed superconducting strings' 
\cite{period}, \cite{Wcon}.

Below we shall construct 
electroweak vortices showing  
a compact superconducting core embedded into a symmetric phase region that is 
wrapped in a cylindrical shell where the Higgs field   
interpolates between the symmetric and broken phases.  This reminds very much of the 
`W-dressed superconducting strings', up to the fact  that our 
 strings are of purely electroweak origin and have nothing to do
with the GUT physics. 
They can exist for any value of $\beta$, 
which is why the Bogomol'nyi 
equations are of little use for us.

The above arguments go differently in the 
semilocal limit, where $g=0$ (see Sec.\ref{sem}), since 
 the area term in \eqref{eB1} then 
vanishes and the magnetic flux through the whole plane 
is no longer constrained to be infinite. 
The flux
is then quantized as  
$\Psi_B=4\pi n $ and the energy 
lower bound is 
$2\pi n $. Eq.\eqref{b1} implies then that 
$\WW^a_{ik}=0$ so that the Yang-Mills field 
can be gauged away, while the 
remaining Bogomol'nyi equations \eqref{b2},\eqref{b3} admit 
vortex solutions
for any given $n\in\mathbb{Z}$ \cite{achuc1}, \cite{skyrmions}.

%It is unclear whether there exists another set
%of Bogomol'nyi equations whose solutions would have giving finite energy
%non-periodic solutions. 

%As was discussed above, the electroweak solutions carry two 
%different gauge-invariant
%fluxes and not just one.  It would then be natural to try and minimize
%the energy keeping both of them fixed and also 
%try using another variables, for example the ad joint variables discussed
%above. Although we have not been able to find anything new in this way,
%it is instructive to see what happens within the adjoint variable %%
%description. 

\section{Axial symmetry} 

\setcounter{equation}{0}

Let us now further restrict our consideration to 
fields which are invariant with respect to 
rotations in the $x^1,x^2$ plane. 
This means that in addition to the two Killing vectors \eqref{Kil}
there exists a third one. 
Choosing the polar coordinates 
in the  plane, $x^1=\rho\cos\varphi$,  $x^2=\rho\sin\varphi$,
it is given by 
\be
K_{(\varphi)}=\frac{\partial}{\partial\varphi}\,.
\ee
Since this vector commutes with the other two, 
for fields respecting all three symmetries at the same time 
there exists 
a gauge where they  do not depend neither on $x^\alpha$ nor on $\varphi$. 
They are then given by Eq.\eqref{03} with all the functions 
depending only on $\rho$. %Let us call this gauge `gauge I'. 
The field equations \eqref{eee}
become in this case a system of 16 ordinary differential equations
for the 16 independent functions left in the ansatz \eqref{03}
after imposing the condition \eqref{YY}. 

Within the gauge chosen one can still 
perform gauge transformations \eqref{gauge} with $\Theta,\theta^a$ 
depending only on $\rho$.
We use this freedom to set to zero the radial components of the gauge 
fields,
\be                           \label{radial}
\A_\rho=0,~~~~\WW^a_\rho=0.
\ee
This reduces the number of independent functions to 12. 
Now, the inspection of the field equations reveals that one can 
 consistently set to zero the imaginary components of the fields.
These include the $\tau^2$ projection of the Yang-Mills field 
and also the imaginary part of the Higgs field. We can therefore set on-shell
\be                           \label{2}
\WW^2_\mu=0,~~~\Im(\Phi)=0.
\ee   
As a result, we are left with only 8 independent real functions, in which case 
we can parametrize the remaining non-trivial components in  
\eqref{03}
as 
\begin{align}                           \label{003}
{\cal W}&=\Y(\rho)\,\om_\alpha dx^\alpha -
\Z(\rho)\,d\varphi               
+
{\tau}^1\,
[\Om_1(\rho)\,\om_\alpha dx^\alpha - \W_1(\rho)\, d\varphi] \nonumber \\
&+
\tau^3\,
[\Om_3(\rho)\,\om_\alpha dx^\alpha - \W_3(\rho)\, d\varphi],
~~~~~~~~
\Phi=\left(\begin{array}{c}
\f(\rho) \\
\p(\rho)
\end{array}\right),         
\end{align}
where $\alpha=0,3$ and $\f,\p$ are real. 
This ansatz has the following properties. 

{\bf a.} It is invariant under spacetime symmetries generated by 
\be
\frac{\partial}{\partial x^0},~~~
\frac{\partial}{\partial x^3},~~~
\frac{\partial}{\partial \varphi}.
\ee

{\bf b.} It is invariant under complex conjugation,
\be
{\cal W}={\cal W}^\ast,~~~~~\Phi=\Phi^\ast.
\ee

{\bf c.} It keeps its form under Lorentz rotations in the $x^0, x^3$ plane,
whose effect on \eqref{003} is to 
Lorentz-transform the components of the (co)vector $\om_\alpha$:
\be                 \label{s1}
\sigma_0\to\sigma_0\cosh\alpha-\sigma_3\sinh\alpha,~~~~
\sigma_3\to\sigma_3\cosh\alpha-\sigma_0\sinh\alpha,
\ee
where $\alpha$ is the boost parameter. The norm $\sigma^2=\sigma_3^2-\sigma_0^2$
is Lorentz-invariant, and we shall call the vortex  magnetic if $\sigma^2>0$,
electric for $\sigma^2<0$, and chiral if $\sigma^2=0$ \cite{Carter}. 

{\bf d.} It keeps its form under gauge 
transformations \eqref{gauge} generated by  
U=$\exp\{-\frac{i}{2}\Gamma \tau^2\}$ with constant $\Gamma$,
whose effect is to rotate the field amplitudes, 
\begin{align}                            \label{s2}  
\f&\to\f\cos\frac{\Gamma}{2}-\p\sin\frac{\Gamma}{2},~~~~~
\p\to\p\cos\frac{\Gamma}{2}+\f\sin\frac{\Gamma}{2}, \nonumber \\
\Om_1&\to\Om_1\cos\Gamma+\Om_3\sin\Gamma,~~~~~~
\Om_3\to\Om_3\cos\Gamma-\Om_1\sin\Gamma,\nonumber \\
\W_1&\to\W_1\cos\Gamma+\W_3\sin\Gamma,~~~~~~
\W_3\to\W_3\cos\Gamma-\W_1\sin\Gamma,
\end{align}
whereas $\Y,\Z$ rest invariant. These transformations can also 
be written in a compact form using the complex variables:
\be                     \label{comp}
(\f+i\p)\to e^{\frac{i}{2}\Gamma}(\f+i\p),~~~~~
(\Om_1+i\Om_3)\to e^{-i\Gamma}(\Om_1+i\Om_3),~~~~
(\W_1+i\W_3)\to e^{-i\Gamma}(\W_1+i\W_3),~~~~
\ee
This symmetry can be fixed by 
requiring that 
\be                       \label{W10}
\W_1(0)=0.
\ee

{\bf e.} The ansatz does not change if we 
multiply the amplitudes $\Y,\Om_1,\Om_3$ by a constant
dividing at the same time $\sigma_\alpha$ by the same constant.
To fix this symmetry, we impose the condition 
\be                      \label{om3}
\Om_3(0)=1,
\ee
which is possible if $\Om_3(0)\neq 0$.

It should be said that the gauge \eqref{003}
is very convenient for calculations, since 
everything depends only on $\rho$, but 
it
is not completely satisfactory, because,
as we shall see below, the functions $\Z$,  
$\W_1$, $\W_3$ do not vanish at $\rho=0$ and so the vector fields 
are not globally defined in this gauge. 
This problem can be cured by passing to 
another gauge (Eq.\eqref{003a}).

\subsection{Field equations}

With the parametrization \eqref{003} the U(1) equations 
\eqref{e1}, \eqref{e3} reduce to 
\begin{align}                                           \label{ee1}
\frac{1}{\rho}(\rho\Y')'&=\left.\left.\frac{g^{\prime\,2}}{2}
\right\{(\Y+\Om_3)\f^2+2\,\Om_1^{}\f^{}\p^{}+(\Y-\Om_3)\p^2\right\},
%\equiv D_2\Y,
\\
\rho\left(\frac{\Z^\prime}{\rho}\right)^\prime&=\left.\left.\frac{g^{\prime\,2}}{2}
\right\{(\Z+\W_3)\f^2+2\,\W_1^{}\f^{}\p^{}+(\Z-\W_3)\p^2\right\},
%\equiv D_2\Z,  
\label{ee2}  
\end{align}
where the prime denotes differentiation with respect to $\rho$.
% and 
%the abbreviations $D_2\Y,D_2\Z$ 
%have been introduced to designate the polynomial
%expressions in the right hand side of the equations. 
The Higgs equations \eqref{e5} become
\begin{align}                                             \label{ee3}
\frac{1}{\rho}(\rho\f^\prime)^\prime&=
\left\{\frac{\om^2}{4}\left[(\Y+\Om_3)^2+\Om_1^2\right]
+\frac{1}{4\rho^2}\left[(\Z+\W_3)^2+\W_1^2\right]
+\frac{\beta}{4}(\f^2+\p^2-1)
\right\}\f\nonumber \\
&+\left(\frac{\om^2}{2}\,\Y\Om_1+\frac{1}{2\rho^2}\,\Z\W_1\right)\p,
%\equiv D_2\f\,,
\\
\frac{1}{\rho}(\rho\p^\prime)^\prime&=
\left\{\frac{\om^2}{4}\left[(\Y-\Om_3)^2+\Om_1^2\right]
+\frac{1}{4\rho^2}\left[(\Z-\W_3)^2+\W_1^2\right]
+\frac{\beta}{4}(\f^2+\p^2-1)
\right\}\p  \nonumber \\
&+\left(\frac{\om^2}{2}\,\Y\Om_1+\frac{1}{2\rho^2}\,\Z\W_1\right)\f.\label{ee4}
%\equiv D_2\p\,.                                                   
\end{align}
The Yang-Mills equations \eqref{e2},\eqref{e4} reduce to  
\begin{align}
\frac{1}{\rho}(\rho\Om_1^\prime)^\prime
&=
-\frac{1}{\rho^2}\left(\W_1\Om_3-\W_3\Om_1\right)\W_3+\frac{g^2}{2}
\left[\Om_1(\f^2+\p^2)+2\Y\f\p\right],            \label{ee5}   \\
\frac{1}{\rho}(\rho\Om_3^\prime)^\prime
&=
+\frac{1}{\rho^2}\left(\W_1\Om_3-\W_3\Om_1\right)\W_1 + \frac{g^2}{2}
\left[(\Om_3+\Y)\f^2+(\Om_3-\Y)\p^2\right],       \label{ee6}\\
\rho\left(\frac{\W_1^\prime}{\rho}\right)^\prime
&=
+\om^2 \left(\W_1\Om_3-\W_3\Om_1\right)\Om_3+\frac{g^2}{2}
\left[\W_1(\f^2+\p^2)+2\Z\f\p\right],             \label{ee7}  \\
\rho\left(\frac{\W_3^\prime}{\rho}\right)^\prime
&=
-\om^2 \left(\W_1\Om_3-\W_3\Om_1\right)\Om_1+\frac{g^2}{2}
\left[(\W_3+\Z)\f^2+(\W_3-\Z)\p^2\right].         \label{ee8}
\end{align}
In addition, a careful inspection reveals that,
although we have set to zero the radial and the second isotopic components of the 
Yang-Mills field, the Yang-Mills equation \eqref{e4} with $a=2$ and 
$k=\rho$ is not satisfied identically but gives the condition 
\be                                \label{CONS}
\Lambda=0,
\ee
where  
\be                                     \label{CONS1}
{\Lambda}\equiv \sigma^2( \Om_1^{}\Om_3^\prime-\Om_3^{}\Om_1^\prime)
+\frac{1}{\rho^2}\,(\W_1^{}\W_3^\prime-\W_3^{}\W_1^{\prime})-
g^2(\f^{}\p^{\prime}-\p^{}\f^{\prime}).
\ee
Differentiating this quantity and using Eqs.\eqref{ee1}--\eqref{ee8} 
we discover the following relation 
\be                            \label{CON1}
{\Lambda}^\prime +\frac{{\Lambda}}{\rho}=0\,,
\ee
such that
\be                        \label{C}
{\Lambda}=\frac{{C}}{\rho}\,,
\ee
where ${C}$ is an integration constant. 
This is a first integral for the above second order equations, so that  
replacing one of them 
by Eq.\eqref{C} would give a completely equivalent system.   
We thus see that Eq.\eqref{CONS} 
plays a role of constraint restricting the value of 
$C$.

The radial energy density for axially symmetric fields is 
$\rho T^0_0=\rho({\cal E}_1+{\cal E}_2)$, where 
\begin{align}                    
{\cal E}_1&=\frac{\om_0^2+\om_3^2}{2}\left\{
\frac{1}{g^{\prime\,2}}\Y^{\prime\,2}
+\frac{1}{g^2}(\Om_1^{\prime\,2}+\Om_3^{\prime\,2})
+\frac{1}{g^2\rho^2}(\W_1\Om_3-\W_3\Om_1)^2 \right.\nonumber \\
&+\left. \frac{1}{2}\left[
((\Y+\Om_3)\f+\Om_1\p)^2
+
((\Y-\Om_3)\p+\Om_1\f)^2
\right]\right\}               \label{EEE1}
\end{align}
and 
\begin{align}                         
{\cal E}_2&=\frac{1}{2\rho^2}\left(
\frac{1}{g^{\prime\,2}}\Z^{\prime\,2}
+\frac{1}{g^{2}}(\W_1^{\prime\,2}+\W_3^{\prime\,2})\right)
+\f^{\prime 2}+\p^{\prime 2}  \label{EEE2} \\ 
&+
\frac{1}{4\rho^2}\left[
((\Z+\W_3)\f+\W_1\p)^2
+
((\Z-\W_3)\p+\W_1\f)^2
\right]
+\frac{\beta}{8}\,(\f^2+\p^2-1)^2\, .  \nonumber    
\end{align}
Using the above equations  
it is not difficult to see that $\rho{\cal E}_1$ is 
actually a total derivative,
\be                    \label{EEE3}          
\rho{\cal E}_1=\frac{\om_0^2+\om_3^2}{2}\left(
\frac{\rho}{g^{\prime\,2}}\,\Y\Y^\prime+
\frac{\rho}{g^{\,2}}\,(\Om_1\Om_1^\prime+\Om_3\Om_3^\prime)\right)^\prime\,,
\ee
in agreement with Eq.\eqref{EN1}. Next, one has 
\begin{align}
\rho{\cal E}_2&=
\frac{1}{2\rho g^{\prime 2}}\,({\cal B}_1)^2
+\frac{1}{2\rho g^{2}}\,[({\cal B}_2)^2+({\cal B}_3)^2]
+\rho[({\cal B}_4)^2+({\cal B}_5)^2]      
+\rho\,\frac{\beta-1}{8}(\f^2+\p^2-1)^2 \nonumber \\
&+\left(\frac{1}{2g^{\prime 2}}\,\Z-(\Z+\W_3)\frac{\f^2}{2}
-(\Z-\W_3)\frac{\p^2}{2}-\W_1\f\p
-\frac{g^2}{8g^{\prime 2}}\,\rho
\right)^\prime \,            \label{EEE4}
\end{align}
with
\begin{align}                                 \label{BBB}
{\cal B}_1&=\Z^\prime+\frac{\rho}{2}\left[g^{\prime 2}(\f^2+\p^2)-1\right],\nonumber \\
{\cal B}_2&=\W_1^\prime+g^{2}\rho\f\p,           \nonumber \\
{\cal B}_3&=\W_3^\prime+g^2\, \frac{\rho}{2}(\f^2-\p^2),     \nonumber \\
{\cal B}_4&=\f^\prime+\frac{1}{2\rho}\,[(\Z+\W_3)\f+\W_1\p],     \nonumber \\
{\cal B}_5&=\p^\prime+\frac{1}{2\rho}\,[(\Z-\W_3)\p+\W_1\f].
\end{align}
It follows that if $\beta>1$ then 
$\int_0^\infty \rho{\cal E}_2d\rho$ is bounded from below
by the integral of the total derivative in the second line in \eqref{EEE4}. 
The bound is achieved for
$\beta=1$ if
${\cal B}_1=\ldots = {\cal B}_5=0$.
Solutions of these Bogomol'nyi equations
automatically fulfill the five second order equations 
\eqref{ee2}--\eqref{ee4},\eqref{ee7},\eqref{ee8} with $\sigma^2=0$.
However,  unless for $g=0$, the Bogomol'nyi configurations 
are not very interesting, 
since, as was discussed above,  they do not approach the
vacuum at infinity and their energy is infinite. In what follows we shall
rather focus on solutions of 
the second order equations \eqref{ee1}--\eqref{CONS} for generic values
of $\beta$ and $g$.

\section{Boundary conditions}
\setcounter{equation}{0}

In order to construct global solutions of  equations \eqref{ee1}--\eqref{CONS}
in the interval $\rho\in[0,\infty)$ we shall need their local solutions
in the vicinity of the singular points, $\rho=0$ and $\rho=\infty$. 
We shall be considering fields that are regular at $\rho=0$ 
and approach the vacuum for $\rho\to\infty$. 
Let us first consider local solutions at small $\rho$.  

\subsection{Boundary conditions at the 
symmetry axis}

Expressions \eqref{EEE1},\eqref{EEE2} for the energy density 
can be used to derive the regularity conditions 
at $\rho=0$. For regular fields 
the energy density at the axis must be bounded, and so the coefficients 
in front of the negative powers of $\rho$ in \eqref{EEE1},\eqref{EEE2}  
should vanish at $\rho=0$. This implies that at $\rho=0$ one should have 
\be                      \label{Z1}
\Z'=\W_1'=\W_3'=0,
\ee
and also 
\begin{align}                   \label{con}
\W_1\Om_3-\W_3\Om_1&=0, \nonumber \\
(\Z+\W_3)\f+\W_1\p&=0, \nonumber \\
(\Z-\W_3)\p+\W_1\f&=0.
\end{align}
Using Eq.\eqref{W10} these conditions reduce to 
\begin{align}                  \label{con1}
\W_3\Om_1&=0, \nonumber \\
(\Z+\W_3)\f&=0, \nonumber \\
(\Z-\W_3)\p&=0.
\end{align}
Let us now remember that the 
azimuthal components of the vector fields should vanish at the symmetry axis
for the fields to be defined there. 
In the gauge \eqref{003} this would require that 
$v(0)=v_3(0)=0$, but imposing such a condition would be 
much too restrictive. 
A more general possibility is to perform a gauge transformation
that gives $\varphi$-dependent phases to the scalar fields  and shifts the azimuthal 
components of the vectors, and only after this
 to impose the regularity conditions for the 
vectors. 
Since the two Higgs field components should be single-valued in the new gauge, their phase
factors should contain integers.   
We therefore apply to the 
ansatz \eqref{003} the gauge transformation 
 generated by 
\be                        \label{g1}
{\rm U}=\exp\left\{
\frac{i}{2}(\eta+\psi\tau^3)\right\}, 
\ee
with
\be                      \label{g1a}
\eta=(2n-\nu)\varphi+\,\sigma_\alpha x^\alpha,~~~~
\psi=\nu\varphi-\sigma_\alpha x^\alpha\,,
\ee
where $n,\nu$ are two integer winding numbers. 
This gives  
\begin{align}                           \label{003a}
{\cal W}&=
\left\{\Y(\rho)+1 +{\tau}^1_\psi\, \Om_1(\rho) 
+\tau^3 [\Om_3(\rho)-1]\right\}\,\om_\alpha dx^\alpha  \\
&+
\left\{2\n-\nu-\Z(\rho)             
-
{\tau}^1_\psi\,\W_1(\rho) 
+\tau^3\, [\nu-\W_3(\rho)]\right\}\, d\varphi,
~~~~ 
\Phi=
\left[\begin{array}{c}
e^{in\varphi}\f(\rho) \\
e^{i(n-\nu)\varphi+i\,\om_\alpha x^\alpha}\p(\rho)
\end{array}\right]       \nonumber
\end{align}
with $
{\tau}^1_\psi={\rm U}\tau^1{\rm U}^{-1}=
\tau^1\cos\psi
 -\tau^2\sin\psi$. 
We can assume without loss of generality that
$n\geq 1 $ and we shall see below that $1\leq\nu\leq  2n$.  
The dependence of the fields on $x^\alpha$
 will be convenient in 
what follows but 
not essential for the regularity issue.

The azimuthal components of the vectors in \eqref{003a} will vanish 
at the axis provided that 
\be                  \label{ccc}
\Z(0)=2n-\nu,~~~~~\W_3(0)=\nu,
\ee
and also if $v_1(0)=0$, as  
required by Eq.\eqref{W10}. 
This gives a larger set 
of the allowed boundary values than in gauge \eqref{003}.
In addition, the regularity of the scalar fields
requires that $f_1(0)=0$ and, unless for  $\nu=n$, that $f_2(0)=0$. 
The conditions \eqref{con1}
now reduce to 
\be                  \label{con2}
\nu\Om_1(0)=0,~~~~
n\f(0)=0,~~~~
(n-\nu)\p(0)=0.
\ee
Summarizing, the boundary 
conditions at $\rho=0$
are given by 
\be                                         \label{ax} 
\Om_1(0)=0,~~\Om_3(0)=1,~~\W_1(0)=0,~~\W_3(0)=\nu,~~\Z(0)=2n-\nu,~~\f(0)=0,
\ee
while $\Y(0)$ and $\p(0)$ can be arbitrary if $\nu=n$,
whereas for $\nu\neq n$ one should have $\p(0)=0$. 

We can now work out the most general local series solutions 
of the equations for small $\rho$.
The corresponding analysis is presented in the Appendix A,
the result is 
\begin{align}                         \label{orig}     
\Y&=a_1+\ldots,~~~       \nonumber \\
\Om_1&=a_2\rho^{\nu}+\ldots,         \nonumber \\
\Om_3&=1+\ldots,                \nonumber \\
v_1&=O(\rho^{\nu+2}),    \nonumber \\ 
\W_3&=\nu+a_3\,\rho^2+\ldots,            \nonumber \\
\Z&=2n-\nu+a_4\,\rho^2+\ldots,  \nonumber \\
\f&=a_5\,\rho^{n}+\ldots ,        \nonumber \\              
\p&=q\rho^{|n-\nu|}+\ldots,~~~                                        
\end{align}          
where $a_1,a_2,a_3,a_4,a_5$ and $q$ are six 
integration constants. The dots here stand for the subleading term.
As explained in the Appendix, 
all such terms
will be taken into account in our numerical scheme.

\subsection{Boundary conditions at infinity}

We want the fields for $\rho\to\infty$ to approach vacuum configurations with zero 
energy density. The energy density ${\cal E}_1+{\cal E}_2$ is given by 
Eqs.\eqref{EEE1},\eqref{EEE2} and can be decomposed into 
the kinetic energy part containing the derivatives of the field amplitudes 
and the potential energy
part  that contains no derivatives. 
The potential energy is a sum of perfect squares that will vanish if only
each term in the sum vanishes. For example, the last term in  
\eqref{EEE2} will vanish if only $\f^2+\p^2=1$ and so 
\be                                 \label{fp}
\f(\infty)=\cos\frac{\gamma}{2},~~~~\p(\infty)=\sin\frac{\gamma}{2}.
\ee 
The value of $\gamma$ is an essential parameter characterising the solutions,
we shall call it vacuum angle. 
It is worth noting that the global symmetry \eqref{s2} acts as 
$
\gamma\to\gamma+\Gamma,
$ 
but this cannot be used to change the value of $\gamma$, since 
this symmetry has already been fixed by the condition \eqref{W10} 
at the origin.  

At the same time, 
just for discussing the local solutions at large $\rho$, 
it is convenient to temporarily  
set $\gamma=0$ to simplify the analysis,
since afterwords one can apply the transformation \eqref{s2}
to restore the generic value of $\gamma$. 
We therefore assume for the time being that at infinity
one has 
\be                         \label{i1}
\f=1,~~~~~~~\p=0.
\ee
With this, the potential energy part of  
Eqs.\eqref{EEE1},\eqref{EEE2} will vanish if only 
\be                         \label{i2}
\Om_1=\W_1=0 
\ee
and
\be                         \label{i3}
\Y=-\Om_3,~~~~~
\Z=-\W_3.
\ee
Under these conditions the field equations \eqref{ee1}--\eqref{ee8} reduce to
\be                                     \label{i30}
(\rho \Y^\prime)^\prime=0,~~~~~~~\left(\frac{\Z^\prime }{\rho}\right)^\prime =0,
\ee 
from where 
\be                          \label{i4}
\Y=c_1+Q\ln(\rho),~~~~~~~\Z=c_2+A\rho^2,
\ee
where $c_1,c_2,Q,A$ are integration constants. One should 
set $A=0$ since otherwise the kinetic part in the energy density 
diverges at infinity. However, one should  keep the logarithmic term 
in \eqref{i4}, since the energy density approaches zero for $\rho\to\infty$
even if $Q\neq 0$. As a result, the conclusion is  that 
for large $\rho$ the fields 
approach 
the following exact solution of the field equations, 
\be                           \label{030}
{\cal W}=(1-\tau^3) \left((c_1+Q\ln\rho )\,\sigma_\alpha dx^\alpha -
c_2\,d\varphi\right),~~
~~
\Phi=
\left[\begin{array}{c}
1 \\
0
\end{array}\right] .
\ee
Computing the $A,Z$ fields defined by Eq.\eqref{az1} gives
\be
{A}_\mu dx^\mu=\frac{1}{gg^\prime}
\left((c_1+Q\ln\rho )\,\sigma_\alpha dx^\alpha-c_2\,d\varphi\right),
~~~~~~~Z_\mu=0.
\ee
This is the electromagnetic  Biot-Savart 
solution describing fields outside  an uniformly charged electric wire. 
The charge and current are obtained with 
Eqs.\eqref{cur1},\eqref{cur2}, 
\be                                   \label{line}
I_\alpha=\int \partial^\mu F_{\mu\alpha} d^2x
=
-\oint \epsilon_{ks}\partial_k A_{\alpha} dx^s
=-2\pi\,\frac{Q}{gg^\prime}\,\sigma_\alpha\,,
\ee
assuming that the charge/current distributions are smooth inside the wire
so that the Gauss theorem applies. In addition, there is also a magnetic 
flux along the wire, 
\be                                   \label{fluxes1}
\Psi_F=
\oint A_{k} dx^k=-\frac{2\pi}{gg^\prime}\,c_2.
\ee  
In what follows we shall construct smooth vortex 
solutions in which the wire is represented by a regular distribution 
of massive non-linear fields, while in the far field zone the massive modes 
die out and everything reduces to the Biot-Savart 
configuration \eqref{030}. 
The energy density at large $\rho$ will then be proportional to   
$Q^2/\rho^2$ and the total energy per
unit vortex length will be logarithmically divergent, just as for the 
ordinary infinitely long electric wire. 
The energy of finite vortex pieces, as for example vortex loops,
will be finite.

%\input{asympt.tex}
%\vspace{1 cm}

Let us now consider the deviations from the 
electromagnetic configuration \eqref{030}  
within the full system of equations \eqref{ee1}--\eqref{ee8} in order to determine 
how the solutions approach their asymptotic form.  
The corresponding analysis in the linear approximation is carried out in the Appendix B
and 
the result is given by Eq.\eqref{dev1}.  
We only need to apply to Eq.\eqref{dev1} the phase rotation \eqref{s2}
to restore the generic value of $\gamma$ in Eq.\eqref{fp}. 
This gives the large $\rho$ behaviour  of the solutions, 
\begin{align}                         \label{inf}
\Y&=Q\ln\rho+c_1+ \frac{c_3g^{\prime 2}}{\sqrt{\rho}}\,e^{-\mz\rho}
+\ldots \nonumber \\
\Z&=c_2+ c_4g^{\prime 2}\sqrt{\rho}\,e^{-\mz\rho}+\ldots \nonumber \\
\Om_1+i\Om_3&=e^{-i\gamma}\left\{\frac{c_7}{\sqrt{\rho}}\,e^{-\int\,m_\om d\rho}+i\,[
-Q\ln\rho-c_1+ \frac{c_3g^{2}}{\sqrt{\rho}}\,e^{-\mz\rho}
]\right\}+\ldots
\nonumber \\
\W_1+i\W_3&=e^{-i\gamma}\left\{c_8\sqrt{\rho}\,e^{-\int\,m_\om d\rho}
+i\,[
-c_2+ c_4g^{2}\sqrt{\rho}\,e^{-\mz\rho}]\right\}+\ldots \nonumber \\
\f+i\p&=e^{\frac{i}{2} \gamma}\left\{1 + \frac{c_5}{\sqrt{\rho}}\,e^{-\mh\rho}+i\,
\frac{c_6}{\sqrt{\rho}}\,e^{-\int\, m_\om d\rho}\right\}+\ldots
\end{align}
These local solutions contain  
10 independent parameters  $c_1,\ldots ,c_8$, $\gamma$, $Q$
and they approach (modulo the $\gamma$-phases)  
the Abelian configuration \eqref{030}   
exponentially fast as $\rho\to\infty$, with the rates 
determined by the masses $\mz,\mh$ and $m_\om$. Here
\be                                   \label{m-sigma}
m_\om=\sqrt{\mw^2+\om^2(Q\ln\rho+c_1)^2}
\ee 
reduces to the W boson mass $\mw=g/\sqrt{2}$ if $\sigma^2=0$.  
For $\sigma^2\neq0$ this could be viewed as the W boson mass `screened' ($\sigma^2<0$)
or `dressed' ($\sigma^2>0$) by the interaction of charged W bosons with the long-ranged 
Biot-Savart  field \eqref{030}. 
We also note that the solutions  \eqref{inf} satisfy the field
equations \eqref{ee1}--\eqref{ee8} but not the constraint \eqref{CONS}. 
In fact, the latter  has already been imposed on the local solutions \eqref{orig}
at {\it small $\rho$}, and since it
`propagates', it will be automatically enforced by extending the local solution 
 \eqref{orig} to large values of $\rho$. 
As a result, it would be redundant 
to impose it again.

We can now outline our strategy for solving the field equations. 
The local solutions for small and large $\rho$ are given, respectively,
by Eqs.\eqref{orig} and \eqref{inf}. 
Within the numerical  multiple shooting method \cite{Press}
we extend these asymptotic solutions to the intermediate values of $\rho$ 
and match them. 
We use the multi-zone version of the method \cite{Stoer}, with many 
zones whose number and sizes are adjusted  to make sure that
for  large $\rho$ 
the asymptotic solution \eqref{inf} is a good approximation. 
In order to illustrate the procedure, let us describe the simplest case
where there are only two zones and one matching point. 
%This gives global solutions in the interval $\rho\in[0,\infty)$. 
To match the solutions at this  point 
we have to fulfill 
the 16 matching conditions for the 8 field amplitudes 
and for their first derivatives via adjusting the 
free parameters in Eqs.\eqref{orig},\eqref{inf}. It is therefore essential 
to have enough of parameters, since otherwise the matching conditions
will be incompatible. Counting the parameters we discover that there are
17 of them: 
6 integration constants $a_1,\ldots ,a_5$ and $q$ in \eqref{orig}, then 
10 integration constants $c_1,\ldots ,c_8$, $Q$, $\gamma$ in \eqref{inf},
and finally $\sigma^2$. This is enough  to fulfill the 16
matching conditions and to have one parameter left free after
the matching. This parameter will label the resulting global solutions
in the interval $\rho\in[0,\infty)$. 
It is convenient to choose it 
to be $q$ -- the coefficient 
in front of the amplitude $\p$ in \eqref{orig}. 

Before integrating the equations, some preliminary 
considerations are in order. 

%\subsection{Recapitulating the boundary conditions}

\section{Conserved quantities}
\setcounter{equation}{0}
Keeping only the leading terms, 
the boundary conditions for the field amplitudes 
for $0\leftarrow \rho\to\infty$ are  
\begin{align}               \label{rec}
a_1\leftarrow\, &\Y\to c_1+Q\ln\rho\,,    \nonumber \\
2n-\nu\leftarrow\, &\Z\to c_2\,,            \nonumber \\  
0\leftarrow\, & \Om_1 \to -(c_1+Q\ln\rho)\sin\gamma\, \,, \nonumber \\ 
1\leftarrow\, & \Om_3 \to -(c_1+Q\ln\rho)\cos\gamma\, \,, \nonumber \\  
0\leftarrow\, & \W_1 \to - c_2\sin\gamma\,,              \nonumber \\ 
\nu\leftarrow\, & \W_3 \to -c_2\cos\gamma\,,              \nonumber \\ 
a_5 \rho^{\n}\leftarrow\, & \f \to \cos\frac{\gamma}{2}\,,          \nonumber \\  
q\,\rho^{|\n-\nu|}\leftarrow\, & \p \to \sin\frac{\gamma}{2}\,,   
\end{align}
  The knowledge of these boundary conditions is sufficient to calculate 
most  of the global physical quantities  associated to the solutions. 

\subsection{Fluxes and current \label{flux-cur}}
Since the fluxes and currents are gauge invariant, 
one can compute them in the  gauge \eqref{003}, in which case the calculations
are simpler.
The non-zero components of the field tensors then read
\begin{align}
\F_{\rho\alpha}&=\sigma_\alpha\Om^\prime,~~~~
\F_{\rho\varphi}=-\W^\prime\,,~~~\nonumber \\
\WW^a_{\rho\alpha}&=\sigma_\alpha\Om_a^\prime,~~~~
\WW^a_{\rho\varphi}=-\W_a^\prime\,,~~~~a=1,3;\nonumber \\
\WW^2_{\alpha\varphi}&=(\Om_1\W_3-\Om_3\W_1)\sigma_\alpha.
\end{align}
Introducing the function $\Omega(\rho)$ defined by
\be                                   \label{Omega}
\cos{\Omega}=\frac{\f^2-\p^2}{{\f^2+\p^2}},~~~~ 
\sin{\Omega}=\frac{2\f\p}{{\f^2+\p^2}},
\ee
such that $\Omega(\infty)=\gamma$, 
one can define three orthogonal unit isovectors 
\begin{align}                                  \label{n1}
n^a&=(\sin\Omega,0,\cos\Omega),   \notag \\
k^a&=(\cos\Omega,0,-\sin\Omega),   \notag \\
l^a&=(0,1,0).
\end{align}
Here $n^a$ corresponds to the definition \eqref{n}.
The Nambu electromagnetic and Z fields \eqref{Nambu}
then read 
\begin{align}                               \label{FZ}
F_{\rho\alpha}&=\left(\frac{g}{g^\prime}\,\Om^\prime-
\frac{g^\prime}{g}\,( \Om_1^\prime \sin\Omega
+ \Om_3^\prime \cos\Omega)\right)\sigma_\alpha \,,  \nonumber \\
F_{\rho\varphi}&=-\frac{g}{g^\prime}\,\W^\prime+
\frac{g^\prime}{g}\,( \W_1^\prime \sin\Omega
+ \W_3^\prime \cos\Omega)\,,  \nonumber \\
Z_{\rho\alpha}&=(\Om^\prime+\Om_1^\prime \sin\Omega
+ \Om_3^\prime \cos\Omega)\sigma_\alpha \,,  \nonumber \\
Z_{\rho\varphi}&=-\W^\prime-\W_1^\prime\sin\Omega-\W_3^\prime\cos\Omega\,,
\end{align}
whereas the 't Hooft \eqref{Hoft} fields have the total derivative structure, 
\begin{align}
F^{\rm H}_{\rho\alpha}&=\left(\frac{g}{g^\prime}\,\Om-
\frac{g^\prime}{g}\,( \Om_1 \sin\Omega
+ \Om_3 \cos\Omega)\right)^\prime\sigma_\alpha \,,  \nonumber \\
F^{\rm H}_{\rho\varphi}&=\left(-\frac{g}{g^\prime}\,\W+
\frac{g^\prime}{g}\,( \W_1 \sin\Omega
+ \W_3 \cos\Omega)\right)^\prime\,,  \nonumber \\
Z^{\rm H}_{\rho\alpha}&=(\Om+\Om_1 \sin\Omega
+ \Om_3 \cos\Omega)^\prime\sigma_\alpha \,,  \nonumber \\
Z^{\rm H}_{\rho\varphi}&=-\left(\W+\W_1\sin\Omega+\W_3\cos\Omega\right)^\prime\,.
\end{align}
The 
electromagnetic current density is
\be                          \label{Io}
J^\alpha=\partial_\mu F^{\mu\alpha}=
\frac{1}{\rho}\,(\rho F^{\rho\alpha})^\prime
\ee
and 
integrating over the $\rho,\varphi$ plane gives the same result
for the both definitions of the fields,  
\be                                          \label{I}
I_\alpha=-2\pi\int_0^\infty(\rho\,F_{\rho\alpha})^\prime d\rho=
-\frac{2\pi Q}{gg^\prime}\,\sigma_\alpha\,.
\ee
For most of the solutions considered below the vector 
$\sigma_\alpha=(\sigma_0,\sigma_3)\equiv (\sigma_0,\sigma_z)$ is 
spacelike, so that there is the rest frame where 
$\sigma_\alpha=\sigma \delta_\alpha^3$. The restframe value of the 
current is $I_\alpha= \delta_\alpha^3\,{\cal I}$ with 
${\cal I}=-{2\pi Q\sigma}/({gg^\prime})$.  
The restframe
components of the electromagnetic field strength and Z field strength
 are 
\be                                     \label{magn}
{B}_{\hat{z}}=\frac{1}{\rho}\,F_{\rho\varphi},~~~~~
{B}_{\hat{\varphi}}=-F_{\rho z},~~~~
{H}_{\hat{z}}=\frac{1}{\rho}\,Z_{\rho\varphi},~~~~~~
{H}_{\hat{\varphi}}=-Z_{\rho z},
\ee
such that the magnetic fluxes are
\be                                           \label{fluxNambu}
\Psi_F=2\pi\int_0^\infty \rho\,{B}_{\hat{z}}\,d\rho\,,~~~~
\Psi_Z=2\pi\int_0^\infty \rho\,{H}_{\hat{z}}\,d\rho\,. 
\ee
In the Nambu case these integrals have to be computed, while 
in the 't Hooft case they
evaluate to 
\begin{align}                                  \label{fluxHoft}
\Psi_F^{\rm H}&=
-\frac{2\pi}{gg^\prime}\,(c_2+(\nu-2\n)g^2
+\nu g^{\prime 2}\cos\Omega(0)), \nonumber \\
\Psi_Z^{\rm H}&=
2\pi (2\n+\nu(\cos\Omega(0)-1)).
\end{align}
One can also define the W-condensate components
by projecting the restframe values of $\WW^a_{\mu\nu}$ onto  
the unit isovectors $k^a,l^a$ defined in \eqref{n1},
\be                                         \label{confld}
{w}_{\hat{z}}=\frac{1}{\rho}\,k^a\WW^a_{\rho \varphi},~~~~
{w}_{\hat{\varphi}}=-k^a\WW^a_{\rho z},~~~~
{w}_{\hat{\rho}}=-\frac{1}{\rho}\,l^a\WW^a_{z \varphi}.~~~~
\ee

\subsection{Energy, momentum, angular momentum}
Using Eqs.\eqref{EEE2},\eqref{EEE3} and the boundary conditions \eqref{rec}, 
the total energy evaluates to  
\be                              \label{EE}
E=\pi\,\frac{\sigma_0^2+\sigma_3^2}{g^2 g^{\prime 2}}\,Q\Y(\infty)+
2\pi\int_0^\infty\rho\,{\cal E}_2 d\rho\,.
\ee 
Here the first term on the right is the contribution of the total derivative in 
Eq.\eqref{EEE3}. 
Comparing Eqs.\eqref{T03},\eqref{EN1} and \eqref{EEE3} one can see 
that the same total derivative 
determines the value of the momentum, 
\be                         \label{PP}
P=2\pi\,\frac{\sigma_0\sigma_3}{g^2 g^{\prime 2}}\,Q\Y(\infty)\,.
\ee 
We finally notice that, since in the axially symmetric case there is an 
additional Killing vector $K_{(\varphi)}$, associated to this there is the 
conserved angular momentum
\be
M=\int T^0_{~\nu} K^\nu_{(\varphi)}d^2x=\int T^0_{~\varphi} d^2x\,.  \label{M} 
\ee
This also has a total derivative structure \cite{VW}, since one has 
\begin{align}                      \label{T0phi}
T^0_{~\varphi}&=
-\frac{1}{g^2}\,\WW^{a 0 \rho}\WW^a_{~\varphi\rho}
-\frac{1}{g^{\prime\,2}}B^{0\rho}B_{\varphi\rho}
+(D^0\Phi)^\dagger D_\varphi\Phi
+(D_\varphi\Phi)^\dagger D^0\Phi  \nonumber \\
&=\sigma^0\left(\frac{1}{g^{\prime\,2}}\,\Y^\prime\Z^\prime+
\frac{1}{g^{\prime\,2}}\,\Om_a^\prime \W_a^\prime\right)
+2\sigma^0\Re\left(\Phi^\dagger(\Y+\Om_a\tau^a)(\Z+\W_b\tau^b)\Phi\right)
\end{align}
with $a=1,3$.
Using the field equations \eqref{ee1},\eqref{ee5},\eqref{ee6} gives
\be
\rho T^0_\varphi=\sigma^0\left(\frac{\rho}{g^{\prime\,2}}\,\Z\Y^\prime+
\frac{\rho}{g^{\prime\,2}}\,(\W_1\Om_1^\prime+\W_3\Om_3^\prime)\right)^\prime
\ee
so that 
\be                         \label{MM}
M=2\pi\,\frac{\sigma_0}{g^2 g^{\prime 2}}\,Q\W(\infty)=
\frac{2\pi Q \sigma_0 c_2}{g^2 g^{\prime 2}}. 
\ee 
Summarizing, the fluxes,
current, energy, momentum and angular momentum 
are given by the above expressions, and in most  cases they 
are totally determined by the boundary conditions \eqref{rec}.

\section{Known solutions}
\setcounter{equation}{0}

The only known  
 solutions of Eqs.
\eqref{ee1}--\eqref{CONS}
for generic values of $g,g^\prime$ are the embedded ANO vortices. 
These exist in two different versions,
called Z strings \cite{Z} and W strings \cite{W}, corresponding to two 
nonequivalent embeddings of U(1) to SU(2)$\times$U(1).

\subsection{Z strings}

These solutions are obtained by setting 
$\Om_3=-\Y=1$ and $\p=\Om_1=\W_1=0$ after which  
equations \eqref{ee1}--\eqref{CONS} reduce to 
\begin{subequations}
\begin{align}                                           
\frac{1}{\rho}(\rho\f^\prime)^\prime&=
\left(
\frac{1}{4\rho^2}\,(\Z+\W_3)^2
+\frac{\beta}{4}(\f^2-1)
\right)\f \,,                                    \label{eee1} \\
 \rho\left(\frac{\Z^\prime}{\rho}\right)^\prime&=\frac{g^{\prime\,2}}{2}
(\Z+\W_3)\f^2,                                  \label{eee2} \\
\rho\left(\frac{\W_3^\prime}{\rho}\right)^\prime
&=\frac{g^2}{2}
(\W_3+\Z)\f^2.                                      \label{eee3}
\end{align}
\end{subequations}
With 
\be
{\cal A}=\frac12\,(g^2\Z-g^{\prime 2}\W_3),~~~~
v_{\mbox{\tiny ANO}}=\frac12\,(\Z+\W_3),~~~
f_{\mbox{\tiny ANO}}=\f
\ee
these equations reduce to the ANO system 
\begin{subequations}                    \label{ANOeqs}
\begin{align}                                           
\frac{1}{\rho}(\rho f_{\mbox{\tiny ANO}}^\prime)^\prime&=
\left(
\frac{v^2_{\mbox{\tiny ANO}}}{\rho^2}
+\frac{\beta}{4}(f_{\mbox{\tiny ANO}}^2-1)
\right) f_{\mbox{\tiny ANO}}\,,                       \label{ano1} \\
 \rho\left(\frac{v_{\mbox{\tiny ANO}}^\prime}{\rho}\right)^\prime&
=\frac{1}{2}\,
f_{\mbox{\tiny ANO}}^2\,v_{\mbox{\tiny ANO}},               \label{ano2} 
\end{align}
\end{subequations}
and  to 
\be                                  \label{AAA}
\left(\frac{{\cal A}^\prime}{\rho}\right)^\prime=0.
\ee
Solutions of the ANO equations \eqref{ANOeqs} can  be found numerically, 
they comprise a 
family labeled by $n=1,2,\ldots$ with the boundary conditions
\be
 0\leftarrow f_{\mbox{\tiny ANO}}\to 1,~~~~
n\leftarrow v_{\mbox{\tiny ANO}}\to 0
\ee
for 
$0\leftarrow \rho\to \infty$. 
The only bounded solution for ${\cal A}$
is the constant whose value is
determined by the boundary conditions \eqref{rec}
at the origin, ${\cal A}=\n g^2-\nu/2$. 

Summarizing, Z string profiles are given by 
\begin{align}
\Y&=-1,~~~~\Z=2g^{\prime 2}(v_{\mbox{\tiny ANO}}-\n)+2\n-\nu,~~~~
\Om_1=0,~~~~\Om_3=1, \nonumber \\
\W_1&=0,~~~~~~
\W_3=2g^{2}(v_{\mbox{\tiny ANO}}-\n)+\nu,~~~~
\f=f_{\mbox{\tiny ANO}},~~~~\p=0,               \label{Zsol}
\end{align}
from where one can read off the values of the asymptotic 
parameters \eqref{rec},
\be
a_1=c_1=-1,~~Q=0,~~
c_2=2\n g^2-\nu,~~\Omega=\Omega_0=\gamma=0.
\ee
Calculating the fluxes 
gives the same result for the Nambu and  't Hooft definitions, 
\be                          \label{flZ}
\Psi_{Z}=4\pi\n,~~~~
\Psi_{F}=0.
\ee  
while the current, momentum and 
angular momentum vanish, so that the energy is finite. 
 The dependence of the amplitudes
in \eqref{Zsol} on the second winding number 
$\nu$ is pure gauge for these solutions,
since in the gauge  
\eqref{003a} 
it disappears:
\be                           \label{003aZ}
{\cal W}=
2(g^{\prime 2}+g^2\tau^3)(\n-v_{\mbox{\tiny ANO}}(\rho))\,d\varphi,      
~~~~ 
\Phi=
\left(
\begin{array}{c}
e^{in\varphi}f_{\mbox{\tiny ANO}}(\rho)\\
0 
\end{array}  
\right).       
\ee

\subsection{W strings} 
These 
are obtained by setting in Eqs.
\eqref{ee1}--\eqref{CONS}   
\be                                \label{Wsol}
\Y=\Z=\Om_1=\Om_3=\W_3=\p=0,~~~\W_1=2v_{\mbox{\tiny ANO}}(\rho),~~~
\f=f_{\mbox{\tiny ANO}}(\rho).
\ee
Writing this solution in the gauge \eqref{003} and applying the gauge 
transformation generated by U$=\exp(i\n\varphi\tau^1)$ gives 
\be                           \label{003aW}
{\cal W}=
2\tau^1(\n-v_{\mbox{\tiny ANO}}(\rho))\,d\varphi,      
~~~~ \phi_1+i\phi_2=f_{\mbox{\tiny ANO}}(\rho)e^{i\n\varphi}~~~~
%\Phi=f_{\mbox{\tiny ANO}}(\rho)
%\left(
%\begin{array}{c}
%\cos(\n\varphi)\\
%i\sin(\n\varphi)
%\end{array}  
%\right),       
\ee
which is the parametrization used in Ref. \cite{W}. 
However, throughout this paper we are using different gauge conventions,
in particular we impose the condition \eqref{W10}. We therefore return to 
\eqref{Wsol} and perform the global gauge rotation \eqref{s2} with $\Gamma=-\pi/2$,
which gives 
\be
\Y=\Z=\Om_1=\Om_3=\W_1=0,~~~\W_3=2v_{\mbox{\tiny ANO}}(\rho),~~~
\f=-\p=\frac{1}{\sqrt{2}}\,f_{\mbox{\tiny ANO}}(\rho).
\ee
From here 
we can read off the values of the asymptotic parameters,
\be
\nu=2\n,~~Q=c_1=c_2=0,~~\gamma=\Omega=\Omega_0=-\frac{\pi}{4}.
\ee
Calculating the fluxes \eqref{fluxNambu},\eqref{fluxHoft} gives, in 
the both definitions, 
\be                                \label{flW}
\Psi_{Z}=\frac{4\pi\n}{\sqrt{2}},~~~~
\Psi_{F}=-\frac{4\pi\n}{\sqrt{2}}\,\frac{g^\prime}{g}\,.
\ee 
These values are not the same as in the Z string case,
in particular the electromagnetic flux is now different from zero.  
W strings are therefore physically different from Z strings.

\section{Small current limit -- bound states around Z strings \label{weak}}
\setcounter{equation}{0}

Our goal is to construct solutions of equations 
\eqref{ee1}--\eqref{CONS} more general that the embedded ANO vortices. 
Our strategy was briefly summarized above: we 
numerically extend the asymptotic solutions \eqref{orig} and \eqref{inf}
to the intermediate region and impose there the 16 matching conditions, which 
 can be fulfilled 
by adjusting the 17 free parameters in the local solutions. 
This leaves one extra parameter, $q$, which determines the value of the 
lower  component of the Higgs field at the origin. 

The matching conditions are resolved iteratively, within the standard 
method described in \cite{Stoer}.  
A good choice of the initial configuration is then important, since otherwise 
the iterations will not converge. The idea is therefore to start the iterations 
in the vicinity of an already known solution, for which 
we can only choose the Z string (it is unclear whether W strings can
also be used). 
Let us therefore choose the Z string solution \eqref{Zsol}
as the starting point of our analysis. One has in this case 
$\p(\rho)=0$ so that $q=0$.

Suppose now that $0<q\ll 1$. 
It is then natural to expect the 
solution to be a slightly deformed  Z string. 
It can therefore be represented in the form 
\begin{align}
\Y&=-1+\delta\Y,~~~~
\Z=v_{\mbox{\tiny Z}}+\delta\Z,~~~~
\Om_1=\delta\Om_1,~~~~\Om_3=1+\delta\Om_3, \nonumber \\
\W_1&=\delta\W_1,~~~~~~
\W_3=v_{\mbox{\tiny Z}3}+\delta\W_3,~~~~
\f=f_{\mbox{\tiny Z}}+\delta\f,~~~~
\p=\delta\p,               \label{Zpert}
\end{align}
where $v_{\mbox{\tiny Z}}\equiv 2g^{\prime 2}(v_{\mbox{\tiny ANO}}-\n)+2\n-\nu$, 
$v_{\mbox{\tiny Z}3}\equiv 2g^{2}(v_{\mbox{\tiny ANO}}-\n)+\nu$,
$f_{\mbox{\tiny Z}}\equiv f_{\mbox{\tiny ANO}}$ and 
$\delta\Y,\ldots ,\delta\p$ are small 
deformations. 
Inserting this to Eqs.\eqref{ee1}--\eqref{ee8} and linearising 
with respect to the deformations, the resulting equations split into three 
independent groups: two coupled equations for $\delta\Y, \delta\Om_3$
plus 
three coupled equations for $\delta\f, \delta\Z,\delta\W_3$ plus 
three coupled equations for $\delta\p, \delta\Om_1,\delta\W_1$.  
The first two groups do not admit interesting solutions,
so that we can set 
$\delta\Y=\delta\Om_3=\delta\f=\delta\Z=\delta\W_3=0$. 

\begin{figure}[h]
\hbox to\linewidth{\hss%
  \psfrag{lnx}{$\ln(1+\rho)$}
 \psfrag{y}{}
  \psfrag{df2}{$\delta\p$}
  \psfrag{dv1}{$\delta \W_1$}
  \psfrag{du1}{$\delta\Om_1$}
  \psfrag{4}{$\nu=1,~{\sigma^2}>0$}
  \resizebox{8cm}{7cm}{\includegraphics{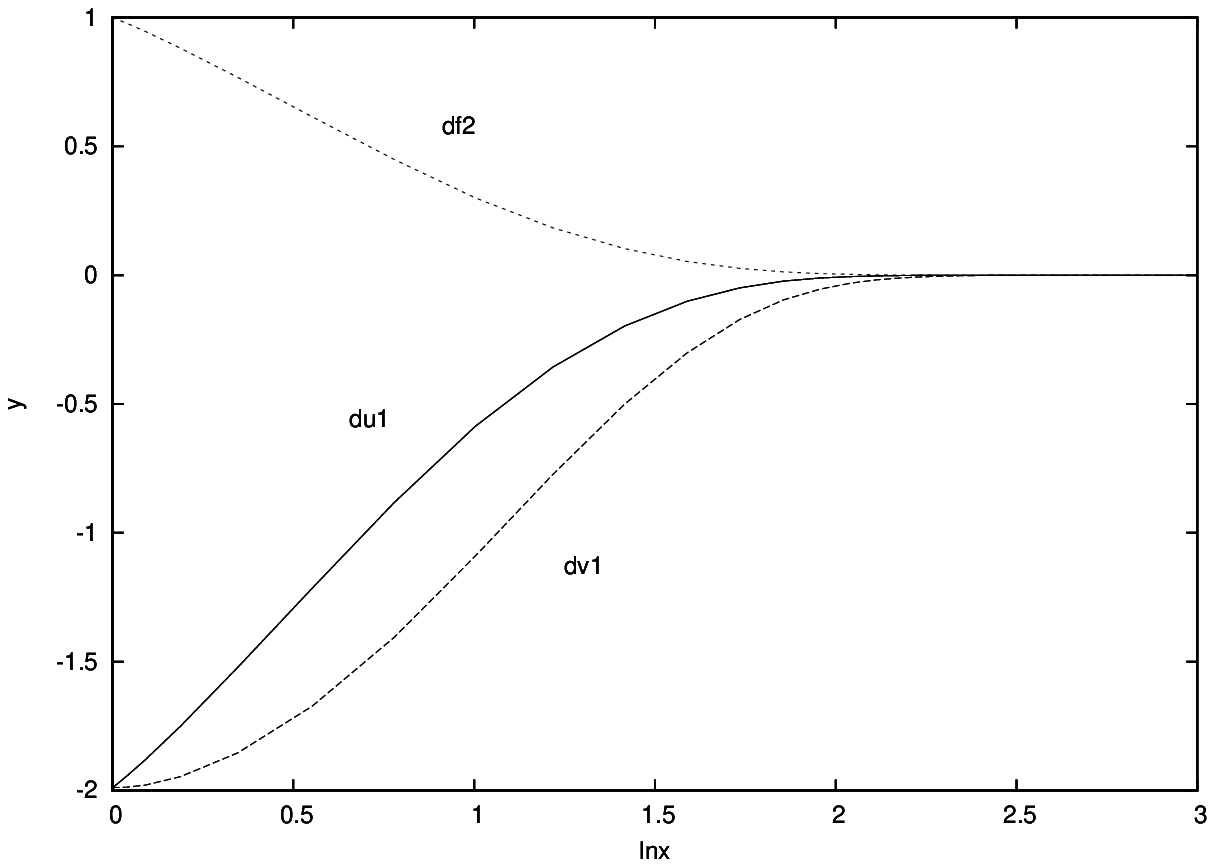}}%
\hspace{2 mm}
  \psfrag{x}{$\ln(1+\rho)$}
  \psfrag{1a}{$\delta\p$}
  \psfrag{dv1overnu}{$\delta \W_1/\nu$}
  \psfrag{2a}{$\delta\Om_1$}
  \psfrag{4a}{$\nu=2,~{\sigma^2}<0$}
  \resizebox{8cm}{7cm}{\includegraphics{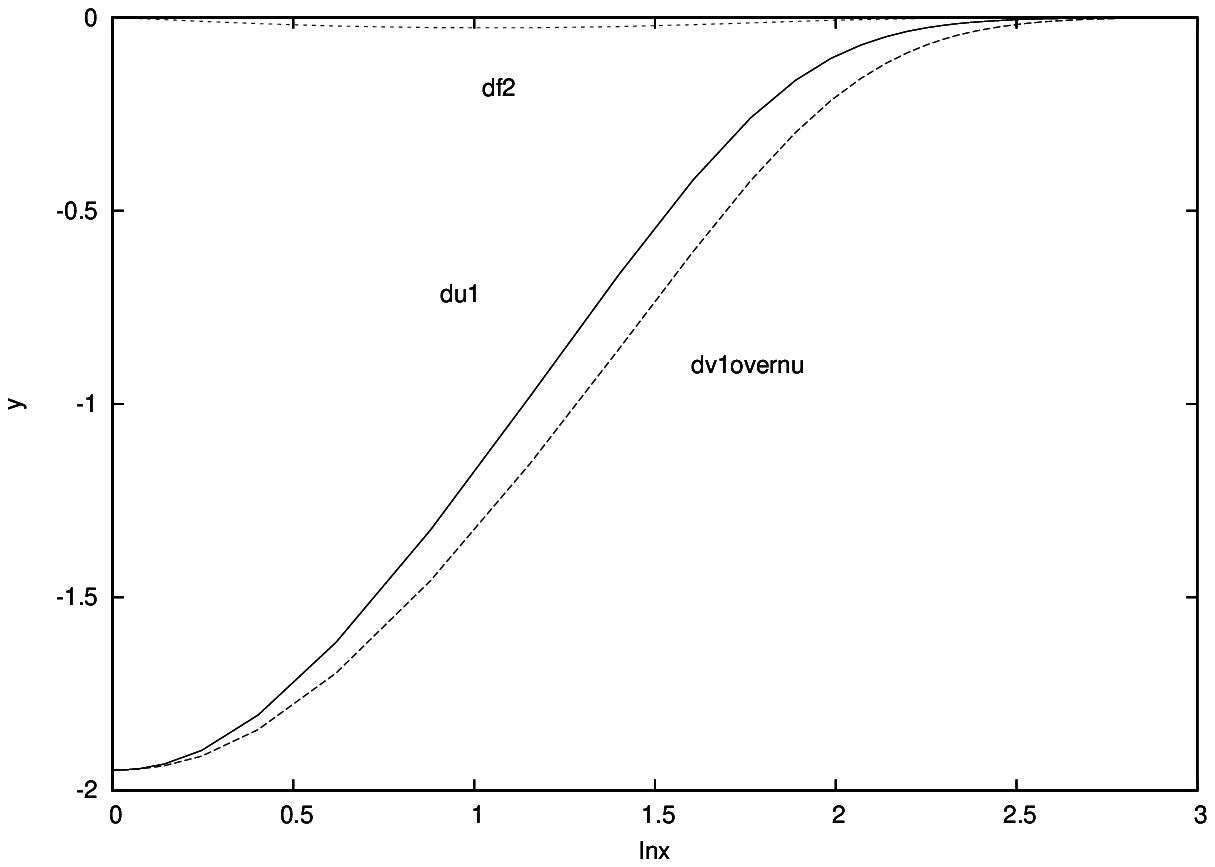}}%
\hss}
\caption{The bound state solutions of Eqs.\eqref{eqs1}
with  $\sin^2\thetaw=0.23$, $\beta=2$ for $n=\nu=1$, 
$\sigma^2=0.5036$ (left) and for $n=1$, $\nu=2$, $\sigma^2=-0.0876$ (right).  
}
\label{FIG1}
\end{figure}

The remaining three equations describe a slightly  deformed 
Z string: ${\cal W}+\delta{\cal W}$, $\Phi+\delta\Phi$, 
where ${\cal W}$ and $\Phi$ are given by Eq.\eqref{003aZ} and   
\be                           \label{003aZ1}
\delta{\cal W}=
\frac{\tau^{+}}{2}\,e^{i\psi}
[\delta\Om_1(\rho) \sigma_\alpha dx^\alpha
-\delta\W_1(\rho) d\varphi] +\text{h.c.}~, 
~~~~ 
\delta\Phi=
\left(
\begin{array}{c}
0\\
e^{i(\n\varphi-\psi)}\delta\p(\rho)
\end{array}  
\right),       
\ee
with $\psi=\nu\varphi-\sigma_\alpha x^\alpha$ and   
$\tau^{+}=\tau^1+i\tau^2$. Here  h.c. stands for Hermitean conjugation. 
The amplitudes $\delta\p,\delta\W_1,\delta\Om_1$ fulfil  the equations 
\begin{subequations}                \label{eqs1}
\begin{align}
\frac{1}{\rho}\left(\rho\delta\p^\prime\right)^\prime&=
\left(\frac{(\Z_{\mbox{\tiny Z}}-\W_{{\mbox{\tiny Z}}3})^2}{4\rho^2}+\frac{\beta}{4}\,
(f_{\mbox{\tiny Z}}^{2}-1)+
\sigma^2\right)\delta\p
+\frac{\Z_{\mbox{\tiny Z}} f_{\mbox{\tiny Z}}}{2\rho^2}\,\delta\W_1
-\frac{\sigma^2}{2}\,f_{\mbox{\tiny Z}}\,\delta\Om_1\,,\\
\frac{1}{\rho}\left(\rho\delta\Om_1^\prime\right)^\prime&=
\left(\frac{\W_{{\mbox{\tiny Z}}3}^2}{\rho^2}+\frac{g^2}{2}\,f_{\mbox{\tiny Z}}^2
\right)\delta\Om_1
-\frac{\W_{\mbox{\tiny Z}3} }{\rho^2}\,\delta\W_1
-g^2\,f_{\mbox{\tiny Z}}\,\delta\p\,, \\
\rho\left(\frac{\delta\W_1^\prime}{\rho}\right)^\prime&=
\left(\frac{g^2}{2}\,f_{\mbox{\tiny Z}}^2+\sigma^2\right)\delta\W_1-
\sigma^2\W_{\mbox{\tiny Z}3}\,\delta\Om_1
+g^2\Z_{\mbox{\tiny Z}} f_{\mbox{\tiny Z}}\,\delta\p\,.
\end{align}
\end{subequations}
This can be viewed as a spectral problem
with the eigenvalue $\sigma^2$.
In fact, by suitably redefining the variables one 
can rewrite the equations in the form 
\be                               \label{Psi}
\Psi''=(\sigma^2+V[\beta,\thetaw,\n,\nu,\rho])\Psi,
\ee
where $\Psi$ is a 3 component vector 
and $V$ is a 
symmetric potential energy  matrix depending on the parameters
of the background Z string solution, $\beta$, $\thetaw$, $\n$,
and also on $\nu$. Although the dependence on $\nu$ for the Z string
background is pure gauge, this is not so for the deformations of this 
background, since for different values of $\nu$ in \eqref{eqs1} 
one obtains different results. 

\begin{figure}[h]
\hbox to\linewidth{\hss%
  \psfrag{x}{$\beta$}
\psfrag{y}{$\om^2$}
  \psfrag{gp=0.01}{$g^{\prime 2}=0.01$}
  \psfrag{gp=0.99}{$g^{\prime 2}=0.99$}
 \psfrag{title}{}
    \resizebox{8cm}{7cm}{\includegraphics{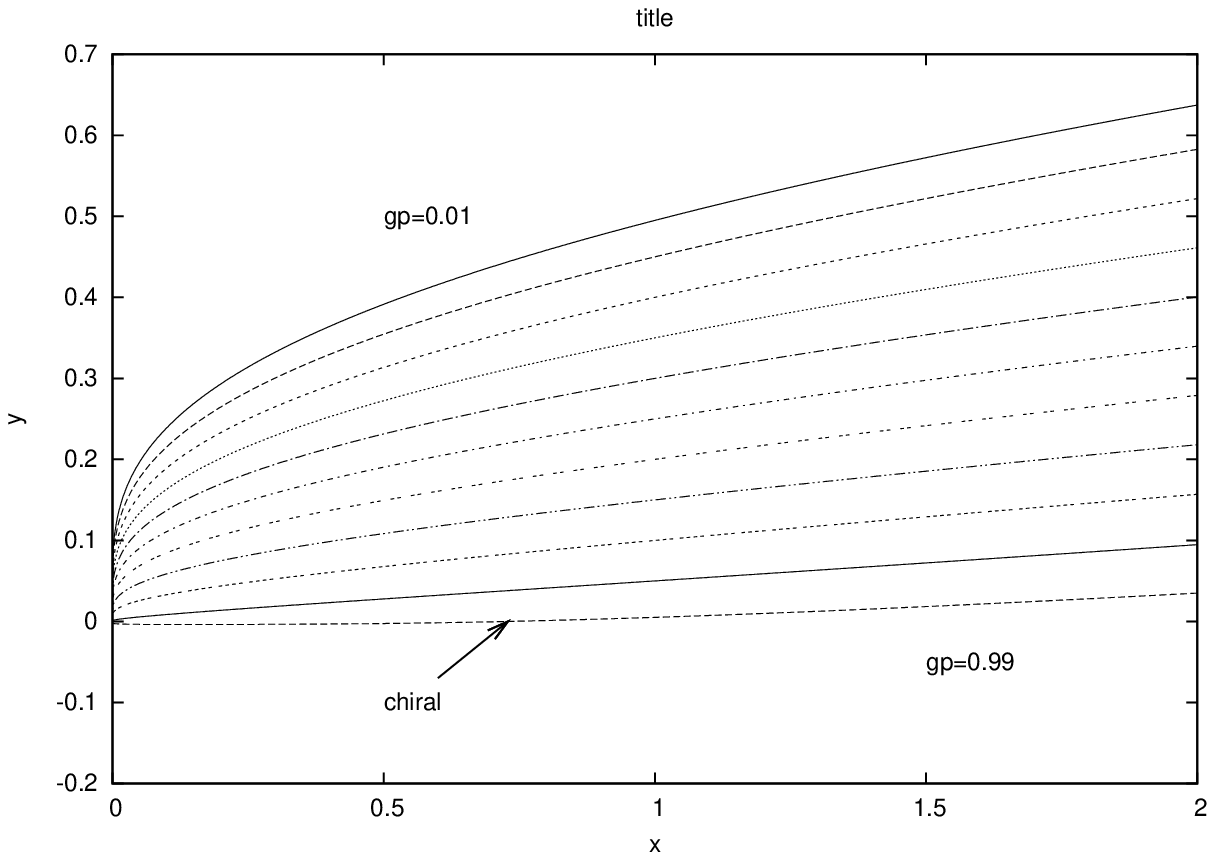}}%
\hspace{2 mm}
  \psfrag{x}{$\beta$}
 \psfrag{y}{$\om^2$}
  \psfrag{gp=0.99}{$g^{\prime 2}=0.99$}
  \psfrag{gp=0.01}{$g^{\prime 2}=0.01$}
  \psfrag{gp=0.6}{$g^{\prime 2}=0.6$}
  \psfrag{gp=0.7}{$g^{\prime 2}=0.7$}
  \psfrag{gp=0.8}{$g^{\prime 2}=0.8$}
  \psfrag{gp=0.9}{$g^{\prime 2}=0.9$}
  \psfrag{gp=0.1}{$g^{\prime 2}=0.1$}
  \psfrag{gp=0.90}{$g^{\prime 2}=0.90$}
  \psfrag{gp=0.2}{$g^{\prime 2}=0.2$}
  \psfrag{gp=0.4}{$g^{\prime 2}=0.4$}
  \psfrag{gp=0.5}{$g^{\prime 2}=0.5$}
 \psfrag{title}{}
 \psfrag{chiral}{chiral}
    \resizebox{8cm}{7cm}{\includegraphics{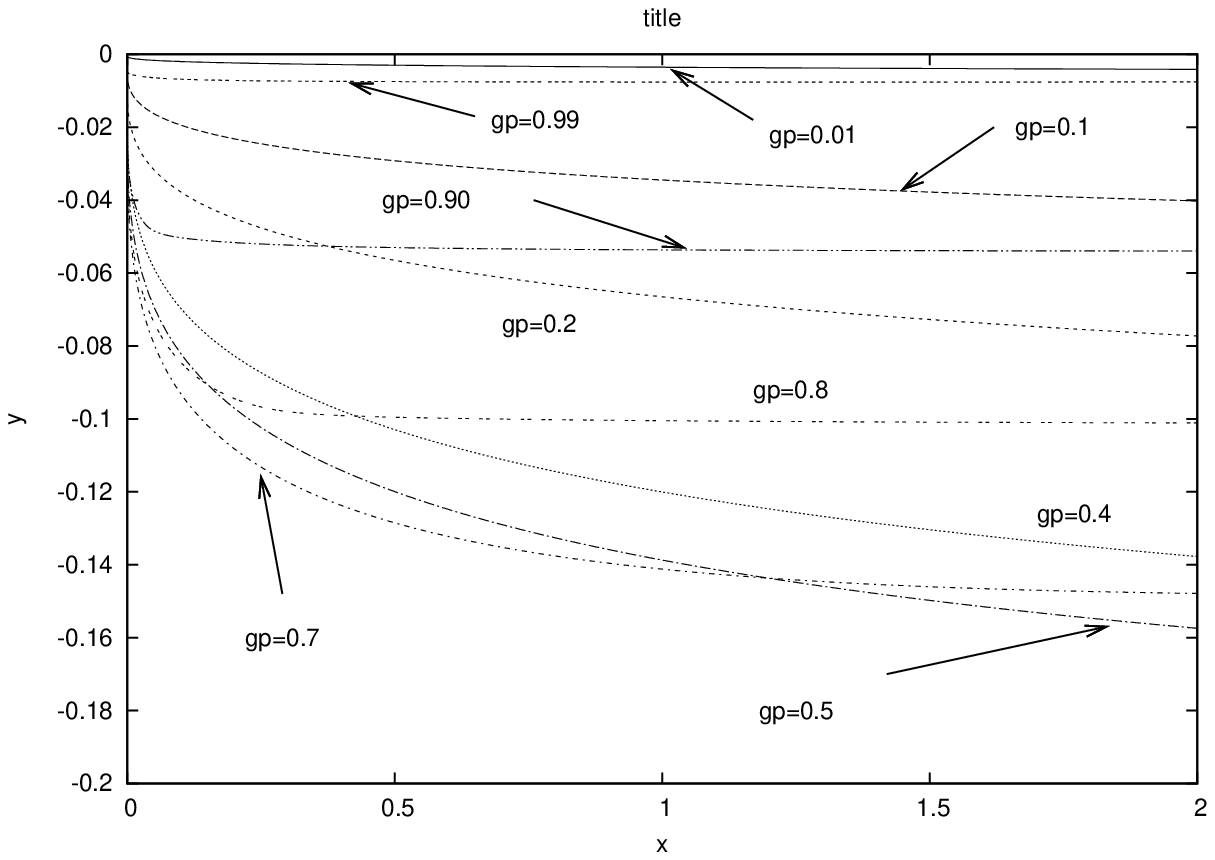}}%
\hss}
\caption{The eigenvalue $\sigma^2$ of the spectral problem \eqref{eqs1}
against $\beta$ for $n=\nu=1$  
(left) and for $n=1$, $\nu=2$ (right) 
 for several values of $g^{\prime 2}=\sin^2\thetaw$. 
}
\label{FIG2}
\end{figure}

We are interested in localized, bound state  solutions   
of Eqs.\eqref{eqs1}.
 These equations admit a global symmetry 
\be                               \label{glob}
\delta\p\to\delta\p+\frac{\Gamma}{2}\,f_{\mbox{\tiny Z}},~~~~~
\delta\Om_1\to \delta\Om_1+\Gamma,~~~~~
\delta\W_1\to\delta\W_1+\Gamma\,\W_{\mbox{\tiny Z}3}\,,
\ee
which is merely the linearised version of \eqref{s2}. Using this symmetry one can 
impose the gauge condition 
\be                  \label{deltap}
\delta\p(\infty)=0, 
\ee
and then the
local behaviour of the solutions at  large $\rho$
can be read off from Eq.\eqref{inf},
\be                                \label{pert0}
\delta\Om_1=\frac{c_7}{\sqrt{\rho}}\,e^{-m_\sigma\rho}+\ldots,~~~
\delta\W_1={c_8}{\sqrt{\rho}}\,e^{-m_\sigma\rho}+\ldots,~~~
\delta\p={c_6}{\sqrt{\rho}}\,e^{-m_\sigma\rho}+\ldots.
\ee 
Here 
\be                         \label{mss}
m_\sigma^2=\mw^2+\sigma^2\,,
\ee
while the logarithm  
present in Eq.\eqref{m-sigma} does not  
appear in the linearized  theory. 
The  local behavior at small $\rho$ can be obtained from Eq.\eqref{orig},
\be                                \label{pertinf}
\delta\Om_1=a_4\rho^\nu+\Gamma+\ldots,~~~
\delta\W_1=\Gamma\W_3+\ldots\,,~~~
\delta\p=q\rho^{|\n-\nu|}+\frac{\Gamma}{2}\,f_{\mbox{\tiny Z}}+\ldots,
\ee 
where we have 
included the free parameter $\Gamma$, 
since the symmetry \eqref{glob} is now fixed
at large $\rho$ and not at $\rho=0$. 
In fact, the gauge condition \eqref{deltap}
imposed at infinity generically implies that  $\delta\W_1(0)\neq 0$,
which does not agree 
with the previously adopted in \eqref{W10}  condition $\W_1(0)=0$.    
However, the advantage
of this gauge is that Z string deformations are clearly localized 
around the string core (see Fig.\ref{FIG1}). One can always apply the symmetry 
\eqref{glob} to return to the gauge where  $\delta\W_1(0)=0$, but then the 
amplitudes $\delta\Om_1,\delta\W_1,\delta p$ will not vanish at infinity. 

The local solutions \eqref{pert0},\eqref{pertinf} contain 6 free parameters:
$c_5,c_6,c_8,a_4,\Gamma,q$. One of them can be absorbed by the overall 
normalization and so there remain only 5, but together with $\sigma^2$ 
their number
is again 6, which is just enough to 
fulfill the 6 matching conditions  to 
construct global solutions of three second order equations \eqref{eqs1}. 
The solutions obtained are characterized by the following properties.  

\begin{figure}[h]
\hbox to\linewidth{\hss%
  \psfrag{x}{$\sin^2\thetaw$}
\psfrag{y}{$\om^2$}
\psfrag{y}{$\om^2$}
  \psfrag{n1nu1}{$(1,1)$}
  \psfrag{n1nu2}{$(1,2)$}
  \psfrag{n2nu1}{$(2,1)$}
  \psfrag{n2nu2}{$(2,2)$}
  \psfrag{n2nu3}{$(2,3)$}
  \psfrag{n2nu4}{$(2,4)$}
 \psfrag{title}{}
 \psfrag{chiral}{chiral}
    \resizebox{8cm}{7cm}{\includegraphics{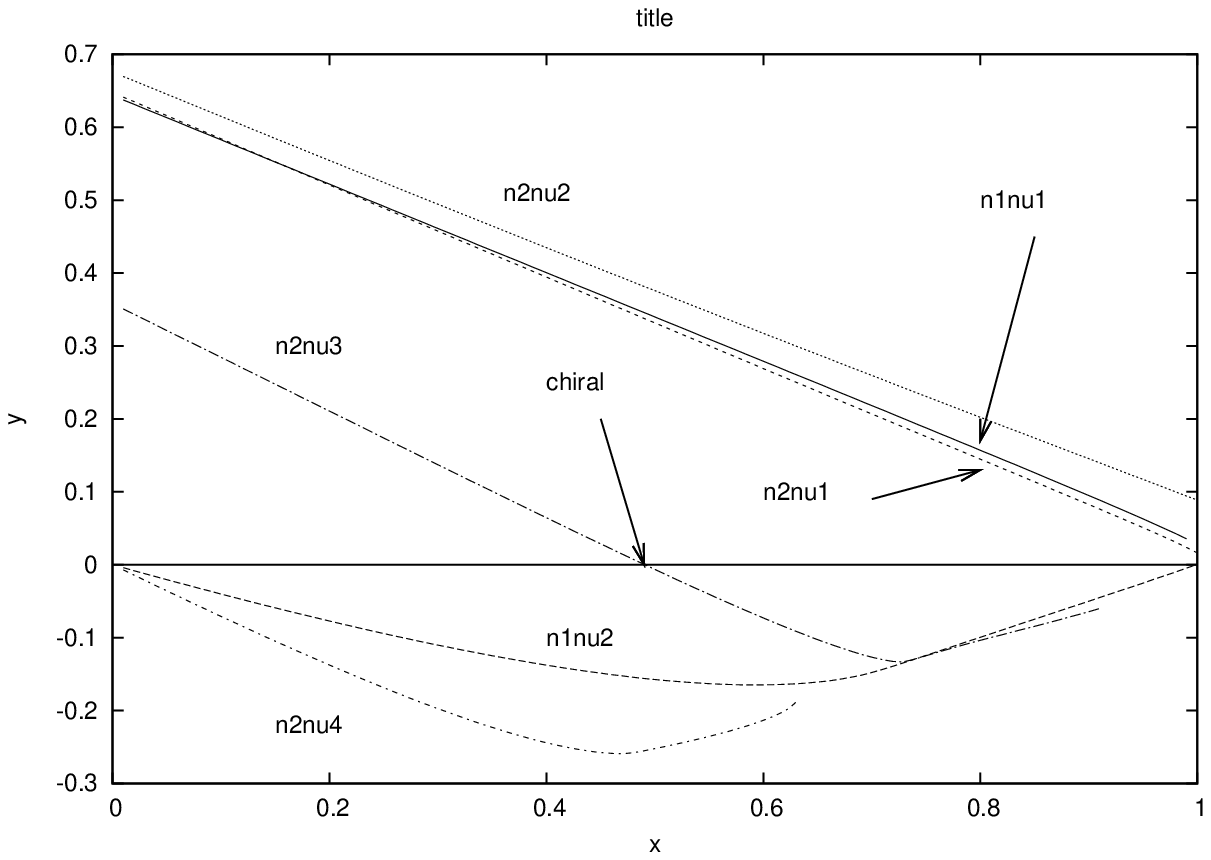}}%
\hspace{2 mm}
  \psfrag{x}{$\sin^2\thetaw$}
 \psfrag{y}{$\om^2$}
  \psfrag{n4nu1}{$(4,1)$}
  \psfrag{n4nu2}{$(4,2)$}
  \psfrag{n4nu3}{$(4,3)$}
  \psfrag{n4nu4}{$(4,4)$}
  \psfrag{n4nu5}{$(4,5)$}
  \psfrag{n4nu6}{$(4,6)$}
  \psfrag{n4nu7}{$(4,7)$}
  \psfrag{n4nu8}{$(4,8)$}
 \psfrag{title}{}
 \psfrag{chiral}{chiral}
    \resizebox{8cm}{7cm}{\includegraphics{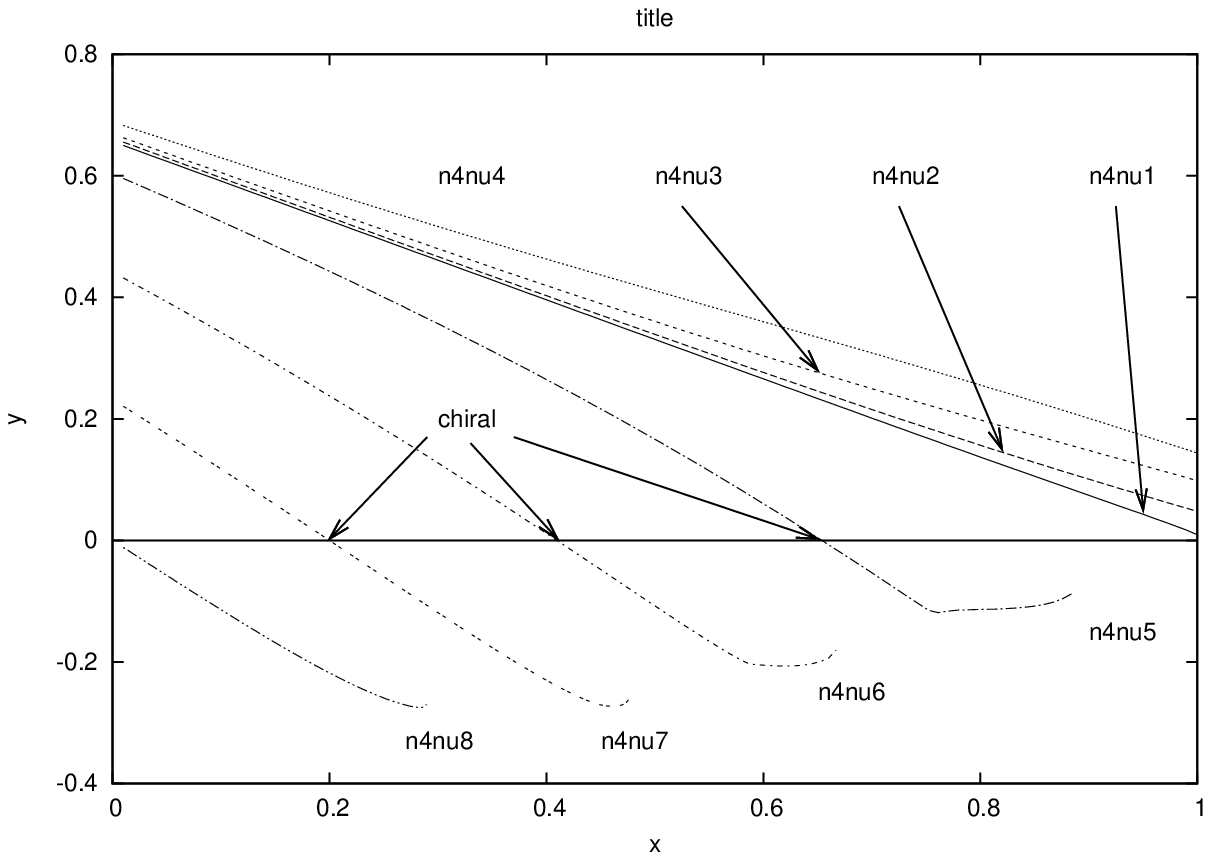}}%
\hss}
\caption{The eigenvalue $\sigma^2$ of the spectral problem \eqref{eqs1}
against $\sin^2\thetaw$ for $\beta=2$. 
The values of $n$ and $\nu$ are shown as $(n,\nu)$.  
In the $\sigma^2<0$ region the curves terminate when the condition 
\eqref{sigmalimit} is violated. The chiral solutions with $\sigma^2=0$ 
are possible only for special values of $\thetaw$. 
}
\label{FIG2a}
\end{figure}

\begin{figure}[t]
\hbox to\linewidth{\hss%
  \psfrag{x}{$\beta$}
 \psfrag{y}{$\sin^2\thetaw$}
  \psfrag{11}{$(1,1)$}
 \psfrag{21}{$(2,1)$}
\psfrag{22}{$(2,2)$}
\psfrag{23}{$(2,3)$}
\psfrag{33}{$(3,3)$}
\psfrag{34}{$(3,4)$}
\psfrag{35}{$(3,5)$}
\psfrag{47}{$(4,7)$}
\psfrag{59}{$(5,9)$}
\psfrag{p}{physical~region}
    \resizebox{14cm}{7cm}{\includegraphics{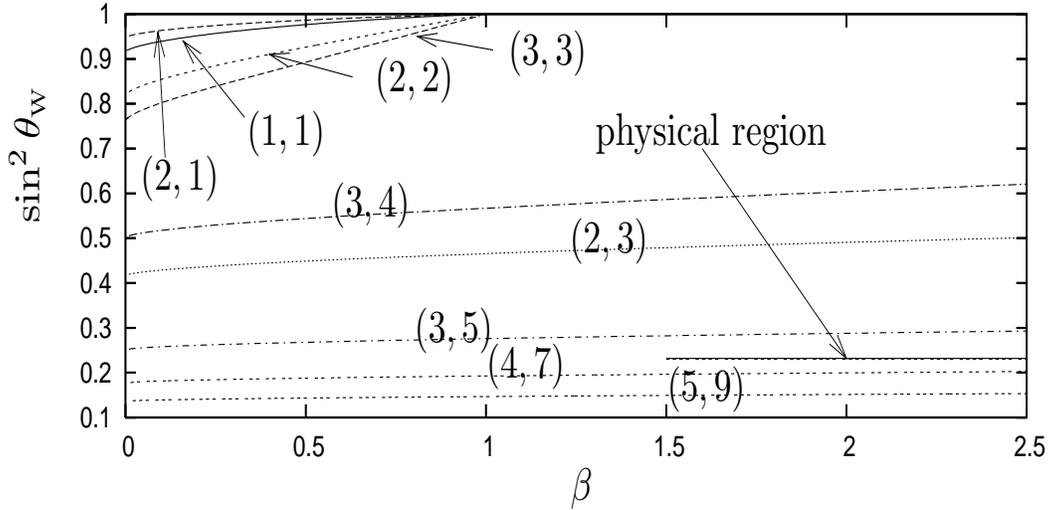}}%
\hss}
\caption{The  
$\sigma^2(\beta,\thetaw,\n,\nu)=0$ curves for several values of $(\n,\nu)$.
Curves with $\nu\leq \n$ are confined in the $\beta\leq 1$ region, while 
those with $\nu>\n$ extend to larger values of $\beta$.  
}
\label{FIG_chiral}
\end{figure}

For given $\beta$ and for small enough $\sin\thetaw$  one finds $2\n$ different
bound states labeled by $\nu=1,2,\ldots ,2\n$ with the eigenvalue 
$\sigma^2=\sigma^2(\beta,\thetaw,\n,\nu)$ (see Fig.\ref{FIG1}). 
These solutions can be interpreted as small deformations
of Z strings by a current $I_\alpha\sim\sigma_\alpha$, 
although the current itself appears only in the next order of perturbation theory. 
If $\beta>1$ then $\n$ of these solutions always have $\sigma^2>0$. For
$\n-1$ solutions the eigenvalue $\sigma^2$ changes 
sign when $\thetaw$ increases, and there is one solution
with  $\sigma^2<0$ for any $\thetaw>0$ (see Figs.\ref{FIG2},\ref{FIG2a}).
If $\sigma^2$ is negative then it cannot be 
too large, since solutions are localized if only 
the mass \eqref{mss} is real, 
 which requires that
\be                            \label{sigmalimit}
\sigma^2>-\mw^2\,.
\ee
Every Z string admits therefore `magnetic' ($\sigma^2>0$) 
and, unless $\thetaw$ is close to $\pi/2$, 
`electric' ($\sigma^2<0$) 
linear deformations. The `chiral' ($\sigma^2=0$) deformations are not generic
and possible only for special values of $\beta,\thetaw$ such that 
\be                      \label{sigma0}
\sigma^2(\beta,\thetaw,\n,\nu)=0.
\ee
This condition 
determines a set of curves in the $\beta,\thetaw$ plane
(see Fig.\ref{FIG_chiral}), let us call them chiral curves.
They coincide with the curves delimiting the parameter regions of 
Z string stability \cite{GH}. 
This can be explained by the fact that 
a simple reinterpretation of the above considerations allows us to
reproduce 
the results of the Z string stability analysis \cite{GH}.
Let
\be                     
(\delta{\cal W},\delta\Phi)\sim \exp\{\pm i(\sigma_0 x^0+\sigma_3 x^3)\}
\ee
be the Z string perturbation eigenvector \eqref{003aZ1} 
for  an eigenvalue $\sigma^2=\sigma_3^2-\sigma_0^2$. 
One has
\be
\sigma_0=\sqrt{\sigma_3^2-\sigma^2}
\ee  
from where it follows that if 
$\sigma^2>0$ then choosing
$\sigma_3<\sigma$ gives  
exponentially growing in time solutions, that is unstable modes,
since the frequency $\sigma_0$ is then imaginary, while 
choosing $\sigma_3>\sigma$ (as was done above) 
gives stationary Z string deformations, since 
$\sigma_0$ is then real.    
If $\sigma^2<0$ then $\sigma_0$ is always real. The sign of $\sigma^2$
therefore determines if the negative modes are present or not. 
As a result, the $\sigma^2(\beta,\thetaw,\n,\nu)=0$ curves separate the 
parameter regions in the $\beta,\thetaw$ plane in which the $\nu$-th deformation
of the $\n$-th Z string is of the
electric/magnetic type, and at the same time the regions
where the $\nu$-th Z string perturbation mode is of stable/unstable type.

\section{Generic superconducting vortices}
\setcounter{equation}{0}

Having constructed the slightly deformed Z strings in the linear theory, 
we can  promote them to solutions of the 
full system of equations \eqref{ee1}--\eqref{ee8}, first for small $q$. 
In the fully nonlinear theory all 8 field amplitudes deviate from their
Z string values, and not only $\Om_1,\W_1,\p$. However, if the 
deviations  $\delta\Om_1,\delta\W_1,\delta\p$ scale as $q$ for $q\ll 1$, those 
for the remaining 5 amplitudes scale as $q^2$. 

Starting from small values of $q$, we increase 
$q$ iteratively, thereby obtaining 
fully non-linear superconducting vortices.  
Equations \eqref{ee1}--\eqref{ee8} can be viewed in this case 
as a non-linear boundary 
value problem with the eigenvalue $\sigma^2$.  
It turns out that only the magnetic ($\sigma^2>0$) 
and chiral ($\sigma^2=0$) solutions are compatible with the 
boundary conditions at infinity. 
The main difference with the linearized case is that, once all  fields
amplitudes are taken into account, the logarithmic term appears in the 
expression \eqref{m-sigma} for the effective mass,
\be                                 \label{mmm}
m_\sigma^2=\mw^2+\sigma^2(Q\ln\rho+c_1)^2.
\ee
This implies that for $\sigma^2<0$ one has the `tachyonic mass'
$m^2_\sigma<0$ at large $\rho$, in which case 
the asymptotic solutions \eqref{inf} oscillate  and do not approach 
the vacuum at infinity (this  was not realised in Ref.\cite{V07}). 
The energy then 
diverges  faster than $\ln\rho$ and    
the interpretation of this is not so clear, since such solutions 
cannot be considered as field-theoretic models of electric wires
or charge distributions. 
We shall therefore not consider the $\sigma^2<0$ 
solutions  in what follows.

For the magnetic solutions with $\sigma^2>0$ the problem does not arise,   
since the logarithmic term in the effective mass \eqref{mmm} only improves
their localization. As a result, the magnetic solutions do generalize 
within the full, non-linear theory.  
The described above multiplet structure of the perturbative solutions persists also at the 
non-linear level, up to the fact that only solutions with $\sigma^2\geq 0$ are now allowed. 
For $n=1$ one finds one such solution, with $\nu=1$, while 
for $n=2$ there are already three: with $\nu=1,2$ and,  
provided that $\thetaw$ is not too large, also with 
$\nu=3$ (see Fig.\ref{FIG2a}). 
The general rule seems to be such that for a given $n$ there are $2n-1$ solutions,
of which those with $\nu=1,2,\ldots n$ exist for any $\thetaw$ while those with 
$\nu=n+1,\ldots,2n-1$ exist if only $\thetaw$ is not too close to $\pi/2$, as shown in 
Figs.\ref{FIG2},\ref{FIG2a}.

The regions of their existence are delimited by the chiral curves in 
Fig.\ref{FIG_chiral}. For each given curve one has 
$\sigma^2(\beta,\thetaw,n,\nu)>0$ in the region 
{\it below} the curve.
Let us call it {\it allowed region}, it corresponds to
magnetic solutions that generalize within the full
non-linear theory. 
The region {\it above} the curve corresponds to the electric 
solutions with $\sigma^2<0$, we therefore call it
{\it forbidden region}. 
The chiral curves with
$\nu\leq n$ are all contained in the upper left corner of the diagram where  
$\beta\leq 1$, while those for $\nu>n$ extend to the region where  $\beta>1$. 
It follows that if $\nu\leq n$ then the allowed region corresponds to 
any $\thetaw$ if $\beta>1$, and to $\thetaw$ that is not too close to $\pi/2$
if $\beta<1$. If $\nu> n$ then the allowed region corresponds 
to the lower part of the diagram located 
below the $(n,\nu)$ chiral curve.

The chiral curves in Fig.\ref{FIG_chiral} are obtained in the limit of vanishing current,
while for finite currents they remain qualitatively similar but shift {\it upward}. 
For large currents they seem to approach the upper boundary of the
diagram, such that the allowed regions increase. Therefore, without entering too much
into details, one can simply
say that the superconducting vortices 
exist for {\it almost all} values of $\beta,\thetaw$.  
  
%%%%%%%%%%%%%%%%%%%%%%%%%%%%%%%%%%%%%%%%%%%%%%%%%%%%%%%%%%%%%%%%%%%%%%%%%%
\begin{figure}[h]
\hbox to\linewidth{\hss%
  \psfrag{x}{$\ln(1+\rho)$}
 \psfrag{lnx}{$\ln(1+\rho)$}
\psfrag{v}{$v$}
\psfrag{v1}{$v_1$}
\psfrag{v3}{$v_3$}
\psfrag{u3}{$u_3$}
\psfrag{u1timesigma}{$\sigma u_1$}
\psfrag{utimesigma}{$\sigma u$}
\psfrag{f1}{$f_1$}
\psfrag{f2}{$f_2$}
\psfrag{title}{}
%  \resizebox{9cm}{7cm}{\includegraphics{sigma_opt_n=1_nu=1_g=0.77_log.eps}}%
% \resizebox{9cm}{7cm}{\includegraphics{unitary_n=1_nu=1_sigma_opt.eps}}%
  \resizebox{8.5cm}{7cm}{\includegraphics{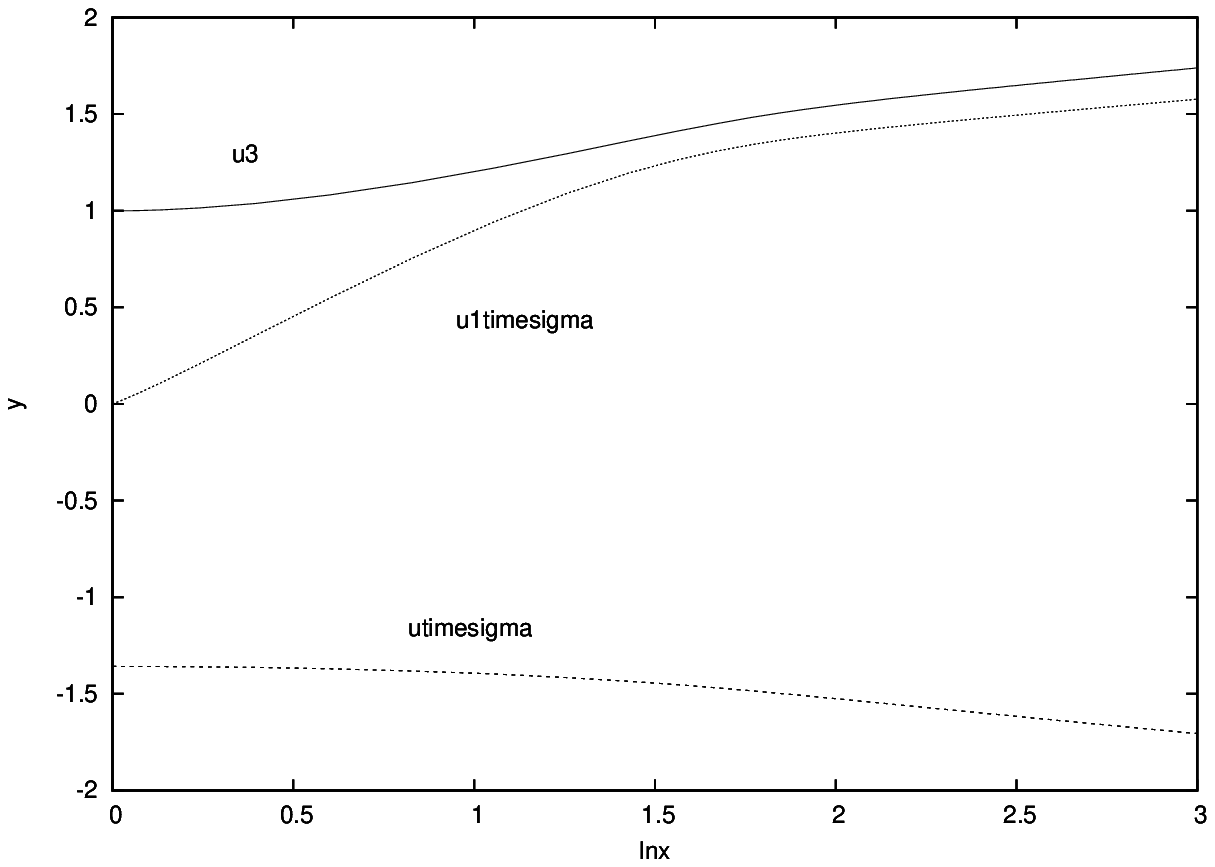}}%
 \resizebox{8.5cm}{7cm}{\includegraphics{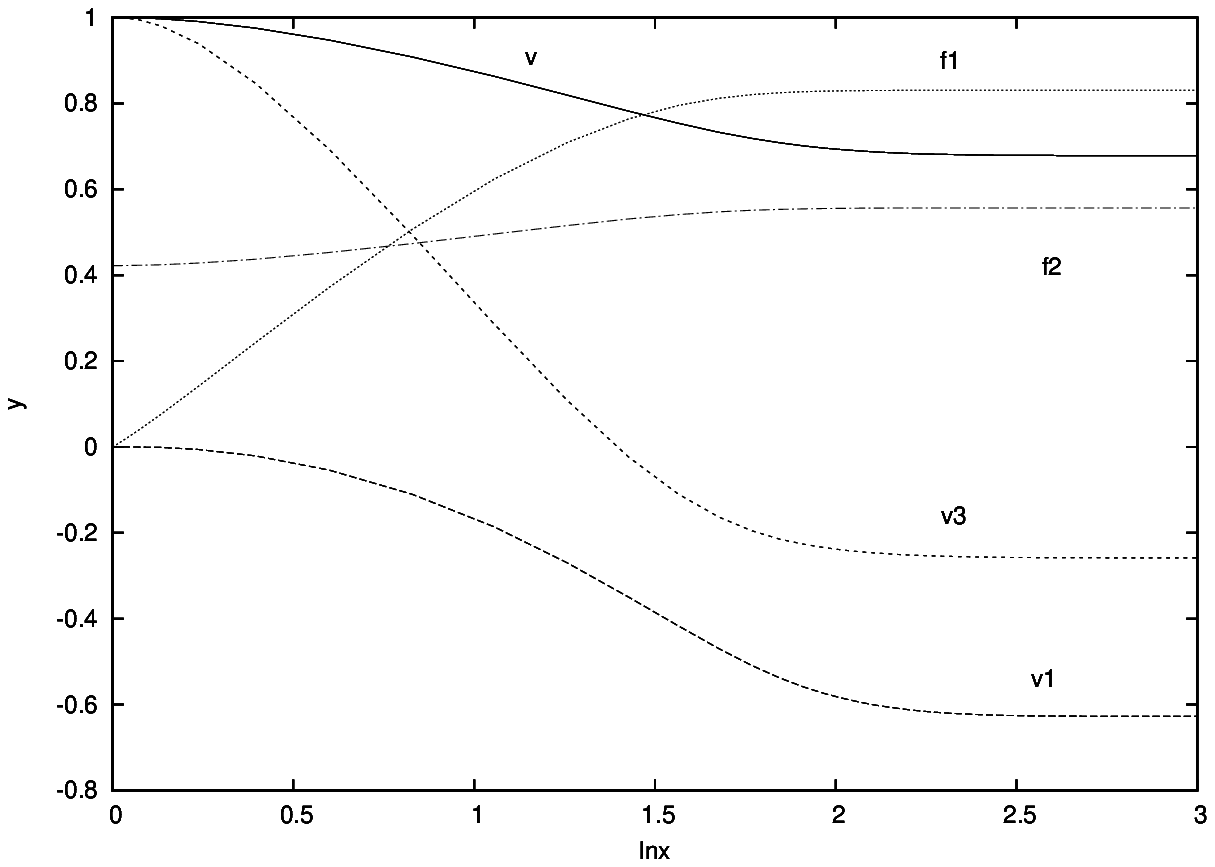}}%
\hss}
\caption{\small  
Profiles of the superconducting vortex solution for 
$\beta=2$, $g^{\prime 2}=0.23$, $\n=\nu=1$ and  
$\sigma=0.373$. The `electric' amplitudes (left) show at large $\rho$ 
the logarithmic growth typical for the Biot-Savart field. 
The remaining amplitudes (right) are everywhere bounded.
}
\label{FIG4}
\end{figure}
%%%%%%%%%%%%%%%%%%%%%%%%%%%%%%%%%%%%%%%%%%%%%%%%%%%%%%%%%%%%%%%%%%%%%%%%%%

The typical solution is shown in Fig.\ref{FIG4}. 
These solutions describe  vortices carrying  an electric current --
smooth, globally regular field-theoretic analogues of electric wires. 
At small $\rho$ they  exhibit a completely regular core 
filled with massive fields 
creating   the  electric charge and current. 
At large $\rho$ the massive fields decay and there remains only 
the Biot-Savart electromagnetic field. 
Integrating the charge and current densities over the vortex cross section  gives the 
vortex `worldsheet current' $I_\alpha\sim\sigma_\alpha$ whose components are
the vortex electric charge (per unit length), $I_0$, 
and the total electric current through the 
vortex cross section, $I_3$. Since 
$\sigma^2\equiv \sigma_3^2-\sigma_0^2>0$, the vector $I_\alpha$ is spacelike, which 
means that there is a comoving reference frame where $\sigma_0=I_0=0$.
However, the current  $I_3$ cannot be boosted away, 
so that this is an essential parameter, which
suggests the term
`superconductivity'. 
In what follows we shall denote by ${\cal I}$ the value of $I_3$ in the
comoving frame. We shall also call $\sigma$ {\it twist}, since in the comoving frame 
one has $\sigma_3=\sigma$, which determines the $z$-dependent relative phase 
of the two Higgs field components  in the ansatz \eqref{003a}.

The solutions exist for any 
$\beta>0$ and for $\thetaw$ belonging to the region below the $(n,\nu)$-chiral curve
in Fig.\ref{FIG_chiral}. Solutions depend on 
$\beta,\thetaw$  and also on
$q,\n,\nu$ so that $\sigma^2$ is in fact a function
of  five arguments, $\sigma^2=\sigma^2(\beta,\thetaw,\n,\nu,q)$. 
In addition, reconstructing the fields \eqref{003a}, 
every solution of the differential equations 
determines actually a whole family of vortices 
with fixed $\om^2=\om_3^2-\om_0^2$ but with different values of  
$(\om_0,\om_3)$ related to each other by Lorentz boosts. 
 For given $\beta,\thetaw$ the superconducting vortices 
therefore comprise a four parameter family that  
can be labeled by $\n,\nu,\sigma_0$ and $q$.  
These parameters determine  
physical quantities associated to the vortex: the electromagnetic and Z fluxes 
$\mbox{ $\Psi$}_{F}$ and 
$\mbox{ $\Psi$}_{Z}$, 
the charge $I_0$ per unit vortex length
and the current $I_3$, 
as well as the momentum along the vortex, $P$, and the angular momentum $M$. 
\begin{figure}[h]
\hbox to\linewidth{\hss%
  \psfrag{x}{{$\q$}}
 \psfrag{y}{${\cal I}$}
 \psfrag{star}{$\sigma_\star^2$}
\psfrag{n=2_nu=1}{$n=2,\nu=1$}
\psfrag{n=2_nu=2}{$n=\nu=2$}
\psfrag{n=1_nu=1}{$n=\nu=1$}
\psfrag{n=2_nu=3}{$n=2,\nu=3$}
\psfrag{title}{}
  \resizebox{8.5cm}{7cm}{\includegraphics{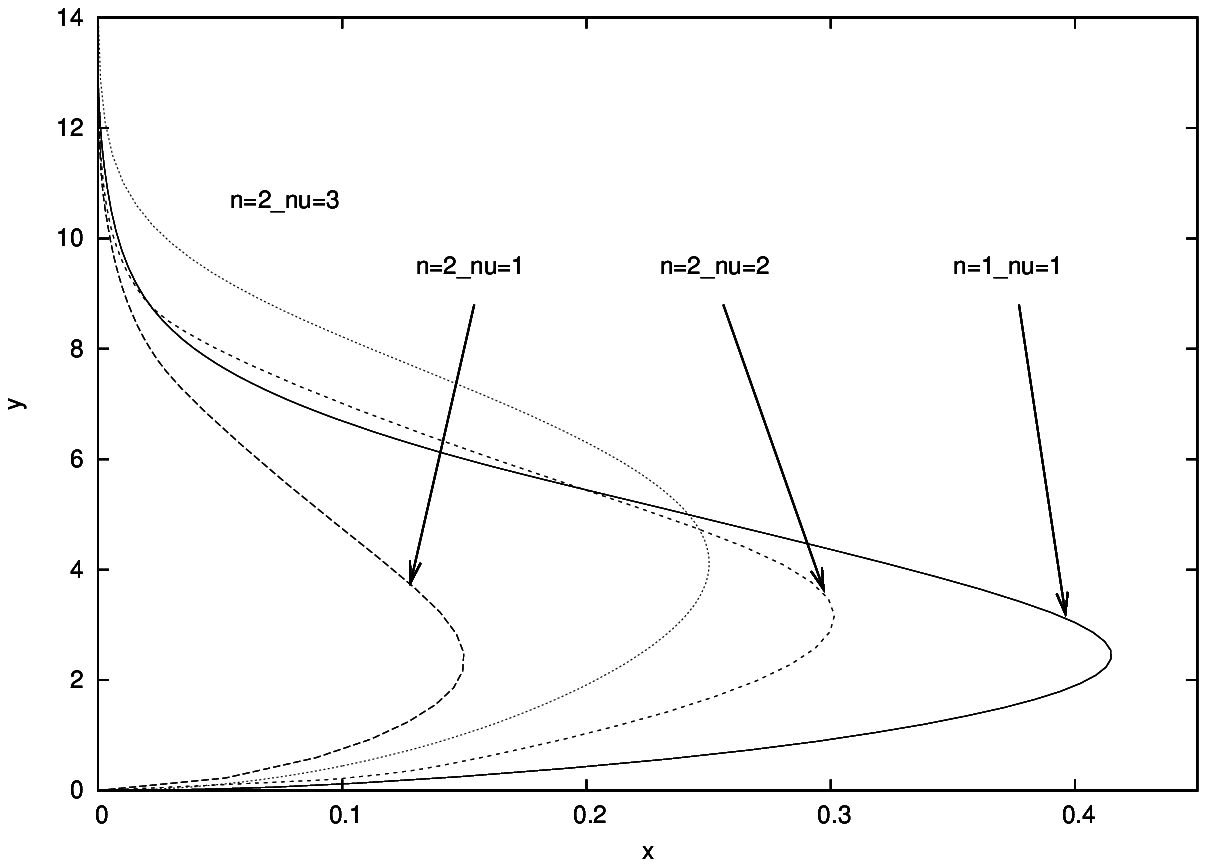}}%
  \psfrag{x}{$\q$}
  \psfrag{y}{$\sigma$}
\psfrag{title}{}
  \resizebox{8.5cm}{7cm}{\includegraphics{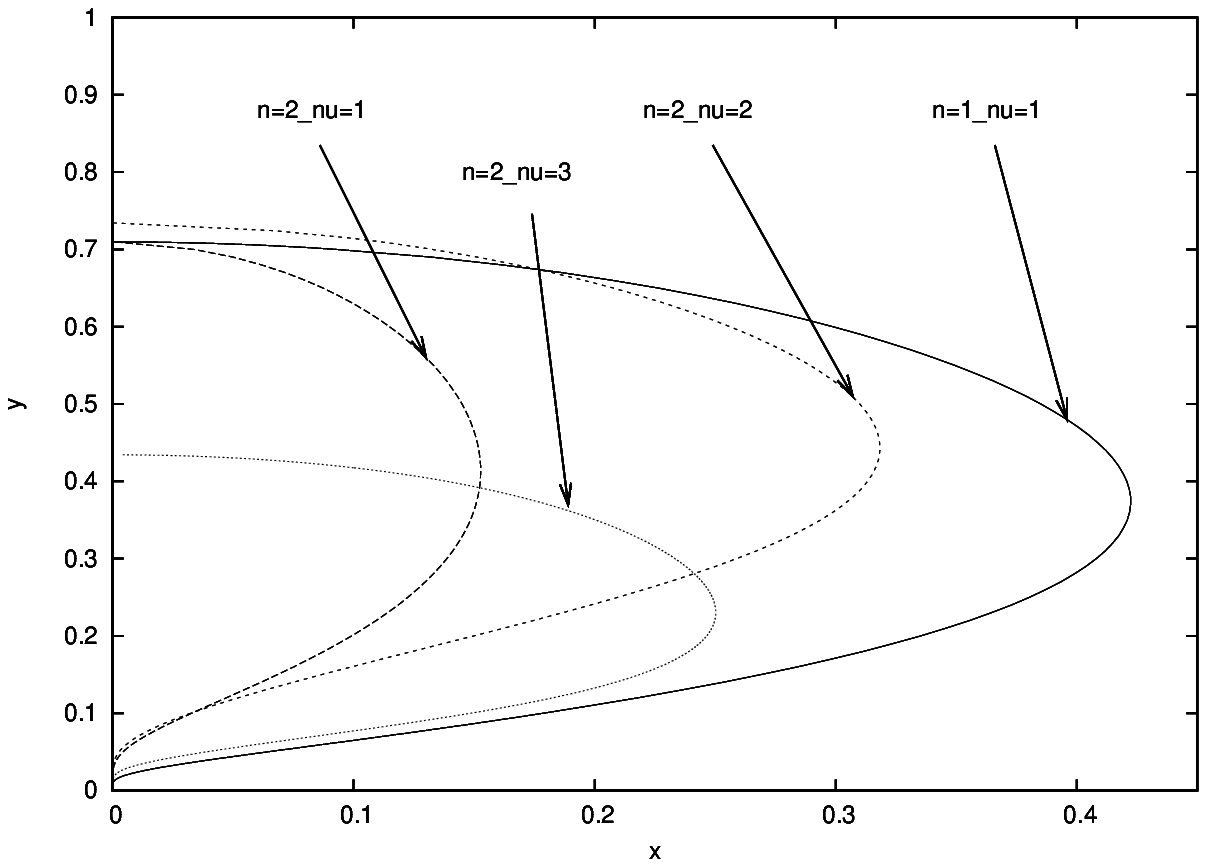}}%
\hss}
\caption{\small 
The restframe current ${\cal I}$ (left) and twist $\sigma$ (right) 
as functions of the parameter $q$ for the 
solutions with 
$\beta=2$, $\sin^2\thetaw=0.23$.    
}
\label{FIG6}
\end{figure}

The parameter $q$ measures the deviation from the Z string limit. 
%Let us call it scalar condensate parameter -- since it determines the 
%`condensate  value' of the lower component of the Higgs field 
%in the vortex core. 
%In particular, for $\nu=n$ one has $q=\p(0)$.  
 Increasing $q$ starting from zero
increases ${\cal I}$. It turns out, however, that the dependence ${\cal I}(q)$
is not one-to-one, since for $q$ not exceeding a certain maximal value 
$q_{\star}(\beta,\thetaw,n,\nu)$ 
one finds {\it two} different solutions with different currents, 
as shown in Fig.\ref{FIG6},  
while for $q>q_{\star}$ there are no solutions at all. 
 The solutions thus show two different branches that merge 
for $q\to\q_\star$ but have different behaviour for $q\to 0$, when  the  
lower branch reduces to the currentless  Z string, while the current 
of the upper branch solutions grows seemingly without bounds.   
It is worth noting that this behaviour  is drastically different
from that found in the model of superconducting strings of Witten (see Fig.\ref{FIG-Wit} 
in the Appendix C), where the maximal value of $q$ corresponds to the `dressed'
currentless string. 

\begin{figure}[ht]
\hbox to\linewidth{\hss%
\psfrag{x}{{${\cal I}$}}
\psfrag{y}{{$\sigma$}}
\psfrag{title}{}
\psfrag{n=1_nu=1}{$n=\nu=1$}
\psfrag{n=2_nu=1}{$n=2,\,\nu=1$}
\psfrag{n=2_nu=2}{$n=\nu=2$}
\psfrag{n=2_nu=3}{$n=2,\,\nu=3$}
	\resizebox{8.5cm}{7cm}{\includegraphics{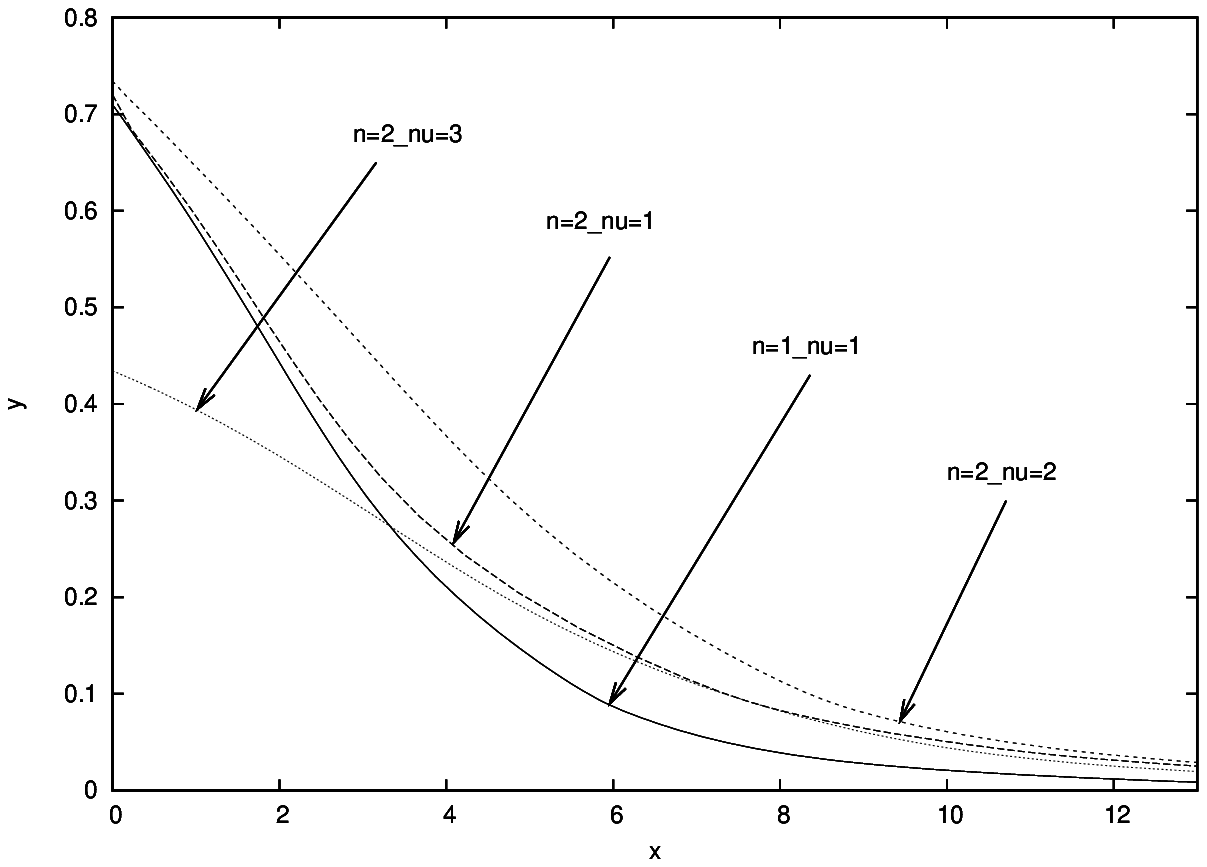}}
\hspace{5mm}%
%\psfrag{x}{{$\sigma$}}
\psfrag{y}{{$\gamma$}}
\psfrag{pi/2}{{$\frac{\pi}{2}$}}
\psfrag{title}{}
        \resizebox{8.5cm}{7cm}{\includegraphics{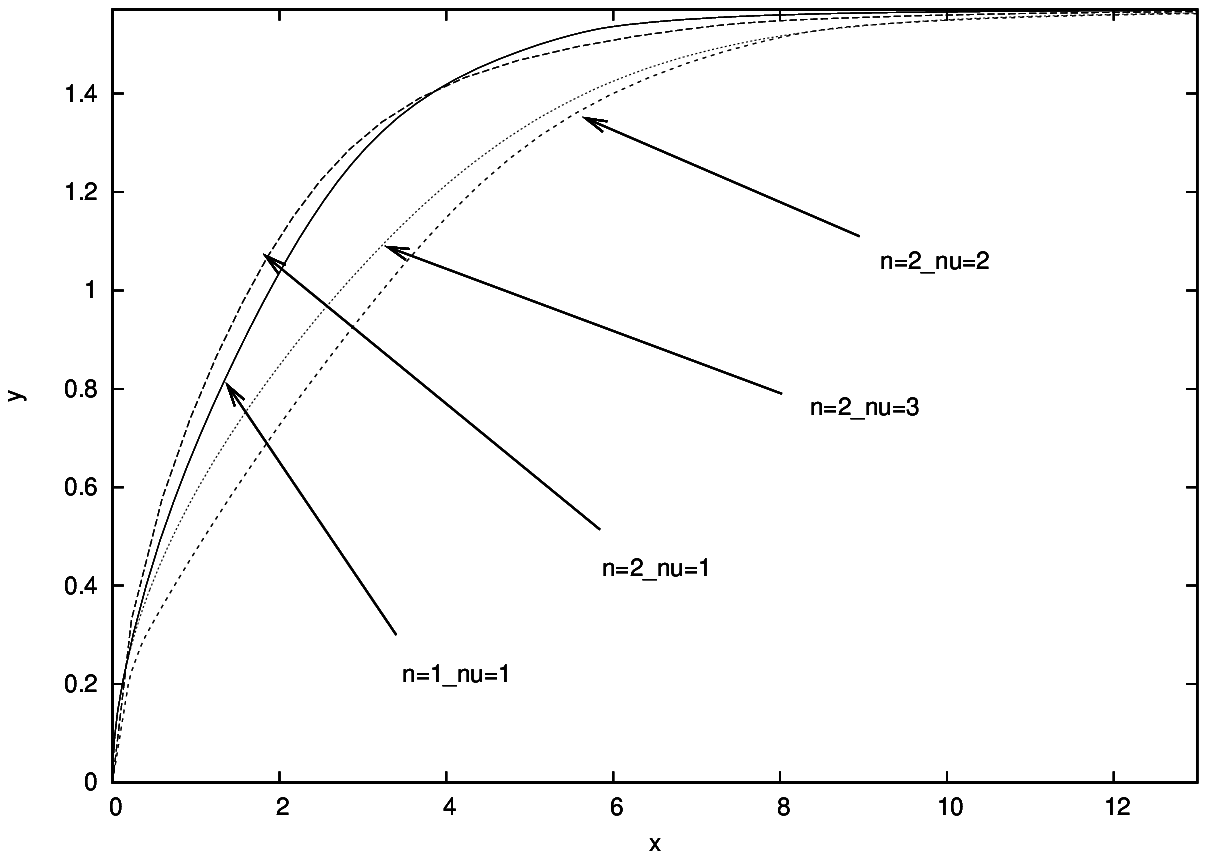}}	
\hss}

\caption{\small  The twist $\sigma$ (left) and vacuum angle $\gamma$ (right) 
against the current ${\cal I}$ for the solutions with 
$\beta=2$, $\sin^2\thetaw=0.23$.  One has 
 $\sigma\to 0$ and $\gamma\to\pi/2$
when ${\cal I}$ grows, while the product $\sigma {\cal I}^3$
approaches a non-zero value, so that $\sigma\sim{\cal I}^{-3}$.   
}
\label{fig3a}
\end{figure}

\begin{figure}[ht]

\hbox to\linewidth{\hss%
\psfrag{x}{{${\cal I}$}}
\psfrag{y}{{$c_1$}}
\psfrag{n=1_nu=1}{$n=\nu=1$}
\psfrag{n=2_nu=1}{$n=2,\,\nu=1$}
\psfrag{n=2_nu=2}{$n=\nu=2$}
\psfrag{n=2_nu=3}{$n=2,\,\nu=3$}
\psfrag{title}{}
	\resizebox{8.5cm}{7cm}{\includegraphics{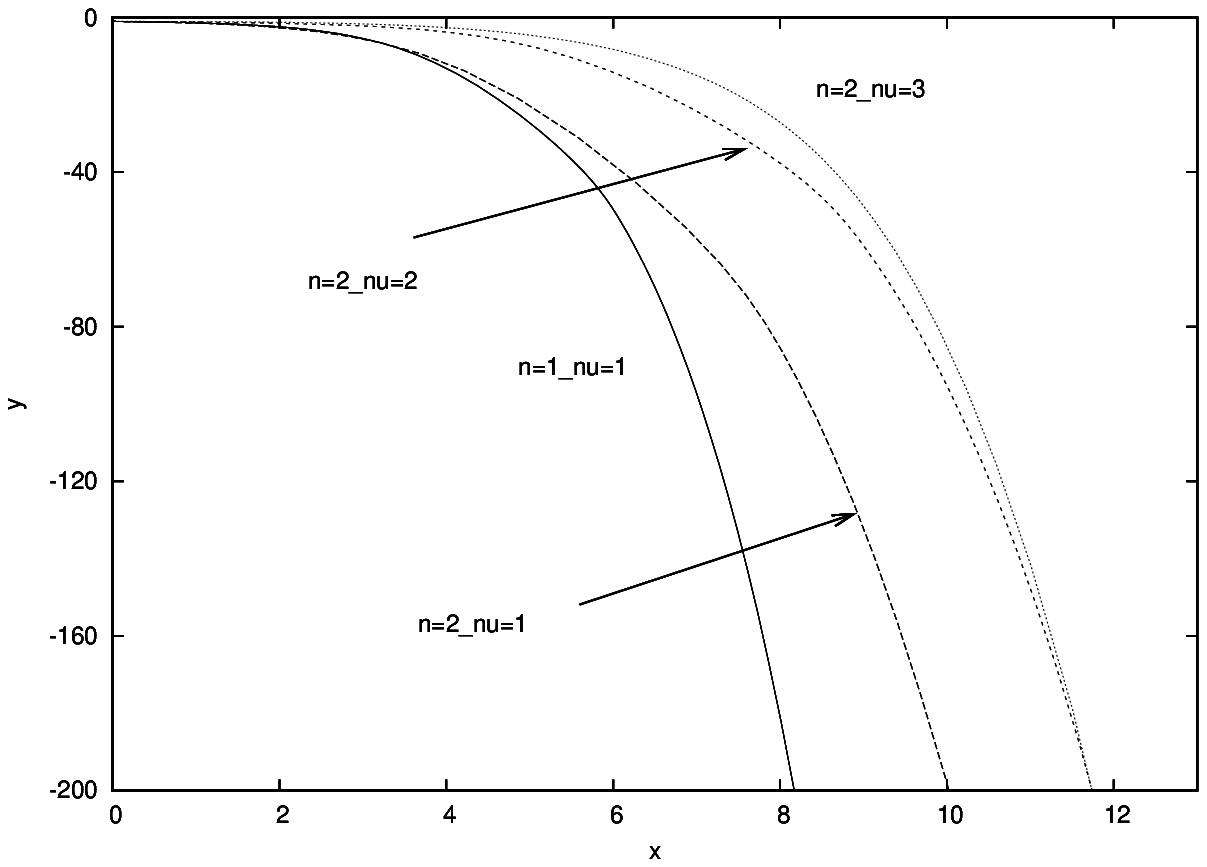}}
\hspace{5mm}%
%\psfrag{x}{{$\sigma$}}
\psfrag{y}{{$c_2/(2\nu-n)$}}
\psfrag{g2}{{$g^2$}}
\psfrag{title}{}
        \resizebox{8.5cm}{7cm}{\includegraphics{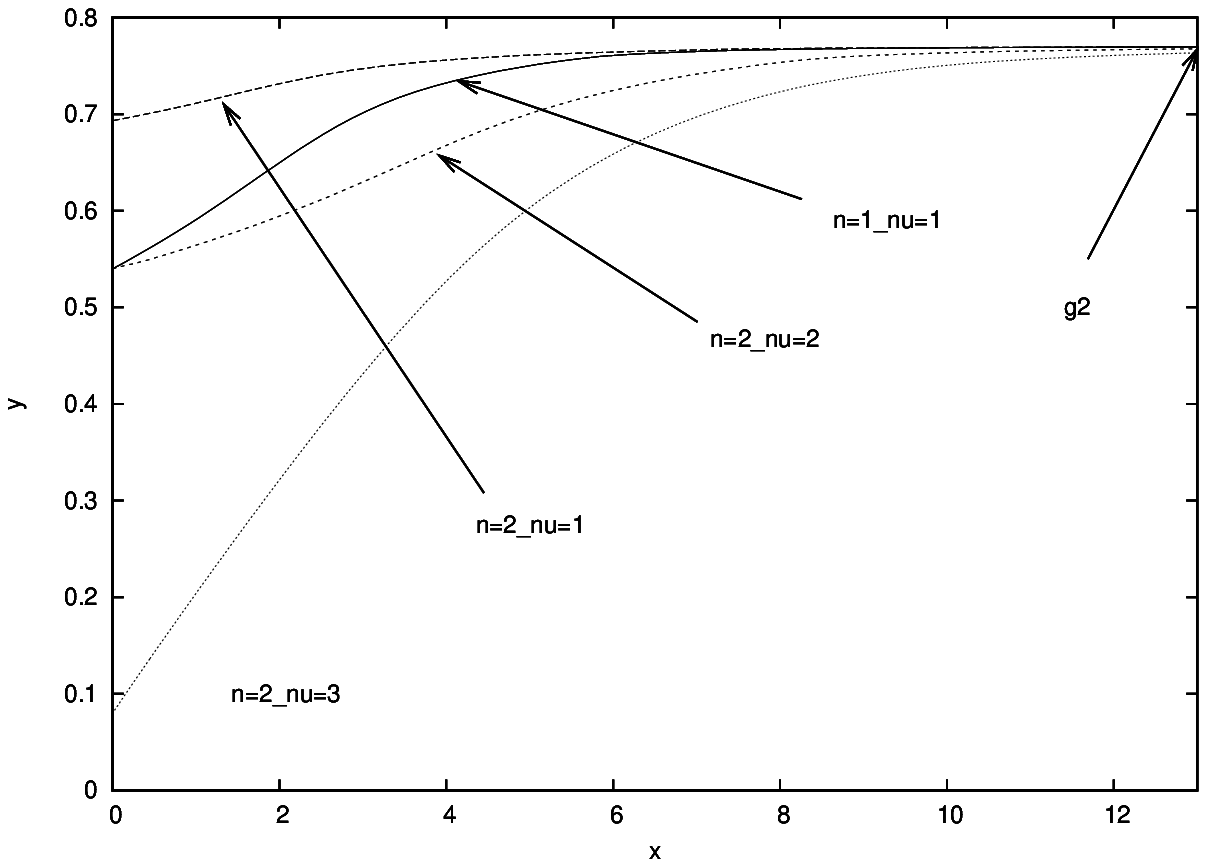}}	
\hss}

\caption{\small  The parameters $c_1$ (left) and $c_2/(2n-\nu)$ (right) 
against ${\cal I}$ for the $n=1,2$ solutions with 
$\beta=2$, $\sin^2\thetaw=0.23$. One has  $c_2\to(2\nu-n)g^2$ 
for ${\cal I}\to\infty$. 
}
\label{fig3b}
\end{figure}

A similar two-branch structure emerges also for other solution parameters 
traced against $q$, as for example for $\sigma(q)$ (see Fig.\ref{FIG6}).
This suggests that solutions should be labeled not by $q$ but by 
a parameter that changes monotonously, as
for example the current ${\cal I}$ or twist $\sigma$
(see Figs.\ref{fig3a},\ref{fig3b}).   

%%%%%%%%%%%%%%%%%%%%%%%%%%%%%%%%%%%%%%%%%%%%%%%%%%%%%%%%%%%%%%ù
%%%%%%%%%%%%%%%%%%%%%%%%%%%%%%%%%%%%%%%%%%%%%%%%%%%%%%

%%%%%%%%%%%%%%%%%%%%%%%%%%%%%%%%%%%%%%%%%%%%%%%%%%%%%%%%%%%%%%ù
%%%%%%%%%%%%%%%%%%%%%%%%%%%%%%%%%%%%%%%%%%%%%%%%%%%%%%

\begin{figure}[h]
\hbox to\linewidth{\hss%
  \psfrag{x}{{${\cal I}$}}
 \psfrag{y}{{$\mbox{ $\Psi$}_{F}/\nu$}}
\psfrag{n=1_nu=1}{$n=\nu=1$}
\psfrag{n=2_nu=1}{$n=2,\,\nu=1$}
\psfrag{n=2_nu=2}{$n=\nu=2$}
\psfrag{n=2_nu=3}{$n=2,\,\nu=3$}
\psfrag{title}{}
  \resizebox{8.5cm}{7cm}{\includegraphics{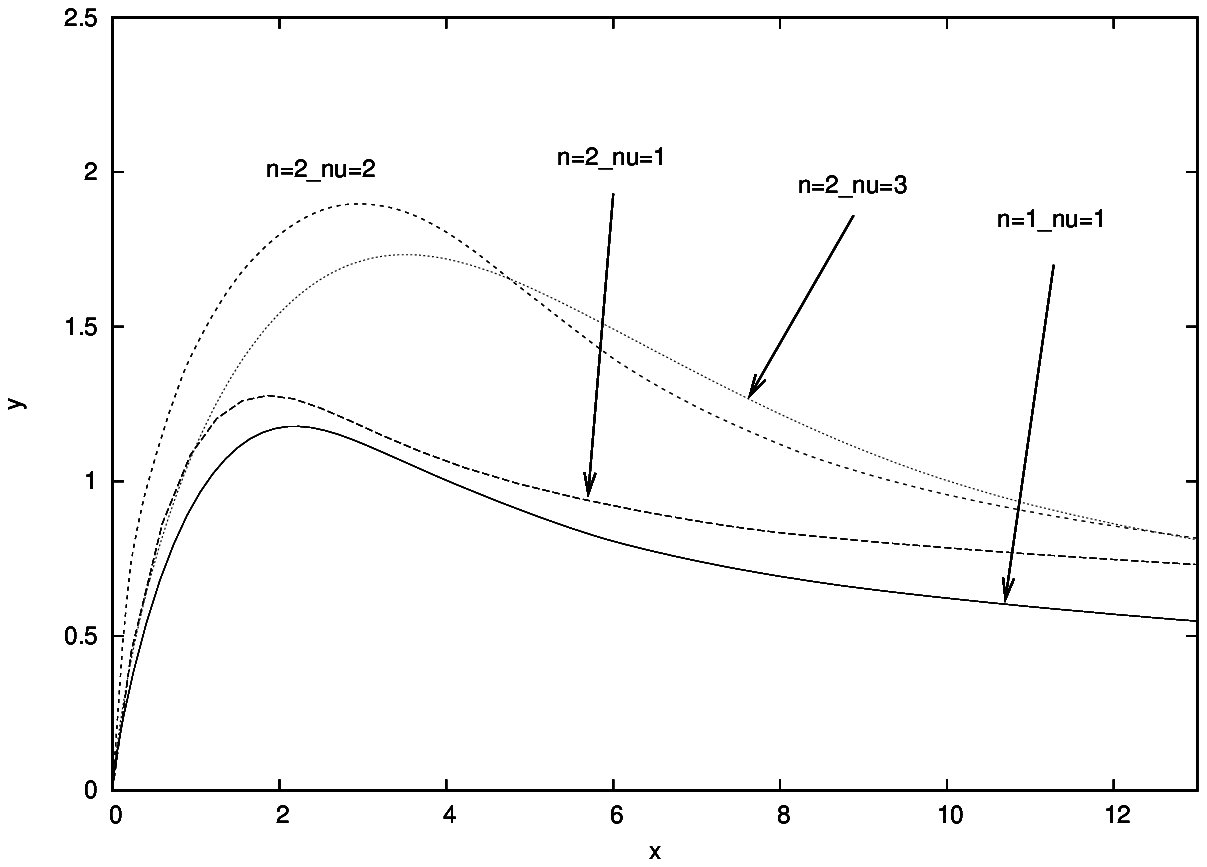}}%
\psfrag{y}{{$\mbox{ $\Psi$}_{Z}/(4\pi n)$}}
  \resizebox{8.5cm}{7cm}{\includegraphics{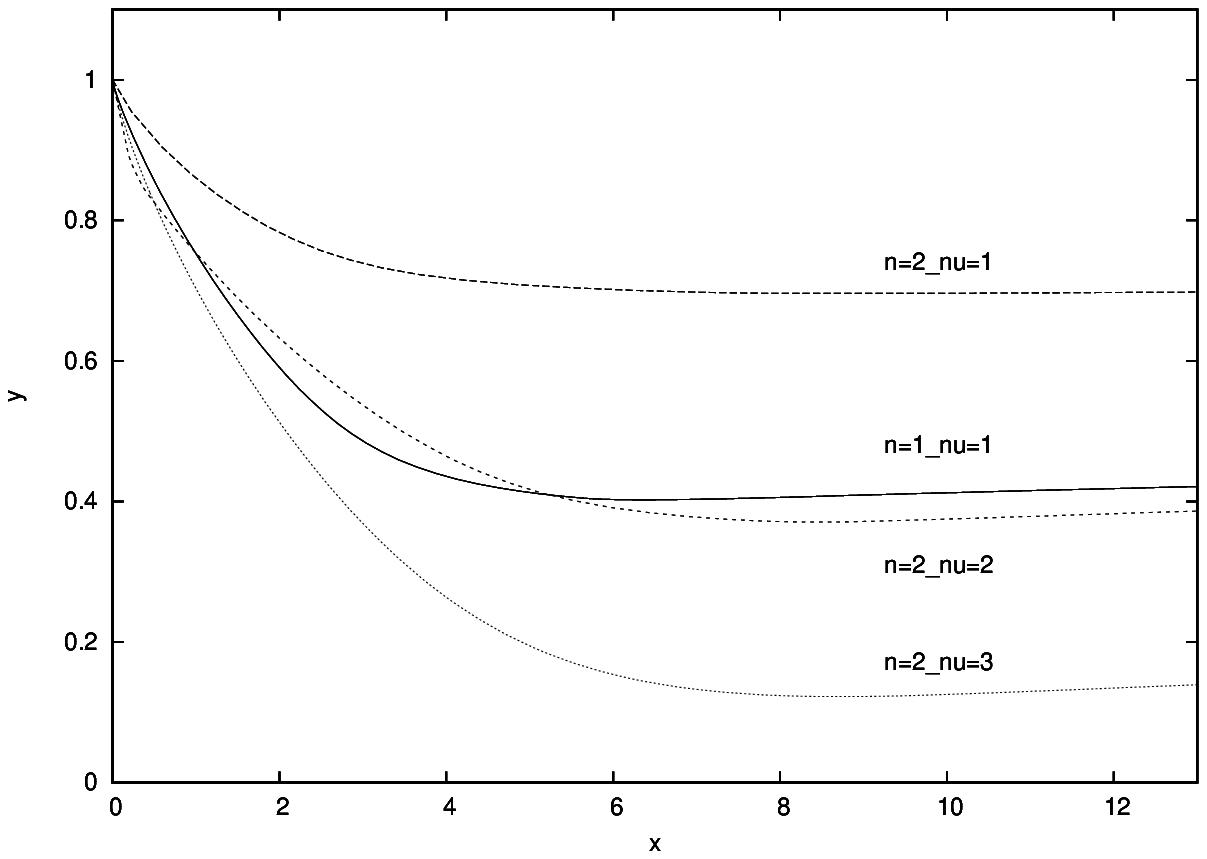}}%
\hss}
\caption{\small 
The Nambu electromagnetic flux   
$\mbox{ $\Psi$}_{F}/\nu$
 (left) and Z flux $\mbox{ $\Psi$}_{Z}/(4\pi n)$ (right)
against the current for the vortices with
$\beta=2$, $\sin^2\thetaw=0.23$.    
}
\label{FIG-fluxes}
\end{figure}

The electromagnetic and Z fluxes calculated according to the Nambu 
definition \eqref{fluxNambu} 
are shown in 
Fig.\ref{FIG-fluxes} as functions of the current. 
We see that in the Z string limit ${\cal I}\to 0$
the Z flux reduces to $4\pi n$
while the electromagnetic flux vanishes,
 as they should. As explained above,
in the full non-Abelian theory 
there is no reason for the fluxes to be quantized, and so 
they vary continuously with ${\cal I}$. 

It is interesting that the fluxes computed according to the 't Hooft 
definition \eqref{fluxHoft} are
completely different and {\it do not} reduce to the 
Z string values in the Z string limit $q\to 0$. 
This is the consequence of the 
fact that the function 
\be
\cos\Omega=\frac{\f^2-\p^2}{\f^2+\p^2}
\ee 
approaches its limiting value for $q\to 0$ 
{\it non-uniformly}. 
Indeed, for $q=0$ one has $\p=0$ and $\cos\Omega=1$ everywhere. 
If $q\ll 1$ then $\cos\Omega\approx 1$ almost everywhere, 
apart from the origin where one has 
$\f\sim\rho^\n$ and $\p\sim q\rho^{|\n-\nu|}$ so that
$\cos\Omega=-1$ at $\rho=0$. 
Using Eqs.\eqref{fluxHoft}  then gives the Z flux
\begin{align}
\mbox{ $\Psi$}^H_{Z}&=4\pi\n,~~~~~~~~~~~~~~~~~~~~~~q=0,\nonumber \\
\mbox{ $\Psi$}^H_{Z}&=4\pi(\n-\nu),~~~~~~~~~~~~~~q\neq 0,~~
\nu=1,\ldots ,2\n-1. 
\end{align}
The formula \eqref{fluxHoft} for the electromagnetic flux contains the parameter $c_2$,
whose value is known only numerically, 
but for $q\to 0$ it reduces
to $c_2=2\n g^2-\nu$, which gives 
\begin{align}
\mbox{ $\Psi$}^H_{F}&=0,~~~~~~~~~~~~~~~~~~~~~~q=0,\nonumber \\
\mbox{ $\Psi$}^H_{F}&=4\pi\nu\,\frac{g^\prime}{g},~~~~~~~~~~~~~~~q\to 0,~~
\nu=1,\ldots ,2\n-1. 
\end{align}
We see that both electromagnetic and Z fluxes computed according to the 't Hooft 
definition are discontinuous for $q\to 0$ and do not reduce to the Z string values 
 when the solutions  approach Z strings in the zero current limit. 
No such problem arises if one uses  the Nambu definition of the fields.   

%%%%%%%%%%%%%%%%%%%%%%%%%%%%%%%%%%%%%%%%%%%%%%%%%%%%%%%%%%%%%%ù
\begin{figure}[ht]

\hbox to\linewidth{\hss%
\psfrag{x}{{${\cal I}$}}
\psfrag{y}{{$Q$}}
\psfrag{title}{}
\psfrag{n=1_nu=1}{$n=\nu=1$}
\psfrag{n=2_nu=1}{$n=2,\,\nu=1$}
\psfrag{n=2_nu=2}{$n=\nu=2$}
\psfrag{n=2_nu=3}{$n=2,\,\nu=3$}
\psfrag{n=1_nu=1}{$n=\nu=1$}
\psfrag{n=2_nu=1}{$n=2,\,\nu=1$}
\psfrag{n=2_nu=2_times0.05}{$n=\nu=2\times 0.05$}
\psfrag{n=2_nu=3_times7e-3}{$n=2,\,\nu=3\times 0.007$}
	\resizebox{8.5cm}{7cm}{\includegraphics{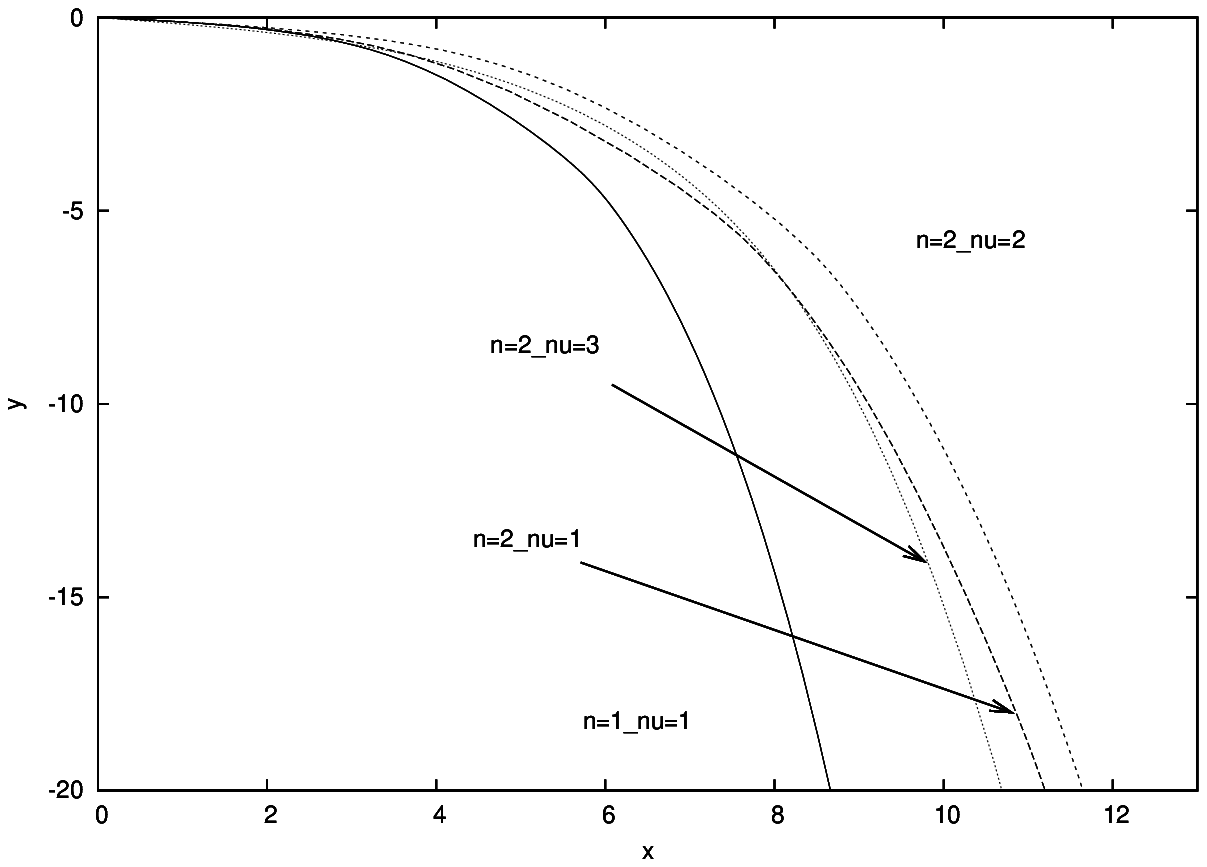}}
\hspace{5mm}%
%\psfrag{x}{{$q$}}
\psfrag{y}{{$E_2/2\pi$}}
\psfrag{n=1_nu=1}{$n=\nu=1$}
\psfrag{n=2_nu=1}{$n=2,\,\nu=1$}
\psfrag{n=2_nu=2}{$n=\nu=2$}
\psfrag{n=2_nu=3}{$n=2,\,\nu=3$}
\psfrag{title}{}
       \resizebox{8.5cm}{7cm}{\includegraphics{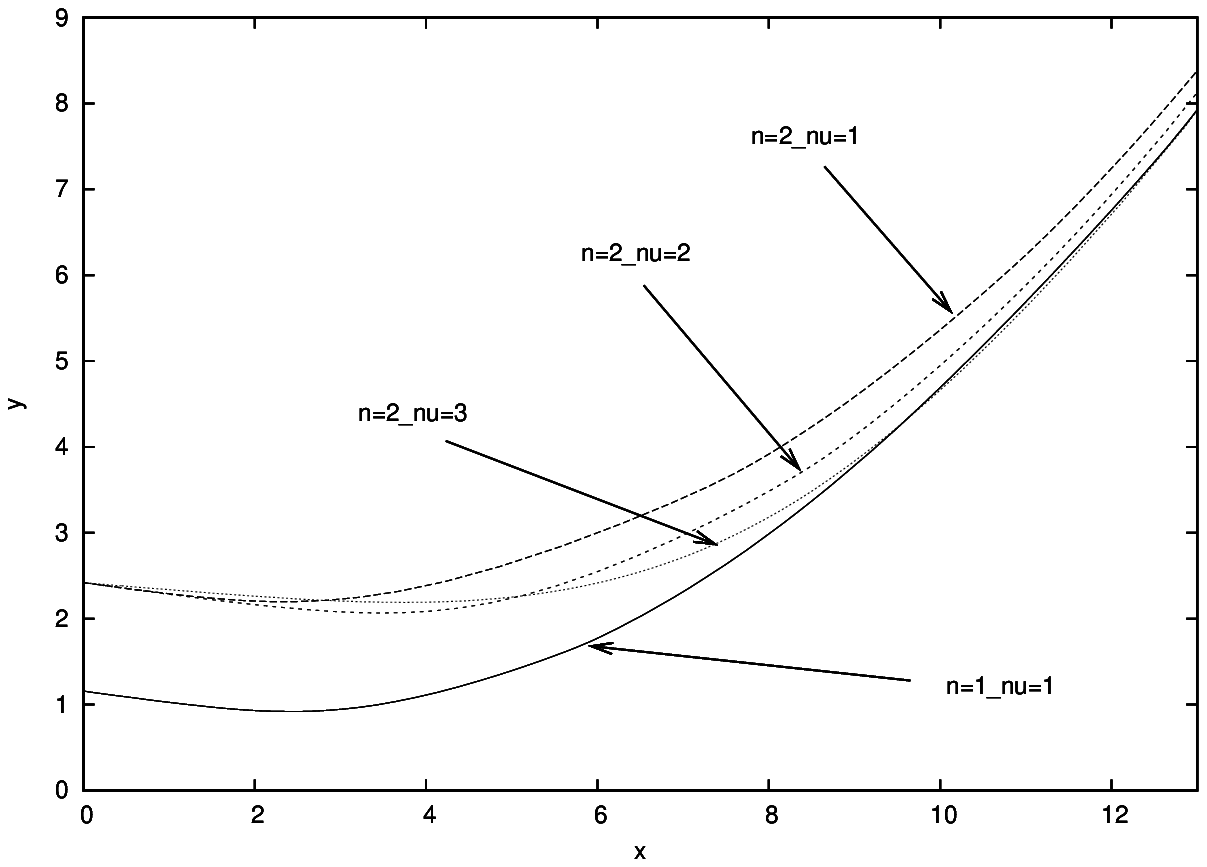}}	
\hss}

\caption{\small  The `Coulombian charge' parameter $Q$ (left) and the 
magnetic energy (right) 
against ${\cal I}$ for the solutions with $\beta=2$, $\sin^2\thetaw=0.23$.  
}
\label{fig35}
\end{figure}
%%%%%%%%%%%%%%%%%%%%%%%%%%%%%%%%%%%%%%%%%%%%%%%%%%%%%%

As the vortices support the long-ranged electromagnetic
field, 
their energy per unit length diverges logarithmically. 
%The same divergence is known for 
%ordinary electrical wires, if they are infinitely long, 
%whereas finite size objects, such as closed vortex loops, 
%will have finite energy. 
The total energy consists of the electric and 
magnetic contributions, $E=E_1+E_2$, and it is  
the electric term that diverges, 
\be
E_1=\pi \frac{\om_0^2+\om_3^2}{g^2g^{\prime 2}}\,Q\Y(\infty),
\ee
since $\Y(\rho)\sim Q\ln\rho$ for $\rho\to\infty$. 
The `Coulombian charge' $Q$ 
 is shown against the current in Fig.\ref{fig35}
and against $\thetaw$ in Fig.\ref{fig36}. 
The magnetic energy $E_2$ is 
finite and reaches its minimal value for $q=q_\star$  as shown in Fig.\ref{fig35}. 
This suggests that the value $q=q_\star$ is in some sense
`energetically preferable'. Since this value determines 
the `bifurcation point' between the two solution branches, 
one can expect that the stability of the solutions 
may change at this point.

\section{Large current limit}

The superconducting 
strings in the U(1)$\times$U(1) model of Witten
exhibit the current quenching  -- 
there is an upper bound for their current 
\cite{Witten,SS,Vilenkin-Shellard}. 
In the Weinberg-Salam theory we ${\it generically}$ 
do not observe this phenomenon. 
Large ${\cal I}$ corresponds to small $\sigma$,
and we were able to descend down to $\sigma\sim 10^{-3}$
(after this the numerics become somewhat involved) 
always finding larger and larger values of ${\cal I}$,
without any tendency for this increase to stop. 

It seems that the reason 
why the current can be unbounded in the 
Weinberg-Salam theory is related to the spin of the current  carries. 
The existence of the critical current in superconductivity  
models usually follows from the  fact that too large currents should produce 
strong magnetic fields which will quench the scalar condensate and 
destroy the superconductivity. 
This is what happens in Witten's model 
where the current is carried by scalars, and so 
quenching the condensate  quenches the current. Specifically, the current defined by  
Eq.\eqref{cur-Witten} of the Appendix C vanishes for $\phi_2\to 0$.  
In the Weinberg-Salam theory, as we shall see below, 
the very strong magnetic field also quenches the Higgs field 
inside the vortex.  However, this does not destroy superconductivity, since 
the current is carried by vectors  and so 
the Higgs field is not the relevant order parameter. 
 In fact, the current defined by 
Eqs.\eqref{Nambu},\eqref{cur1} does not vanish for $\Phi\to 0$.  
For large currents the vector boson condensate can be 
described by the pure Yang-Mills field, whose scale invariance
implies that the current can be as large as one wants.
In the Witten model this does not work since the current carriers are scalars
which break the scale invariance.

At the more technical level  
the current quenching can be related 
to the `dressed' currentless  solutions. 
If they exist, then changing the twist $\sigma$ makes the system interpolate 
between the `bare' and `dressed' strings, and since both of them are currentless,
the current passes through a maximum in between. As we shall see in the 
next section, `dressed' solutions exist in the electroweak theory  
only for special values of $\beta,\thetaw$, which means that current quenching is 
non-generic. For generic parameter values  there are no 
`dressed' solutions, and when one starts from the `bare' 
Z string and decreases the twist,
the current always grows. 

For large currents a simple approximation for the solutions can be found 
that agrees very well with the numerics and suggests that the current can be 
arbitrarily large. 
%A better understanding of the large current limit can be achieved for 
%$\thetaw=\pi/2$, since the solution then  
%can actually be constructed analytically, in terms of an asymptotic series
%\cite{SL}. For $\thetaw<\pi/2$ the   

The large current limit is characterized by the presence of two 
different length scales. One of them does not depend on the current
and is determined by the 
Z boson or Higgs boson masses,
which are both of the same order of magnitude
 (for physical values of $\beta$), 
$
R_{\mbox{\tiny Z}}\sim\mz^{-1}\sim\mh^{-1}\sim 1.
$
Another scale is related to the `dressed' W boson mass \eqref{m-sigma},
which {\it increases} with  current and becomes  large 
for large currents, 
$m_\sigma\sim \sigma Q\sim {\cal I}$, so that the corresponding length
is  small 
$
R_\sigma\sim m_\sigma^{-1}\sim {\cal I}^{-1}.
$
One can therefore expect the solutions to show a small central 
region where  $\rho\leq R_\sigma$ that accommodates  the
very heavy field modes of mass $m_\sigma$.
For $R_\sigma\leq\rho\leq R_{\mbox{\tiny Z}}$ these `supermassive'   
modes die out but 
the other massive modes remain, while the asymptotic region where  
$R_{\mbox{\tiny Z}}\leq\rho$ contains only the massless modes. 

%%%%%%%%%%%%%%%%%%%%%%%%%%%%%%%%%%%%%%%%%%%%%%%%%%%%%%%%%%%%%%%%%%%%%%%%%%
\begin{figure}[h]
\hbox to\linewidth{\hss%
  \psfrag{x}{$\ln(1+\rho)$}
 \psfrag{lnx}{$\ln(1+\rho)$}
\psfrag{v}{$v$}
\psfrag{v1}{$v_1$}
\psfrag{v3}{$v_3$}
\psfrag{u3}{$u_3$}
\psfrag{u1timesigma}{$\sigma u_1$}
\psfrag{utimesigma}{$\sigma u$}
\psfrag{f1=f2}{$f_1,f_2$}
\psfrag{f2}{$f_2$}
\psfrag{title}{}
 \resizebox{8.5cm}{7cm}{\includegraphics{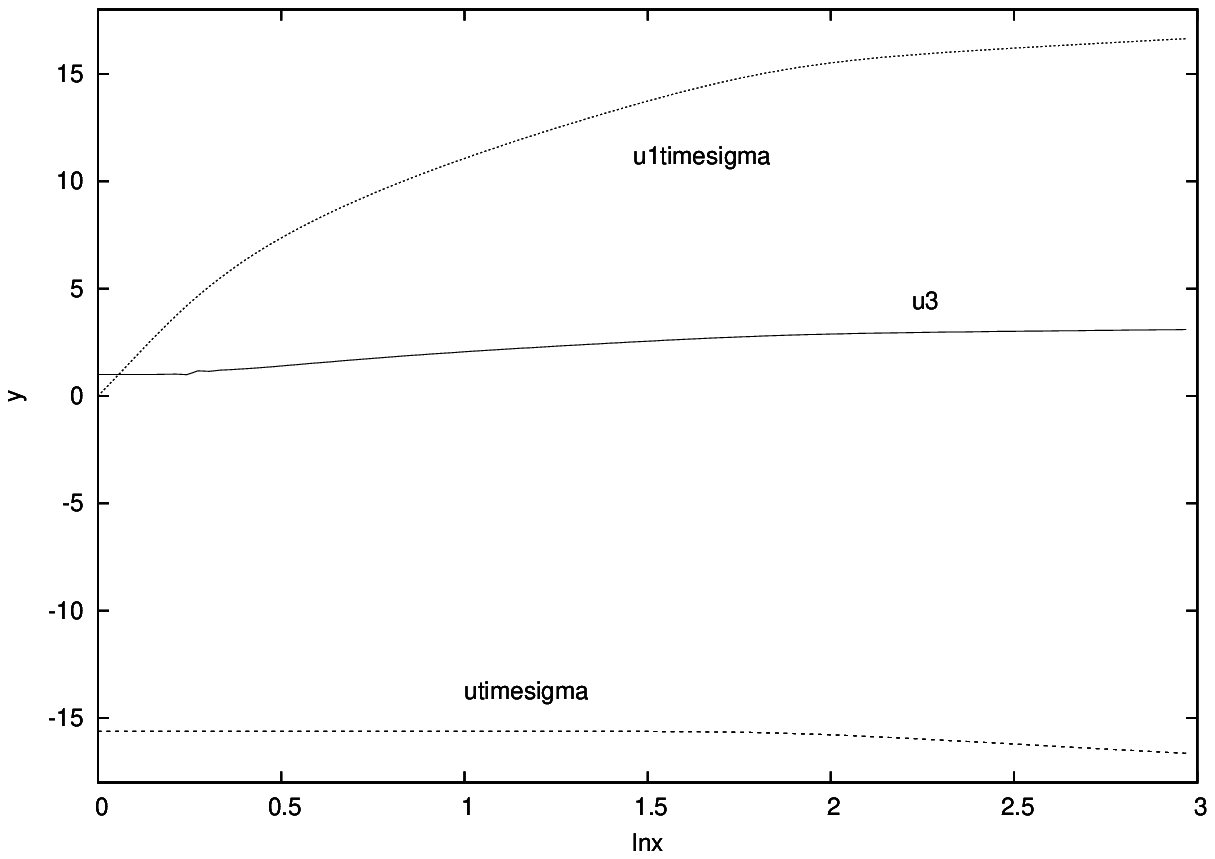}}%
 \resizebox{8.5cm}{7cm}{\includegraphics{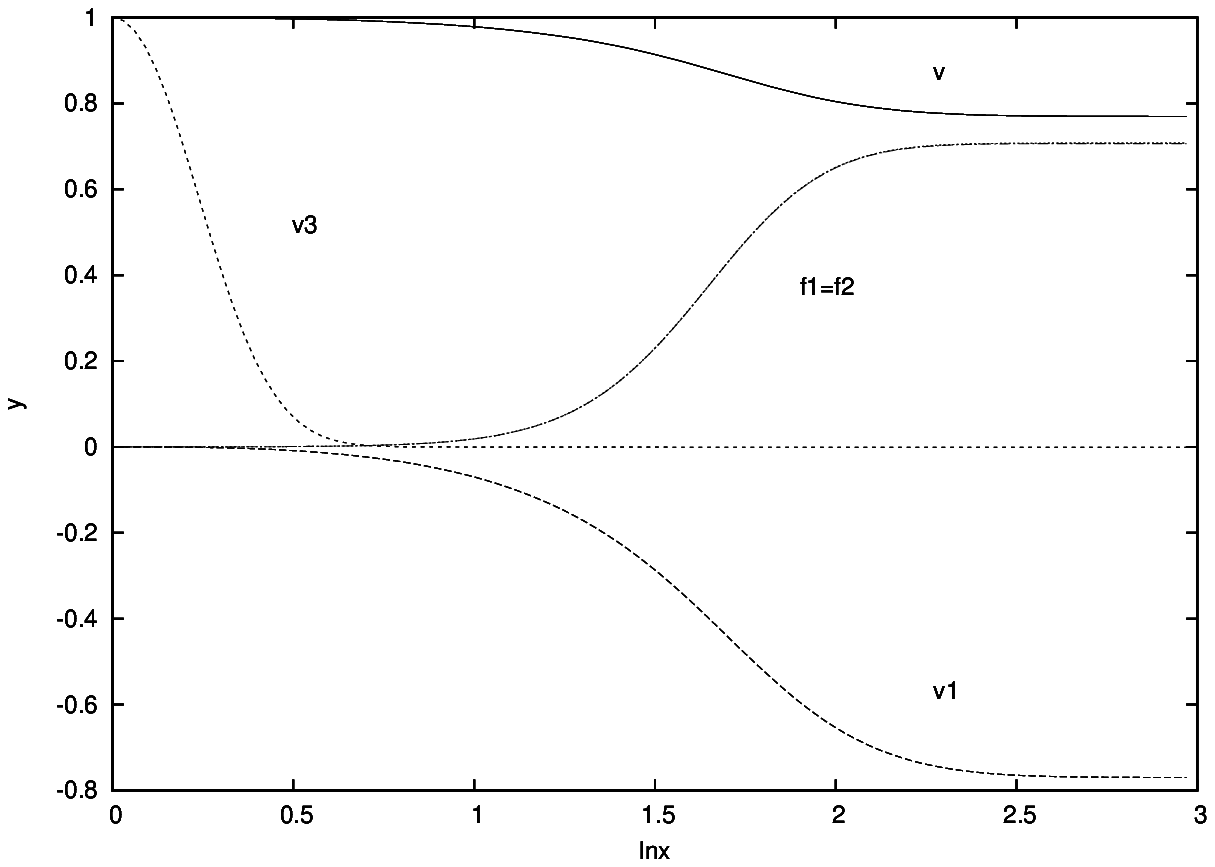}}%
%  \psfrag{y}{$\frac12(u-V_3)$} 
 % \resizebox{10cm}{6cm}{\includegraphics{FIG3.eps}}%
\hss}
\caption{\small Solutions 
profiles for 
$\beta=2$, $g^{\prime 2}=0.23$, $\n=\nu=1$ and  
$\sigma=0.008$. 
}
\label{FIG4a}
\end{figure}
%%%%%%%%%%%%%%%%%%%%%%%%%%%%%%%%%%%%%%%%%%%%%%%%%%%%%%%%%%%%%%%%%%%%%%%%%%

The numerical profiles 
of the large ${\cal I}$ solutions (see Fig.\ref{FIG4a}) 
essentially confirm these expectations, but also reveal  additional 
interesting features. 
For large currents the vacuum angle $\gamma$ tends to $\pi/2$ 
(see Fig.\ref{fig3a}) so that one has $f_1=f_2=1/\sqrt{2}$ for 
$\rho\to\infty$. However, as is seen in  Fig.\ref{FIG4a}, 
one has $f_1\approx f_2$ not only for large $\rho$ but everywhere.   
In particular, there is a region 
where $f_1\approx f_2\approx 0$,
which means that the gauge symmetry is completely 
restored to SU(2)$\times$U(1). In fact, at the origin one still has 
$f_1=a_5\rho$ and $f_2=q$ (see Eq.\eqref{orig}) but the parameters 
$q,a_5$ are extremely small. In addition, one has in this region  
$v_1\approx\sigma u_3\approx 0$, $v\approx 1$, 
$\sigma u\approx const.$, while $v_3$ and $u_1$
change very quickly, and in particular $v_3$ vanishes very rapidly.  
The latter corresponds to freezing away of the 
very massive modes.

 One can wonder how the existence
of the very massive modes can be compatible with the 
complete symmetry restoration. However, these modes 
arise not via the Higgs mechanism but rather due to
the screening by the current, which is somewhat similar to the colour screening 
in pure Yang-Mills systems. 
When $v_3$ vanishes, the Higgs field starts to change and  approaches its 
vacuum value, after which all  massive 
degrees of freedom die out and the solutions reduce 
to the U(1) electric wire
\eqref{030}.

It turns out to be possible to find a relatively 
simple approximation for the large current solutions 
that explains
all their empirically observed features.  
Specifically,  
since the vacuum angle $\gamma$ approaches $\pi/2$ for large currents,
let us set $\gamma=\pi/2$. 
Let us also restrict ourselves to the simplest case where 
$n=\nu=1$. 
The boundary conditions  \eqref{rec} then become
\begin{align}               \label{rec12}
a_1\leftarrow\, &\Y\to c_1+Q\ln\rho\,,~~~~~
0\leftarrow\,  \Om_1 \to -(c_1+Q\ln\rho)\,,~~~~~ 
1\leftarrow\,  \Om_3 \to 0\,,\nonumber \\
1\leftarrow\, &\Z\to c_2\,,~~~~~~~~~~~~~~~~~  
0\leftarrow\,  \W_1 \to - c_2\,,~~~~~~~~~~~~~~~~~~
1\leftarrow\,  \W_3 \to 0\,,\nonumber \\
a_5 \rho\leftarrow\, & \f \to 1/\sqrt{2}\,,~~~~~~~~~~~
q\,\leftarrow\,  \p \to 1/\sqrt{2}\,.  
\end{align}  

\subsection{W-condensate region  -- SU(2) Yang-Mills string} 

Let us approximate the solution at small $\rho$ by 
\be                                   \label{uv}
f_1=f_2=\sigma u_3=v_1=0,~~\sigma u=\sigma a_1,~~~v=1,~~
\sigma u_1(\rho)=\lambda U_1(\lambda\rho),~~v_3=V_3(\lambda\rho),
\ee
where $\lambda$ is a scale parameter. Inserting this to 
\eqref{ee1}--\eqref{CONS} the equations reduce to 
\begin{subequations}                      \label{uv:0}
\begin{align}                                     
\frac{1}{x}(x U_1^\prime)^\prime&=\frac{V_3^2}{x^2}\,U_1,  \\
{x}\left(\frac{V_3^\prime}{x}\right)^\prime&=U_1^2 V_3,\
\end{align}
\end{subequations} 
with $x=\lambda\rho$. 
These equations admit a solution with the following boundary
conditions for $0\leftarrow x\to\infty$, 
\begin{subequations}                      \label{uv1}
\begin{align}
x+\ldots\leftarrow &U_1(x)\to 0.85+0.91\ln(x)+\ldots \\
1-0.45x^2+\ldots\leftarrow &V_3(x)\to 0.32\sqrt{x}\,e^{0.06 x}x^{-0.91x}+\ldots
\end{align}
\end{subequations}
for which $V_3$ rapidly approaches zero 
(see Fig.\ref{fig_approx}). 
 Inserting $U_1(x),V_3(x)$ to \eqref{uv} 
gives a family of solutions distinguished from each other 
by values of the scale parameter $\lambda$.
This 
scaling symmetry arises due to 
the fact that, since the Higgs field is zero, the system is
pure Yang-Mills. 
In fact, Eqs.\eqref{uv:0} coincide with the Yang-Mills equations 
for the pure SU(2)  system,
\be                             \label{LYM}
%\partial^\mu\WW^a_{\mu\nu}=-\epsilon_{abc}\WW^{b\mu}\WW^c_{\mu\nu},
{\cal L}_{SU(2)}=-\frac14\,\WW^a_{\mu\nu}W^{a\mu\nu}
\ee
provided that the Yang-Mills field is cylindrically-symmetric 
\be
\tau^a\WW^a_\nu dx^\mu=\tau^1 \lambda U_1(\lambda \rho) dz
+\tau^3 V_3(\lambda \rho)d\varphi.
\ee
We shall therefore call the solution shown in 
Fig.\ref{fig_approx} Yang-Mills string.

The Yang-Mills strings are made of the pure gauge field
and describe the charged W-condensate 
trapped in the region where 
the amplitude $V_3$ is different from zero. 
Outside this region there remains the long-range  
Biot-Savart field generated by the condensate and  represented by $U_1$. 
The right-hand sides of Eqs.\eqref{uv:0} 
can be viewed as conserved (although not gauge invariant)
current densities. Integrating over the $\rho,\varphi$ plane gives
the current along the $z$-direction, ${\cal I}\sim\lambda$, localized 
in the region of the size $\sim 1/\lambda$  
where $V_3$ is different from zero. 

Yang-Mills strings approximate the central current-carrying core of 
the electroweak vortex. 
Their scale invariance implies  
that there is no upper bound for the current  
${\cal I}\sim\lambda$, since the scale parameter $\lambda$ 
can assume any value. 
%The self-gravitating analogues of the 
%Yang-Mills strings were studied in  Ref. \cite{EYM}. 

\subsection{External U(1)$\times$U(1) region}

\begin{figure}[h]
\hbox to\linewidth{\hss%
  \psfrag{x}{$x$}
 \psfrag{y}{}
\psfrag{v}{$V_3$}
\psfrag{s}{$\sin\Omega$}
\psfrag{v3}{$v_3$}
\psfrag{u3}{$\sigma u_3$}
\psfrag{u1}{$\sigma u_1$}
\psfrag{u}{$U_1$}
\psfrag{f}{$f$}
\psfrag{f2}{$f_2$}
\psfrag{title}{}
 \resizebox{8.5cm}{7cm}{\includegraphics{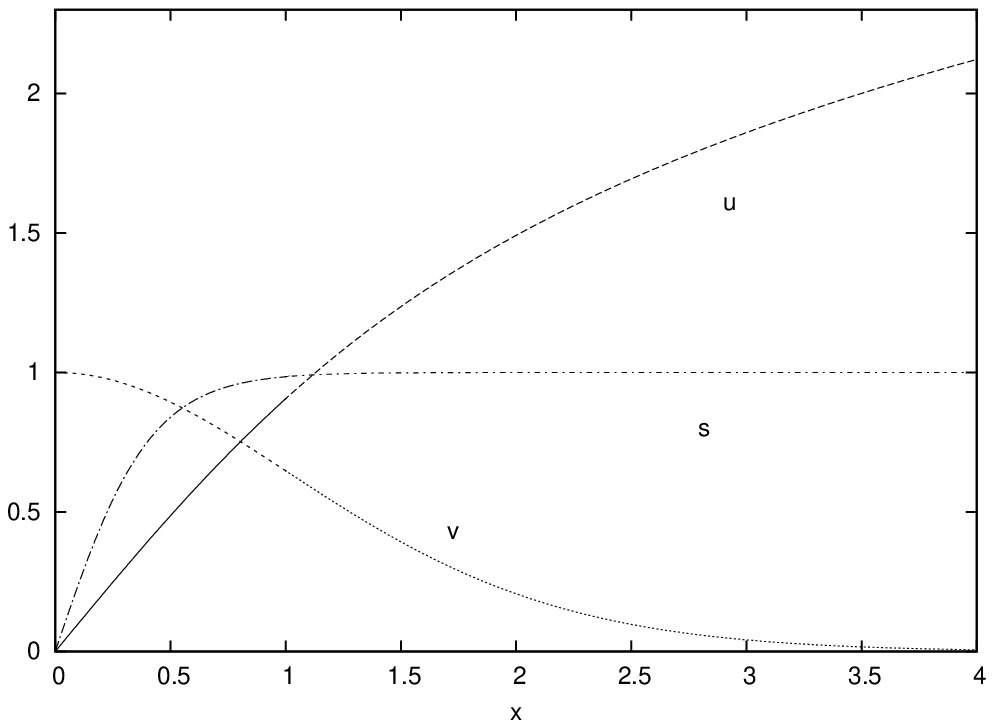}}%
\psfrag{v}{$V$}
\psfrag{s}{$\sin\Omega$}
\psfrag{v3}{$v_3$}
\psfrag{u3}{$\sigma u_3$}
\psfrag{u1}{$\sigma u_1$}
\psfrag{u}{$U$}
\psfrag{f}{$f$}
\psfrag{f2}{$f_2$}
\psfrag{x}{$\ln(1+\rho)$}
 \resizebox{8.5cm}{7cm}{\includegraphics{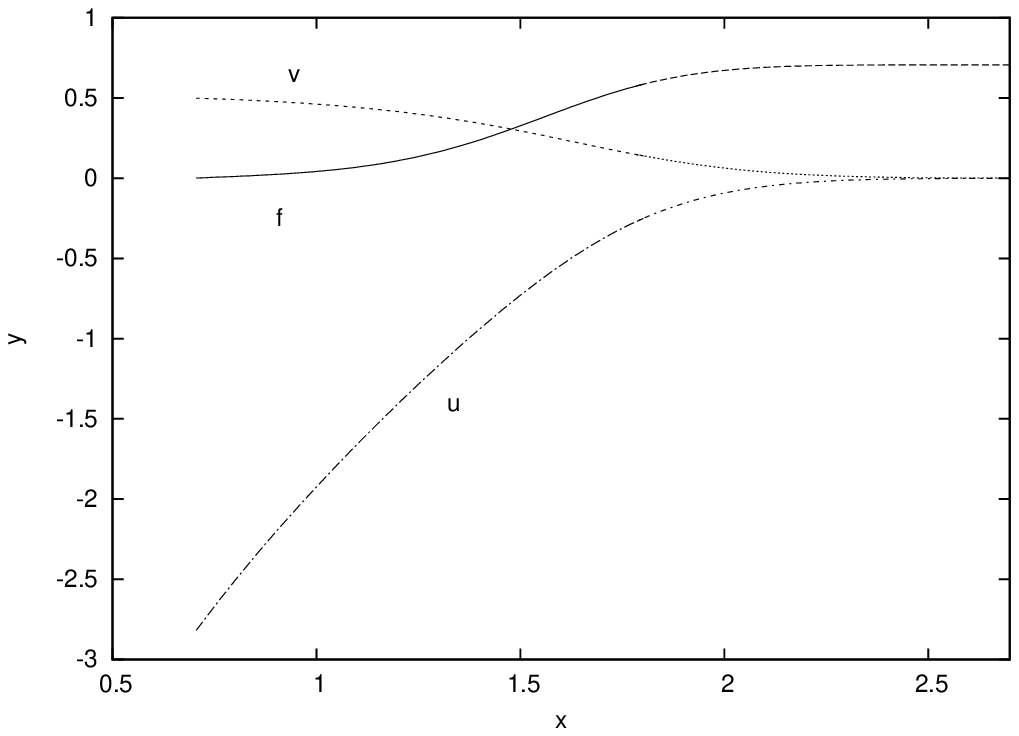}}%
%  \psfrag{y}{$\frac12(u-V_3)$} 
 % \resizebox{10cm}{6cm}{\includegraphics{FIG3.eps}}%
\hss}
\caption{\small Left: solution of Eqs.\eqref{uv:0} with the boundary condition \eqref{uv1}
and the function $\sin\Omega$ defined by Eqs.\eqref{Omega},\eqref{ang}. Right: 
the typical solution of Eqs.\eqref{Zm}. 
}
\label{fig_approx}
\end{figure}

Let us now assume that the Yang-Mills string approximation is valid 
only in the central vortex core, for $x\leq x_0$, with $x_0$  chosen such that  
$V_3(x_0)$ is sufficiently  close to zero, 
for example $x_0=4$ (see Fig.\ref{fig_approx}). 
This determines the upper value 
$\rho_0=x_0/\lambda$.  As we shall see, for large $\lambda$ 
 the precise
knowledge  of $x_0$ is unimportant.

%%%%%%%%%%%%%%%%%%%%%%%%%%%%%%%%%%%%%%%%%%%%%%%%%%%%%%%%%%%%%%%%%%%%%%%%%%
%%%%%%%%%%%%%%%%%%%%%%%%%%%%%%%%%%%%%%%%%%%%%%%%%%%%%%%%%%%%%%%%%%%%%%%%%%
For  $\rho>\rho_0$ we set  
\be                                    \label{uv+}
v_3=\sigma u_3=0,~~~~~f_1=f_2\equiv f/\sqrt{2},
\ee
while the 
remaining field amplitudes can be combined to define
the electromagnetic and Z field variables, 
\bea                                       \label{uv++} 
U&=&\frac{\sigma}{2}(u+u_1),~~~~~~~~V=\frac12(v+v_1),\nonumber \\
U_A&=&\sigma(g^2 u-g^{\prime 2}u_1),~~~~~~V_A=g^2 v- g^{\prime 2}v_1.
\eea
More precisely, using formulas of Sec.\eqref{flux-cur} and Eq.\eqref{uv+}, the 
electromagnetic and Z fields potentials are 
\be
A_{\mu}=\frac{g}{g^\prime}\, \A_{\mu}-\frac{g^{\prime}}{g}\,\WW^1_{\mu},~~~~~
{Z}_{\mu}=\A_{\mu}+\WW^1_{\mu},~~
\ee
the Higgs field components are
$\phi_1=\phi_2\equiv \phi$
while the W-condensate components vanish. 
The original SU(2)$\times$U(1) theory \eqref{L} reduces then to 
the U(1)$\times$U(1) model 
\be                                     \label{U11}
{
{\cal L}_{U(1)\times U(1)}=-\frac14F_{\mu\nu}F^{\mu\nu}-\frac14 Z_{\mu\nu}Z^{\mu\nu}
+({\cal D}_\mu\phi)^\ast{\cal D}^\mu\phi-\frac{\beta}{8}(|\phi|^2-1)^2 
}
\ee
with 
${\cal D}_\mu\phi=(\partial_\mu-\frac{i}{2}\,Z_\mu)\phi$. This 
contains the free Maxwell theory and the Abelian Higgs model.   
The fields are ${gg^\prime}A_\mu dx^\mu=U_A dz+V_A d\varphi$ and 
$2Z_\mu dx^\mu=Udz+Vd\varphi$ while $\phi=f$. Eqs.\eqref{ee1}--\eqref{CONS} 
then reduce to  
the Maxwell equations 
\bea                    \label{Maxwell}
(\rho U_A^\prime)^\prime=0,~~~~~~~~
\left(\frac{V_A^\prime}{\rho}\right)^\prime=0,
\eea
and to the equations of the Abelian Higgs model  
\begin{subequations}                              \label{Zm}
\begin{align}                                  
\frac{1}{\rho}(\rho f^\prime)^\prime&=\left(U^2+\frac{V^2}{\rho^2}
+\frac{\beta}{4}(f^2-1) \right)f,  \label{Zm1}\\
\rho\left(\frac{V^\prime}{\rho}\right)^\prime&=\frac12\,f^2V, \label{Zm2} \\
\frac{1}{\rho}(\rho U^\prime)^\prime&=\frac12\,f^2U.         \label{Zm3}
\end{align}
\end{subequations}
%which is actually the extended version of the ANO system \eqref{ANOeqs}. 
The solution of the Maxwell equations \eqref{Maxwell} is 
\be                              \label{uvA}
U_A=A+B\ln\rho,~~~~V_A=C, 
\ee
with constant $A,B,C$.  
Equations \eqref{Zm} admit a one-parameter family of solutions 
with the boundary conditions 
\be                      \label{uvZ}
U(\rho_0)\leftarrow U\to 0,~~~~
\frac12\leftarrow V\to 0,~~~~
0\leftarrow f\to 1
\ee
for $\rho_0\leftarrow \rho\to\infty$ 
(see Fig.\ref{fig_approx}). The numerics show that 
if $U^\prime(\rho_0)$ is large and positive then 
$U(\rho_0)$ is large and negative and the 
derivatives 
 $f^\prime(\rho_0)$
and $V^\prime(\rho_0)$ tend to zero such that 
there is a neighbourhood of $\rho_0$ where 
\be				\label{appr}
V\approx 1/2,~~~~f\approx 0,~~~~U\approx a+b\ln\rho
\ee
with the parameters $a,b$
related to each other, $a=a(b)$.  
As a result, 
\be
U(\rho_0)=a(b)+b\ln\rho_0,~~~~U^\prime(\rho_0)=b/\rho_0. 
\ee   

\subsection{Matching the solutions}

The solutions obtained  for $\rho\leq\rho_0$
and for $\rho\geq\rho_0$ should agree
at $\rho=\rho_0=x_0/\lambda$ where the functions 
and their first derivatives 
should match. Matching $u,u_1$ and their derivatives gives 
%\bea
%\frac12(\sigma a_1+\gamma U_1(x_0))&=&U_Z(\alpha),~~~~~
%g^2\sigma a_1-g^{\prime 2}\gamma U_1(x_0)
%=A+B\ln\left(\frac{x_0}{\gamma}\right),             \nonumber \\
% \frac12\gamma U^\prime_1(x_0)&=&\alpha,~~~~~~~~~~~~~~~~~~~
%-g^{\prime 2}\gamma U^\prime_1(x_0)
%=B\frac{\gamma}{x_0},        
%\eea  
%with $U_1(x)=0.85+0.91\ln(x)$. These 
four conditions that determine  the 
free parameters $a_1,b,A,B$ in the above solutions,
\bea
B&=&-0.91g^{\prime 2}\lambda,\nonumber \\
b&=&0.45\lambda,\nonumber \\
\sigma a_1&=&2a(b)-0.85\lambda-2b\ln\lambda,~~~~~~~~~\nonumber \\
A&=&B\ln\lambda +g^2 a_1-0.85 g^{\prime 2}\lambda.
\eea
It is worth noting that $x_0$ has dropped from these relations. 
Matching similarly $v,v_1,f$ gives only one condition,  
$C=g^2$, but the derivatives $v^\prime,v_1^\prime,f^\prime$ do not match precisely,
since they vanish for  $\rho\leq\rho_0$ but not for  $\rho\geq\rho_0$. 
However, this discrepancy tends to zero when $\lambda$ grows, since   
$U^\prime(\rho_0)\sim \lambda^2$ then increases thus rendering the approximation
\eqref{appr} better and better. Similarly, although there is a discrepancy in
values of $v_3$, since it  vanishes for  $\rho\geq\rho_0$ but not 
for  $\rho\leq\rho_0$, 
increasing $\lambda$ reduces this discrepancy as well.

As a result, by matching the two solutions 
we can approximate the global solution in the interval 
$0\leq \rho<\infty$. 
The approximate solution depends on $\lambda$ 
and approaches the true solution when $\lambda$ increases.
For $\lambda\to\infty$ the discrepancy of values of  
$v_3,v^\prime,v_1^\prime,f^\prime$ at 
the matching point vanishes and so the approximate  solution becomes exact.   
There only remains 
to relate $\lambda$ to the current. Comparing with \eqref{rec12} we see that 
$\sigma Q=B$ (also $\sigma c_1=A$ and, as noticed in Fig.\ref{fig3b}, $c_2=g^2$)
and since the current ${\cal I}=-2\pi\sigma Q/(gg^\prime)$ it follows that 
\be
\lambda=0.17\,\frac{g}{g^\prime}\,{\cal I}.
\ee
%
%%%%%%%%%%%%%%%%%%%%%%%%%%%%%%%%%%%%%%%%%%%%%%%%%%%%%%%%%%%%%%%%%%%%%%%%%%
\begin{figure}[h]
\hbox to\linewidth{\hss%
  \psfrag{x}{$\ln(1+\rho)$}
 \psfrag{y}{}
\psfrag{v}{$v$}
\psfrag{v1}{$v_1$}
\psfrag{v3}{$v_3$}
\psfrag{u3}{$\sigma u_3$}
\psfrag{u1}{$\sigma u_1$}
\psfrag{u}{$\sigma u$}
\psfrag{f}{$f_1,f_2$}
\psfrag{f2}{$f_2$}
\psfrag{title}{}
 \resizebox{8.5cm}{7cm}{\includegraphics{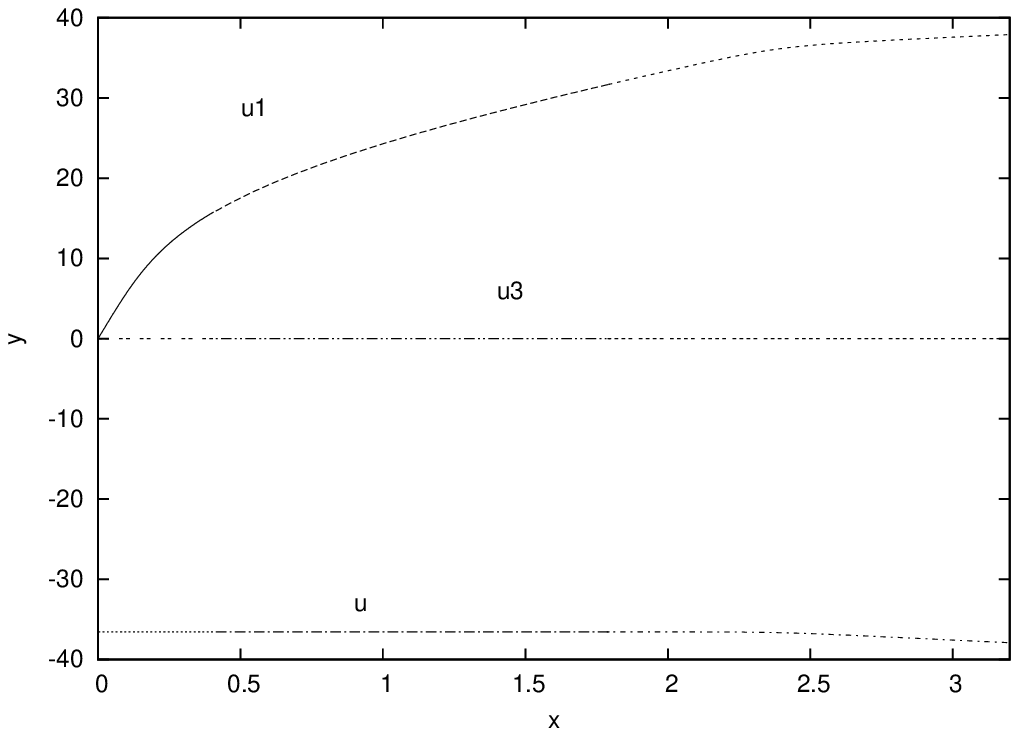}}%
 \resizebox{8.5cm}{7cm}{\includegraphics{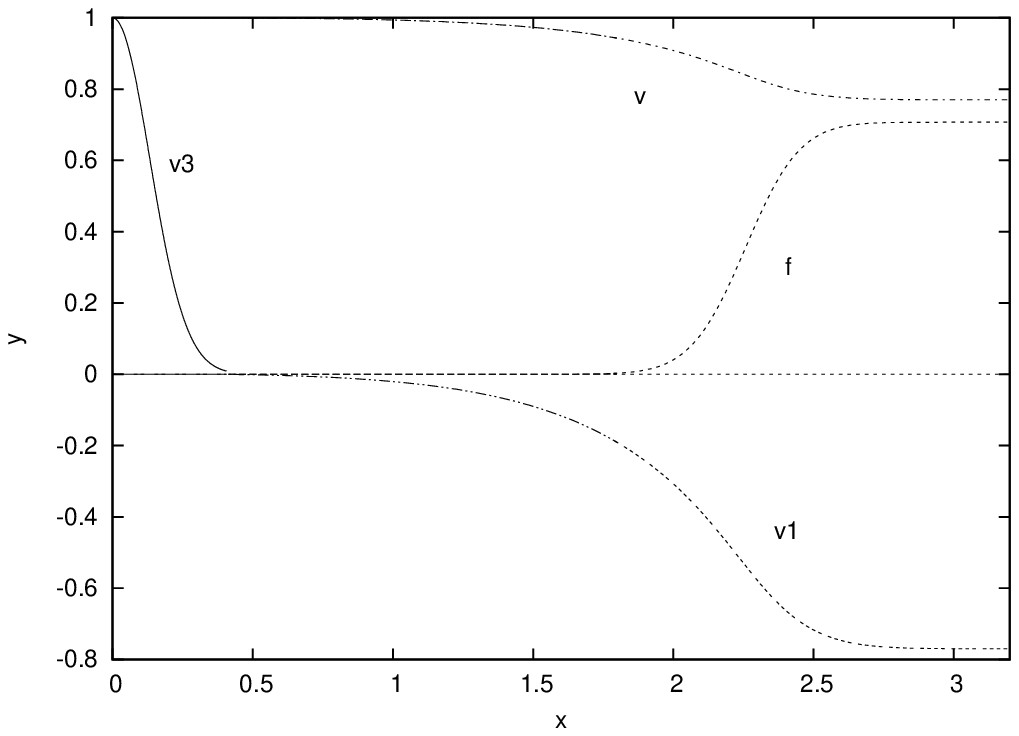}}%
%  \psfrag{y}{$\frac12(u-V_3)$} 
 % \resizebox{10cm}{6cm}{\includegraphics{FIG3.eps}}%
\hss}
\caption{\small The approximate solution 
for 
$\beta=2$, $g^{\prime 2}=0.23$, $\n=\nu=1$ and  
$\lambda=7.6$, ${\cal I}=24.43$.  
}
\label{FIG4aa}
\end{figure}
%%%%%%%%%%%%%%%%%%%%%%%%%%%%%%%%%%%%%%%%%%%%%%%%%%%%%%%%%%%%%%%%%%%%%%%%%%
%
The approximate solution  shown in Fig.\ref{FIG4aa}
clearly exhibit the same structure as the true one in Fig.\ref{FIG4a}.
Although it 
corresponds to not a very large value of the scale parameter, $\lambda=7.6$, 
its current is twice 
as large as compared to that in Fig.\ref{FIG4a}. 
The true solution  in Fig.\ref{FIG4a}
shows a slowly growing amplitude $u_3$ whereas in Fig.\ref{FIG4aa} 
one has $\sigma u_3=0$. 
This difference is a subleading effect, since in Fig.\ref{FIG4a}
the vacuum angle is not exactly $\pi/2$ and so there is a growing 
mode $u_3\sim\cos\gamma\ln\rho$. For ${\cal I}\to\infty$ 
one has $\gamma\to\pi/2$ and this mode disappears, so that 
$u_3$ becomes everywhere bounded, and multiplying it by $\sigma\to 0$ will give
$\sigma u_3=0$, as shown in  Fig.\ref{FIG4aa}.  With some more analysis  such 
subleading effects can be taken into account to determine, for example, 
$\gamma({\cal I})$, $\sigma({\cal I})$, $q({\cal I})$.

\subsection{Symmetry restoration inside the vortex}

\begin{figure}[h]
\hbox to\linewidth{\hss%
 \resizebox{8cm}{8cm}{\includegraphics{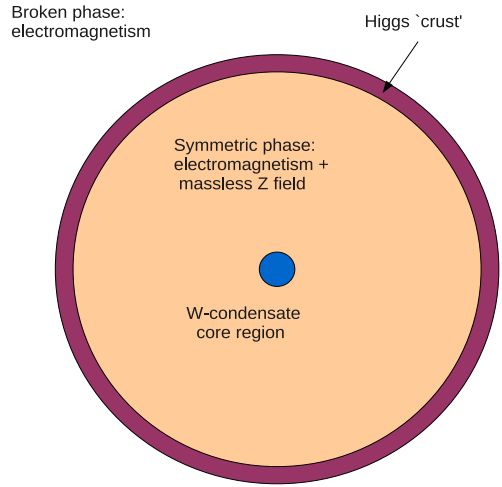}}%
 %\resizebox{8.5cm}{7cm}{\includegraphics{critical.eps}}%
%  \psfrag{y}{$\frac12(u-V_3)$} 
 % \resizebox{10cm}{6cm}{\includegraphics{FIG3.eps}}%
\hss}
\caption{\small The schematic  structure 
of the large current vortex.
}
\label{FIG-LARGE}
\end{figure}

Comparing Figs.\ref{FIG4a} and \ref{FIG4aa} reveals that 
the zero Higgs field region expands when the current increases. 
The inner structure  of the large current vortex can 
be schematically represented by Fig.\ref{FIG-LARGE}. 
The vortex shows a large 
symmetric  phase region of size $\sim{\cal I}$ where 
the Higgs field is very close to zero. In the centre of this region  there is
a compact core  of size  $\sim{\cal I}^{-1}$
containing the charged W-condensate and approximated by the 
Yang-Mills string. Outside the core there live 
the electromagnetic and Z fields.
The symmetric phase is surrounded by a 
`crust' of thickness  $\sim R_{\mbox{\tiny Z}}$ where 
the Higgs field approaches the constant vacuum value 
while the Z field becomes massive and dies away.  
The region outside the `crust' is described by the Maxwell electrodynamics
and is dominated by the Biot-Savart magnetic field 
 generated by the current in the core.   
%%%%%%%%%%%%%%%%%%%%%%%%%%%%%%%%%%%%%%%%%%%%%%%%%%%%%%%%%%%%%%%%%%%%%%%%%%
%%%%%%%%%%%%%%%%%%%%%%%%%%%%%%%%%%%%%%%%%%%%%%%%%%%%%%%%%%%%%%%%%%%%%%%%%%

This picture reminds somewhat the `W-dressed superconducting
string' scenario of Ambjorn and Olesen \cite{period}, 
briefly described in 
Sec.\ref{amjorn} above. In this scenario Witten's superconducting string 
produced at the GUT energies 
acts as  the external source of 
magnetic field that changes the structure of the electroweak vacuum. 
There is a cylindrical layer around the string where the field 
falls within the limits Eq.\eqref{F12} and where the Higgs field is 
non-trivial. Inside this layer the field is supercritical 
and the Higgs field is zero,
while outside  the field is undercritical and the Higgs field 
is in vacuum.  
All this is very similar to what is shown in  Fig.\ref{FIG-LARGE}. 
 This suggests that we have 
a self-consistent electroweak realization of the Ambjorn-Olesen scenario
in which the Witten string is replaced by the Yang-Mills string of 
completely electroweak origin. The following estimates confirm this. 

First of all, the radius of the layer where the Higgs field 
is non-trivial should scale as ${\cal I}$,
and we shall now see that 
the crust radius scales  precisely in this way. Indeed,
inside the crust $f$ is very small and one can neglect the right 
hand sides in \eqref{Zm3} to obtain $U=b\ln(\rho/\rho_\star)$,
where $\rho_\star$ is an integration constant
and 
$b=0.45\lambda$.  
As long as 
$\rho\ll\rho_\star$ the $U^2$ term in \eqref{Zm1} is large and makes $f$
stay very close to zero, since $f$ can increase and pass through the 
inflection point if only $U^2$ is small. The latter follows from the fact that 
the second derivative of $f$ with respect to $\ln\rho$ vanishes if only 
the right-hand side of \eqref{Zm1} vanishes, which is only possible
if $U^2$  is small enough to be canceled by the negative term 
$\frac{\beta}{4}(f^2-1)$. Therefore, $f$ can vary   if only $U\sim\sqrt{\beta}$, 
but as soon as $f$ deviates from zero, it acts as the mass term for $U$
to drive it to zero.  As a result, $f$ interpolates between zero and one
close to the place where $U$ vanishes, for $\rho\sim\rho_\ast$, 
in an interval of size $\delta\rho\sim R_{\mbox{\tiny Z}}$. 
Multiplying \eqref{Zm3} by $\rho$ and integrating from $\rho_0$
to infinity gives 
$
2b=-\int_{\rho_{0}}^\infty \rho f^2U d\rho
$ and  
since the integrand 
is almost everywhere zero apart from a small vicinity of $\rho_\star$,
it follows that $b\sim\sqrt{\beta}\rho_\star$. This gives the size of the 
symmetric phase region (the crust radius) 
\be                                   \label{rho-star}
\rho_\ast=0.28\times\,\frac{g}{g^\prime}\frac{{\cal I}}{\sqrt{\beta}},
\ee
where the coefficient is found numerically. As a result, one can 
approximate the solution of Eqs.\eqref{Zm} by 
\be                                \label{step}
f\approx\Theta(\rho-\rho_\star),~~~~
U\approx b\,(1-f)\ln({\rho}/{\rho_\star}),~~~~
V\approx\frac12(1-f)(1-\left({\rho}/{\rho_\star}\right)^2)
\ee
where 
$\Theta(\rho-\rho_\star)$ is the step-function smoothed
over a finite interval $\delta\rho\sim R_{\mbox{\tiny Z}}$.

Let us now consider the distribution of the fields inside the vortex.  
The fields are defined by Eqs.\eqref{FZ},\eqref{confld} where there is   
a hitherto undetermined  function $\Omega$. Using Eqs.\eqref{Omega},\eqref{uv+} 
one finds that $\Omega=\pi/2$ in the external region, while in the core $\Omega$ 
is not yet defined. Let us calculate 
the first order correction to the core configuration \eqref{uv} assuming that 
$f_1,f_2$ are 
small functions of $x=\lambda\rho$ and that $f_1\sim x$ and 
$f_2=O(1)$ for $x\to 0$. Linearising Eqs.\eqref{ee3},\eqref{ee4} 
gives for large $\lambda$, with $f_\pm=f_1\pm f_2$,  
\be                                                    \label{ang}
\frac{1}{x}(xf^\prime_{\pm})^\prime=\left((\frac{U_1}{2}\mp p)^2+
\frac{1+V_3^2}{4x^2}\right)f_{\pm}+\frac{V_3}{2x^2}\,f_{\mp},
\ee
where $p=-\sigma a_1/(2\lambda)\sim \ln\lambda$ and 
$U_1,V_3$ are given by Eqs.\eqref{uv:0}. Integrating these equations 
and using Eq.\eqref{Omega} 
determines $\Omega$ inside the core 
and shows that it
approaches $\pi/2$ very rapidly
(see Fig.\ref{fig_approx}). 

%%%%%%%%%%%%%%%%%%%%%%%%%%%%%%%%%%%%%%%%%%%%%%%%%%%%%%%%%%%%%%%%%%%%%%%%%%
\begin{figure}[h]
\hbox to\linewidth{\hss%
  \psfrag{x}{$\ln(1+\rho)$}
 \psfrag{y}{}
\psfrag{Bz}{${B}_{\hat{z}}$}
\psfrag{Bp}{$B_{\hat{\varphi}}$}
\psfrag{me}{$\mw^2/e$}
\psfrag{mh}{$\mh^2/e$}
\psfrag{I}{$0.1\times {J}_z$}
\psfrag{T}{$0.05\times T^0_0$}
\psfrag{II}{${J}_z$}
\psfrag{TT}{$T^0_0$}
\psfrag{f}{$f_1,f_2$}
\psfrag{Zz}{$-{H}_{\hat{z}}$}
\psfrag{Zp}{$-{H}_{\hat{\varphi}}$}
\psfrag{BZz}{${B}_{\hat{z}},-{H}_{\hat{z}}$}
\psfrag{W1}{${w}_{\hat{\rho}}$}
\psfrag{W2}{$-{w}_{\hat{\varphi}}$}
\psfrag{W3}{$-{w}_{\hat{z}}$}
\psfrag{title}{}
\psfrag{W12}{$-{w}_{\hat{z}},{w}_{\hat{\rho}}$}
 \resizebox{8.5cm}{7cm}{\includegraphics{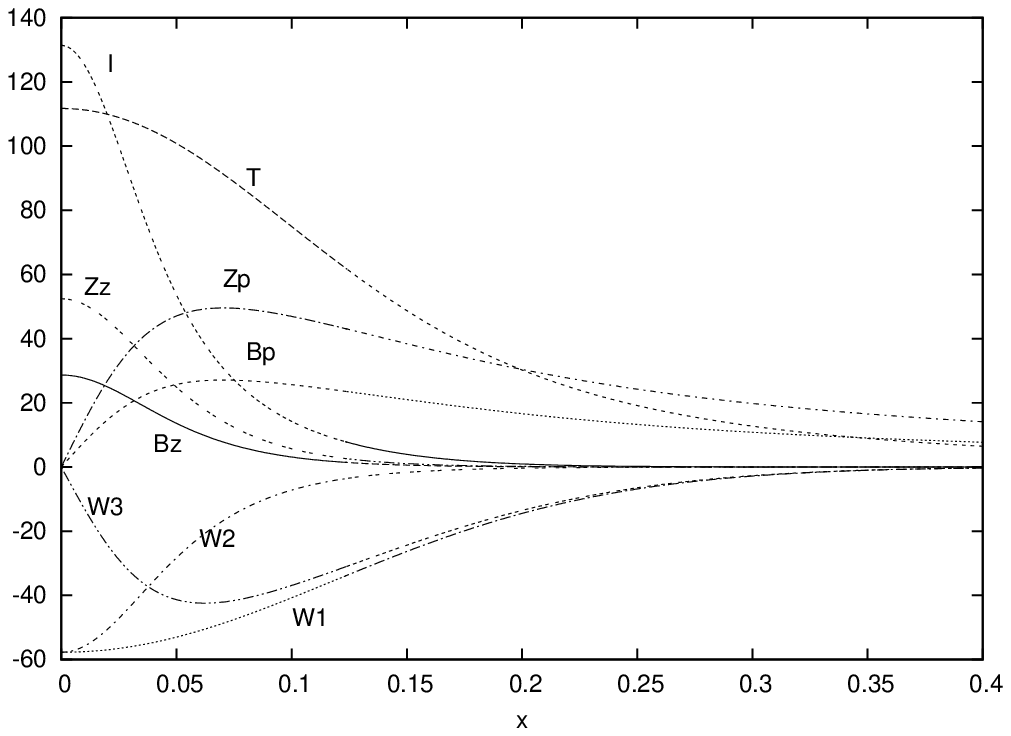}}%
 \resizebox{8.5cm}{7cm}{\includegraphics{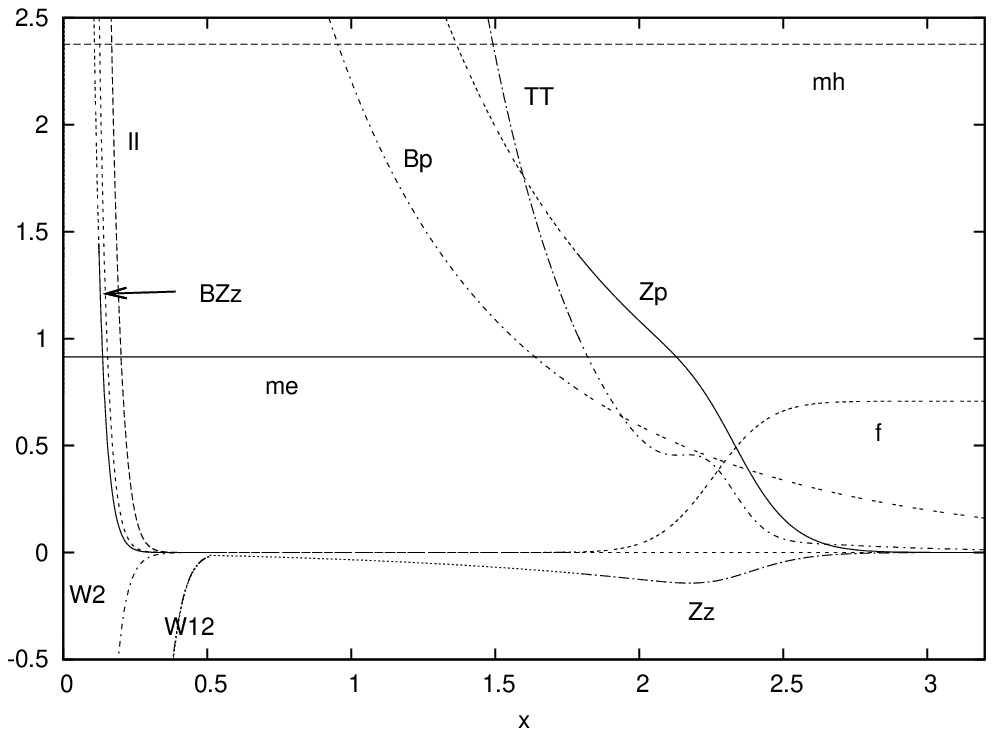}}%
%  \psfrag{y}{$\frac12(u-V_3)$} 
 % \resizebox{10cm}{6cm}{\includegraphics{FIG3.eps}}%
\hss}
\caption{\small The field  distributions inside (left) 
and outside (right) the current-carrying condensate 
 core of the vortex,  for the same solution as in 
Fig.\ref{FIG4aa}. 
}
\label{FIG-current}
\end{figure}
%%%%%%%%%%%%%%%%%%%%%%%%%%%%%%%%%%%%%%%%%%%%%%%%%%%%%%%%%%%%%%%%%%%%%%%%%%
%

We have now  all the ingredients to compute the fields 
\eqref{FZ},\eqref{confld} and the current density \eqref{Io}. The result
presented in Fig.\ref{FIG-current} shows  that the current, W-condensate,  
and the $z$-component of the magnetic field are trapped in the core,
where  
the energy density is maximal. Outside the core 
there extend only the Biot-Savart
magnetic field and the Z field, which are both massless 
as long as the Higgs field stays close to zero. 
At a distance $\sim{\cal I}$ away from the
core 
the system enters the `crust' region where  the Higgs field approaches 
the vacuum value.
This gives rise to a local maximum in the energy density and drives the Z field
very rapidly to zero. Outside the `crust' there remains only 
the Biot-Savart magnetic field.

The most interesting feature is that the `crust' region where the 
Higgs field is non-trivial 
exists only under certain values of the magnetic field.
Outside this region, where the field is either too weak or too strong,
the Higgs field is either in vacuum or vanishes, in agreement with 
the predictions of  \cite{period}. At the same time,
the magnetic field inside the crust does not quite fall within 
the limits $(\mw^2/e,\mh^2/e)$ predicted in \cite{period} 
for the {\it homogeneous} magnetic field with $\beta=1$. 
The electroweak condensate in the latter case  contains the W, Z 
and Higgs fields \cite{period}. In 
our case the crust region does not contain the W field, since it 
develops 
the very large effective mass $m_\sigma$ due to the interaction with the 
inhomogeneous Biot-Savart field, so that it
 can be different form zero only
in the central core region.

We also notice in Fig.\ref{FIG-current} that, since 
$w_{\hat{\rho}},w_{\hat{\varphi}}$ do not vanish at $\rho=0$,
the W-condensate field strength is actually singular at the axis. 
However, this singularity is mild, since the energy density and  
 the field potentials are globally regular. This singularity
is absent for solutions with $\nu> 1$.

\section{Special parameter limits}

The described above properties of the solutions  
remain qualitatively the same for the generic parameter values. 
However, for special values of $\beta,\thetaw$ solutions can 
exhibit new features.  Even though such values do not always belong
to the physical region, where $1.5\leq\beta\leq 3.5$ and  $\sin^2\thetaw=0.23$, 
it is instructive to consider them, since this helps to understand the structure
of the solution space.   

\begin{figure}[ht]

\hbox to\linewidth{\hss%
\psfrag{x}{{$\sin^2\thetaw$}}
\psfrag{y}{{$Q$}}
\psfrag{title}{}
	\resizebox{8.5cm}{7cm}{\includegraphics{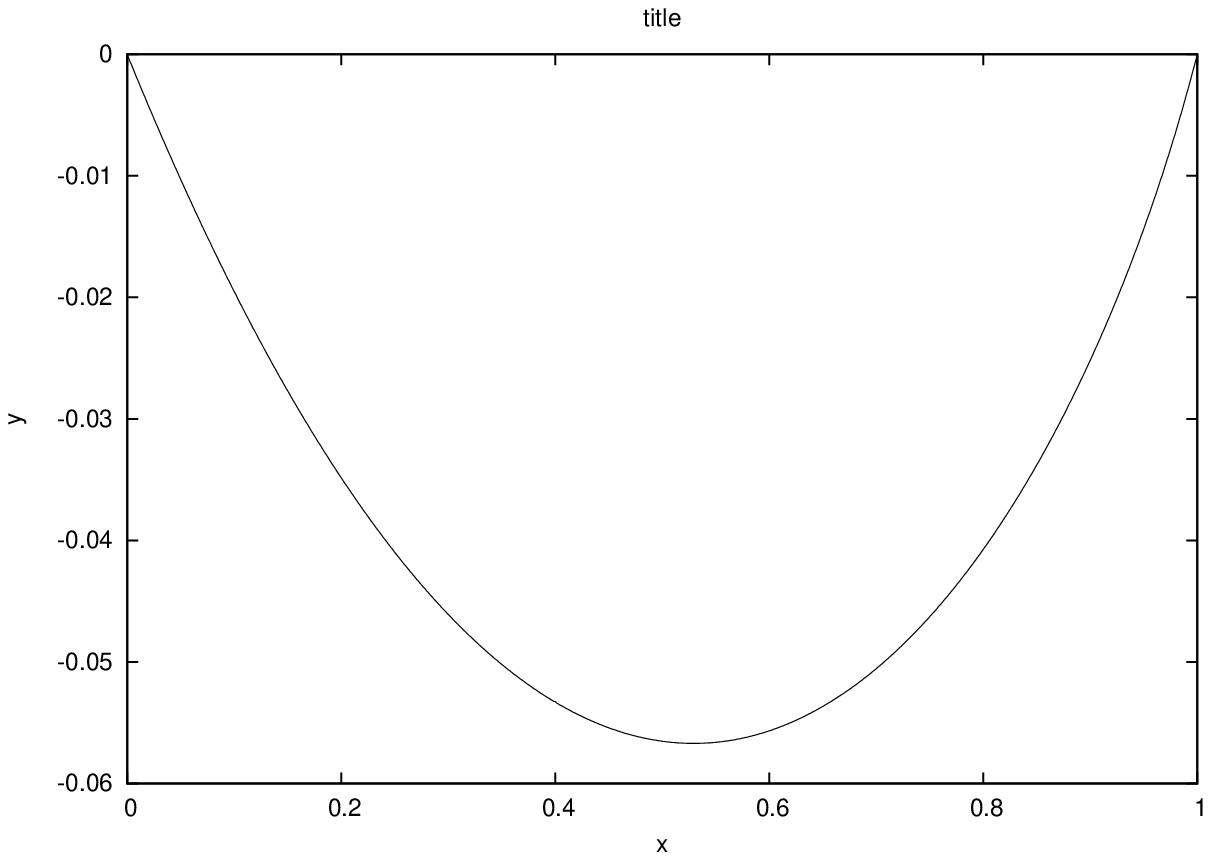}}
\hspace{5mm}%
\psfrag{x}{{$q$}}
\psfrag{y}{{$\sigma$}}
\psfrag{g=0.999}{{$g^2=0.999$}}
\psfrag{g=0.05}{{$g^2=0.05$}}
\psfrag{g=0.025}{{$g^2=0.025$}}
\psfrag{g=0.01}{{$g^2=0.01$}}
\psfrag{g=0.001}{{$g^2=0.001$}}
\psfrag{title}{}
        \resizebox{8.5cm}{7cm}{\includegraphics{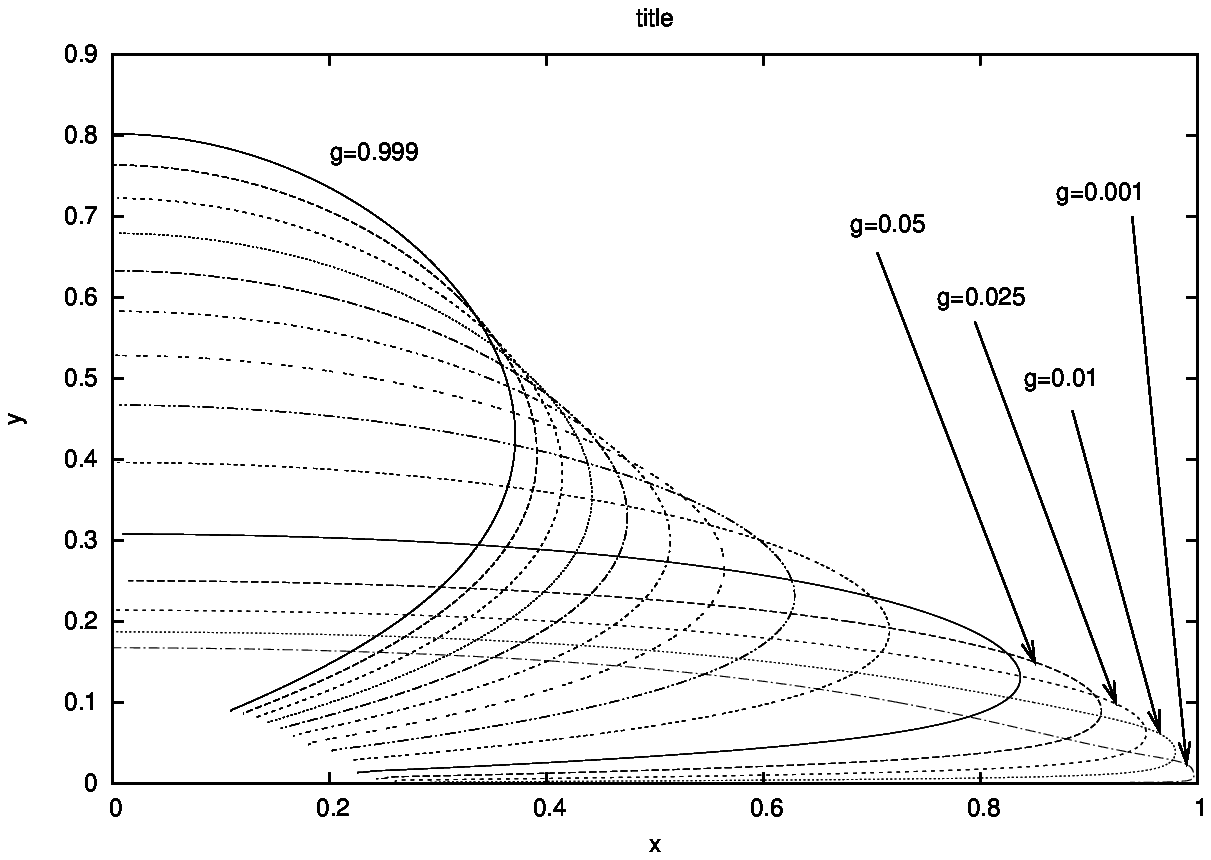}}	
\hss}

\caption{\small  Left: the `Coulombian charge' parameter $Q$ against $\thetaw$ for
the solutions with $\beta=2$, $n=\nu=1$ and $q=0.2$.  
Right: the twist $\sigma$ against $q$ 
for solutions with $n=\nu=1$, $\beta=2$. 
For $g\neq 0$ the curves $\sigma(q)$ start at $q=0$
at their maximum, then go down
towards larger values of $q$, but then always bend back to $q=0$. 
If $g=0$ then $\sigma$ decreases monotonously to zero at $q=1$.   
}
\label{fig36}
\end{figure}

For example, the limits $\thetaw\to 0$ or $\thetaw\to\pi/2$  are special, because 
either the hypercharge field
$\A_\mu$ or isospin field $\WW^a_\mu$ then become massless and 
decouple. The long-range mode $Q\ln\rho$ then disappears since $Q$ vanishes
(see Fig.\ref{fig36}) and  
the total energy of the solutions becomes {\it finite}. 
Solutions for $\thetaw=\pi/2$ have been studied
in the literature \cite{SL}, but it is worth reviewing them in the 
context of our general discussion.  

Another interesting case corresponds to the chiral solutions, which exist only for 
 special values of $\thetaw$ given by a certain function 
$\thetaw(\beta,n,\nu,q)$. This function can be determined  by requiring 
the solutions to respect the $\sigma^2=0$ condition. Finally, we shall consider
infinite Higgs mass limit, $\beta\to\infty$, in which case the theory
reduces to the gauged $\mathbb{CP}^1$ sigma-model.     
%%%%%%%%%%%%%%%%%%%%%%%%%%%%%%%%%%%%%%%%%%%%%%%%%%%%%%%%%%%%%%ù
%%%%%%%%%%%%%%%%%%%%%%%%%%%%%%%%%%%%%%%%%%%%%%%%%%%%%%%%%%%%%%%%%

\subsection{Semilocal limit, $\thetaw=\pi/2$ \label{sem}} 

 Since the right 
hand side of the Yang-Mills equations \eqref{P1} is proportional to $g^2$,  
one has $\WW^a_\mu\sim g^2$ for small $g$ so that the field 
$\WW^a_\mu$ and its kinetic term in the Lagrangian 
%$-\frac{1}{4g^2}\,\WW^a_{\mu\nu}\WW^{a\mu\nu}$
vanish for $g\to 0$ and the theory \eqref{L} reduces to
\be                             \label{LLsemi}
{\cal L}=
-\frac{1}{4}\,{\F}_{\mu\nu}{\F}^{\mu\nu}
+(D_\mu\Phi)^\dagger D^\mu\Phi
-\frac{\beta}{8}\left(\Phi^\dagger\Phi-1\right)^2,
\ee
with $D_\mu\Phi=(\partial_\mu-\frac{i}{2}\,{\A}_\mu)\Phi$.
This reduced theory is sometimes
called semilocal, since it has the local U(1) invariance, 
while the SU(2) symmetry of the 
original theory now becomes global \cite{achuc}. 
The perturbative mass spectrum of the model \eqref{LL} can be obtained 
by setting $g=0$ in
Eq.\eqref{masses}: it contains a massive 
vector boson and a Higgs boson, as well as a pair 
of massless Goldstone scalars. However, the latter acquire a 
finite mass  
via the screening mechanism similar to that described above for the W bosons, 
so that all fields are massive.

The field equations can be obtained by setting in 
Eqs.\eqref{ee1}--\eqref{CONS} the SU(2) field amplitudes to be constant and equal
to their values at the origin,
$\Om_1=\W_1=0$,
$\Om_3=1$, $\W_3=\nu$. 
This solves the Yang-Mills equations \eqref{ee5}--\eqref{CONS} 
and implies the following values of the asymptotic parameters in \eqref{rec}:
$Q=\gamma=0$, $c_1=-1$, $c_2=-\nu$.
In the gauge \eqref{003a}   the Yang-Mills field then vanishes
and the fields read 
\begin{align}                           \label{003-semi}
{\cal W}=(\Y(\rho)+1)\,\om_\alpha dx^\alpha +
(2\n-\nu-\Z(\rho))d\varphi,~~~
\Phi=
\left(
\begin{array}{c}
e^{in\varphi}\f(\rho)  \\
e^{i(n-\nu)\varphi+i\,\om_\alpha x^\alpha}\p(\rho)  
\end{array}
\right).      
\end{align}
Eqs.\eqref{ee5}--\eqref{CONS} reduce to 
\begin{align}                                           \label{se1}
\frac{1}{\rho}(\rho\Y')'&=\left.\left.\frac{1}{2}
\right((\Y+1)\f^2+(\Y-1)\p^2\right),\\
\rho\left(\frac{\Z^\prime}{\rho}\right)^\prime&=\left.\left.\frac{1}{2}
\right((\Z+\nu)\f^2+(\Z-\nu)\p^2\right),                  \label{se2}  \\
\frac{1}{\rho}(\rho\f^\prime)^\prime&=
\left(\frac{\om^2}{4}(\Y+1)^2
+\frac{1}{4\rho^2}(\Z+\nu)^2
+\frac{\beta}{4}(\f^2+\p^2-1)
\right)\f,                                  \label{se3}    \\
\frac{1}{\rho}(\rho\p^\prime)^\prime&=
\left(\frac{\om^2}{4}(\Y-1)^2
+\frac{1}{4\rho^2}(\Z-\nu)^2
+\frac{\beta}{4}(\f^2+\p^2-1)
\right)\p .                                       \label{se4}
\end{align}
The boundary conditions are obtained from Eqs.\eqref{orig},\eqref{inf},
\begin{align}               \label{rec0}
a_1+\ldots \leftarrow\, &u\to -1+ \frac{c_3g^{\prime 2}}{\sqrt{\rho}}\,e^{-\mz\rho}
+\ldots \,,    \nonumber \\
2n-\nu+a_4\rho^2+\ldots\leftarrow\, &\Z\to -\nu+ 
c_4g^{\prime 2}\sqrt{\rho}\,e^{-\mz\rho}+\ldots\,,            \nonumber \\  
a_3 \rho^{\n}+\ldots \leftarrow\, & \f \to 1
+ \frac{c_5}{\sqrt{\rho}}\,e^{-\mh\rho}+\ldots\,,          \nonumber \\  
q\,\rho^{|\n-\nu|}+\ldots\leftarrow\, & 
\p \to \frac{c_6}{\sqrt{\rho}}\,e^{-m_\om \rho}+\ldots\,, 
\end{align}
where $\mz,\mh$ are the same as in Eq.\eqref{masses}, while 
Eq.\eqref{m-sigma} now gives $m_\sigma=\sigma$. 
Since all the fields are massive and approach zero 
exponentially fast, the vortex energy is finite.

\begin{figure}[ht]

\hbox to\linewidth{\hss%
\psfrag{lnx}{{$\ln(1+\rho)$}}
\psfrag{y}{{}}
\psfrag{title}{}
\psfrag{u}{$u$}
\psfrag{v}{$v$}
\psfrag{f1}{$f_1$}
\psfrag{f2}{$f_2$}
	\resizebox{8.5cm}{7cm}{\includegraphics{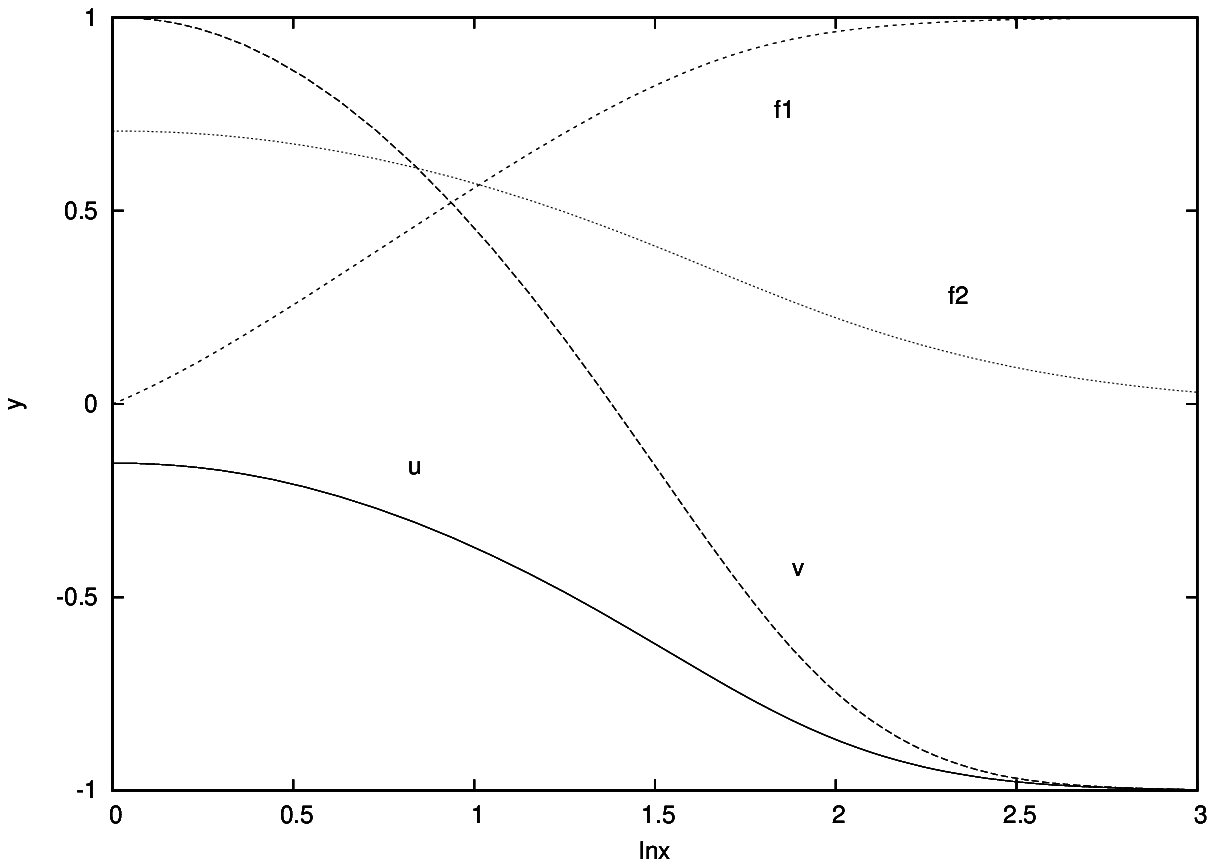}}
\hspace{5mm}%
\psfrag{x}{{${\cal I}$}}
\psfrag{y}{{$E/2\pi n$}}
\psfrag{n1nu1}{{$n=\nu=1$}}
\psfrag{n2nu2}{{$n=\nu=2$}}
\psfrag{n2nu1}{{$n=2,\nu=1$}}
\psfrag{title}{}
        \resizebox{8.5cm}{7cm}{\includegraphics{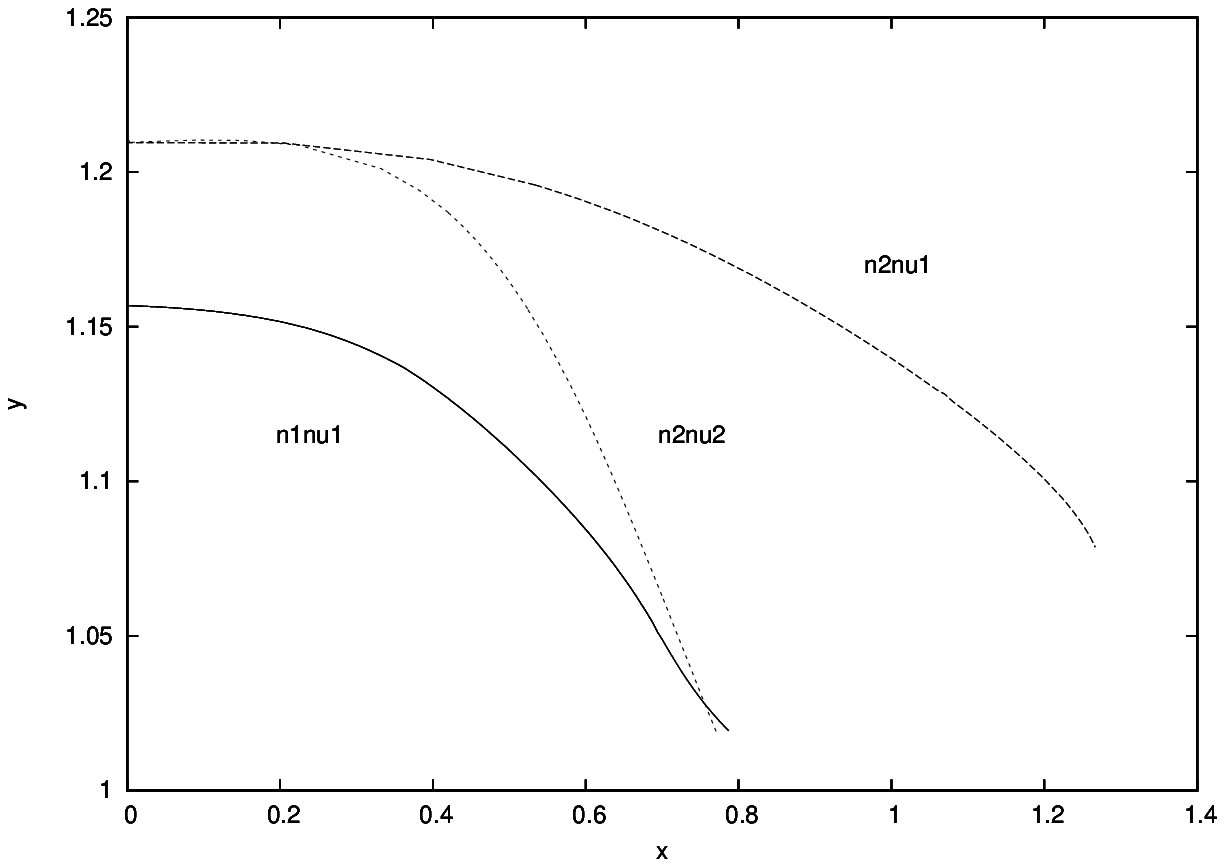}}	
\hss}

\caption{\small  Features of the $n=\nu=1$, $\beta=2$ solutions 
in the semilocal limit, $g=0$. The solution profiles for $\sigma=0.1$ (left)
and the total energy $E/(2\pi n)$
against the restframe current (right). 
}
\label{FIG_semilocal}
\end{figure}

The simplest solution of Eqs.\eqref{se1}--\eqref{se4}
is obtained for $q=0$, 
 this is the Z string \eqref{Zsol}
restricted to $\thetaw=\pi/2$, in which case it is called
{\it semilocal string} \cite{achuc},\cite{achuc1}.
%\be
%\Y=-1,~~~\p=0,~~~\Z=2v_{\mbox{\tiny ANO}}(\rho)-\nu,~~~
%\f=f_{\mbox{\tiny ANO}}(\rho).
%\ee
Choosing $q\neq 0$ gives current-carrying vortices
called  {\it twisted semilocal strings} 
\cite{SL}, a typical solution being shown in Fig.\ref{FIG_semilocal}. 
These solutions exist only for $\beta>1$ and for $\nu=1,\ldots n$. This is due 
to the fact that for $\thetaw=\pi/2$ the allowed region in 
Fig.\ref{FIG_chiral} is $\beta>1$ if $\nu\leq n$ while for 
$\nu>n$ the allowed region is empty. For given $\beta,n,\nu$
the solutions comprise a family that can be labeled by $q\in[0,1)$. 

%%%%%%%%%%%%%%%%%%%%%%%%%%%%%%%%%%%%%%%%%%%%%%%%%%%%%%%%%%%%%%ù
%%%%%%%%%%%%%%%%%%%%%%%%%%%%%%%%%%%%%%%%%%%%%%%%%%%%%%
 
 The current in this case is not local but global, 
related to the SU(2) invariance
of the model \eqref{LL}. It is determined by $J^a_\mu$ in Eq.\eqref{P1},
\be                                \label{Isemi}
I^{{\rm SU(2)}}_\alpha=\int J^3_\alpha\, d^2x=
\pi\sigma_\alpha\int_0^\infty\left\{(u+1)f_1^2
+(1-u)f_2^2\right\}\rho d\rho,
\ee
while the similar integrals of $J^1_\alpha$ and $J^2_\alpha$ vanish. 
The energy $E=2\pi\int_0^\infty({\cal E}_1+{\cal E}_2)\rho d\rho$
is finite and 
can be obtained form Eqs.\eqref{EEE1},\eqref{EEE2} by setting
the SU(2) amplitudes to the specified above values and omitting terms with 
their derivatives. For $\beta>1$ and $g=0$ there is a non-trivial lower bound for the 
energy determined by the total derivative in the second line of 
Eq.\eqref{EEE4}, 
\be                                    \label{BPS1}
E\geq E_{\rm min}= 2\pi\int_0^\infty \frac12\left\{\Z-(\Z+\nu)\f^2
-(\Z-\nu)\p^2 \right\}^\prime d\rho=2\pi n\,.
\ee
It turns out that the energy of the `twisted strings' monotonously 
{\it decreases} with current (see Fig.\ref{FIG_semilocal})
and approaches 
the lower bound $2\pi n$ when the current tends to infinity
\cite{SL}. The twist $\sigma$ also decreases with current, 
as for solutions with $g\neq 0$ (see Fig.\ref{fig3a}).  
The parameter $q$ increases monotonously 
  and approaches unit value  
when the current tends to infinity
\cite{SL}, while for solutions with $g\neq 0$ it always passes
through a maximum and then tends to zero for large currents,
as shown in Fig.\ref{FIG6}. If one plots $q$ against $\sigma$ (or against the current)
for different $g$, then it turns out that the curves approach
the limiting curve for $g=0$ pointwise but not uniformly, since the 
boundary condition at $\sigma=0$ are different, 
depending on whether 
$g=0$ or $g\neq 0$ (see Fig.\ref{fig36}).

The lower energy bound $E_{\rm min}$ in Eq.\eqref{BPS1} is attained
for $\beta=1$ if the first order Bogomol'nyi equations
${\cal B}_s=0$ with  ${\cal B}_s$
defined by Eq.\eqref{BBB} are fulfilled. For $g=0$ there are 
only three such equations,
\bea                 \label{BBB1}
\Z^\prime+\frac{\rho}{2}\left[(\f^2+\p^2)-1\right]&=&0,~~\notag \\ 
\f^\prime+\frac{1}{2\rho}\,(\Z+\nu)\f&=&0,~~ \notag \\
\p^\prime+\frac{1}{2\rho}\,(\Z-\nu)\p&=&0. ~~~
\eea
Their solutions 
comprise a family labeled by $q\in[0,1)$ \cite{skyrmions},
they also fulfill the three second order equations 
\eqref{se2}--\eqref{se4} with $\sigma^2=0$, 
while the amplitude $u$ can be obtained from Eq.\eqref{se1}. 
Reconstructing the fields  \eqref{003-semi}, 
solutions with $\sigma_\alpha=0$  
are  sometimes called `skyrmions',
while those with $\sigma_0=\pm\sigma_3\neq 0$ 
can be called `spinning skyrmions'  \cite{Abraham}. 
Their profiles
are qualitatively similar to those for the twisted strings, 
but since $m_\sigma=\sigma=0$ in this case, they show polynomial tails
at large $\rho$. It turns out that the twisted strings with 
$\beta>1$  approach the `spinning skyrmion' configurations for ${\cal I}\to\infty$
(modulo an infinite rescaling) \cite{SL}.

\subsection{Isospint limit, $\thetaw=0$} 

\begin{figure}[ht]

\hbox to\linewidth{\hss%
\psfrag{lnx}{{$\ln(1+\rho)$}}
\psfrag{y}{{}}
\psfrag{title}{}
\psfrag{u1}{$u_1$}
\psfrag{u3}{$u_3$}
\psfrag{v1}{$v_1$}
\psfrag{v3}{$v_3$}
\psfrag{f1}{$f_1$}
\psfrag{f2}{$f_2$}
	\resizebox{8.5cm}{7cm}{\includegraphics{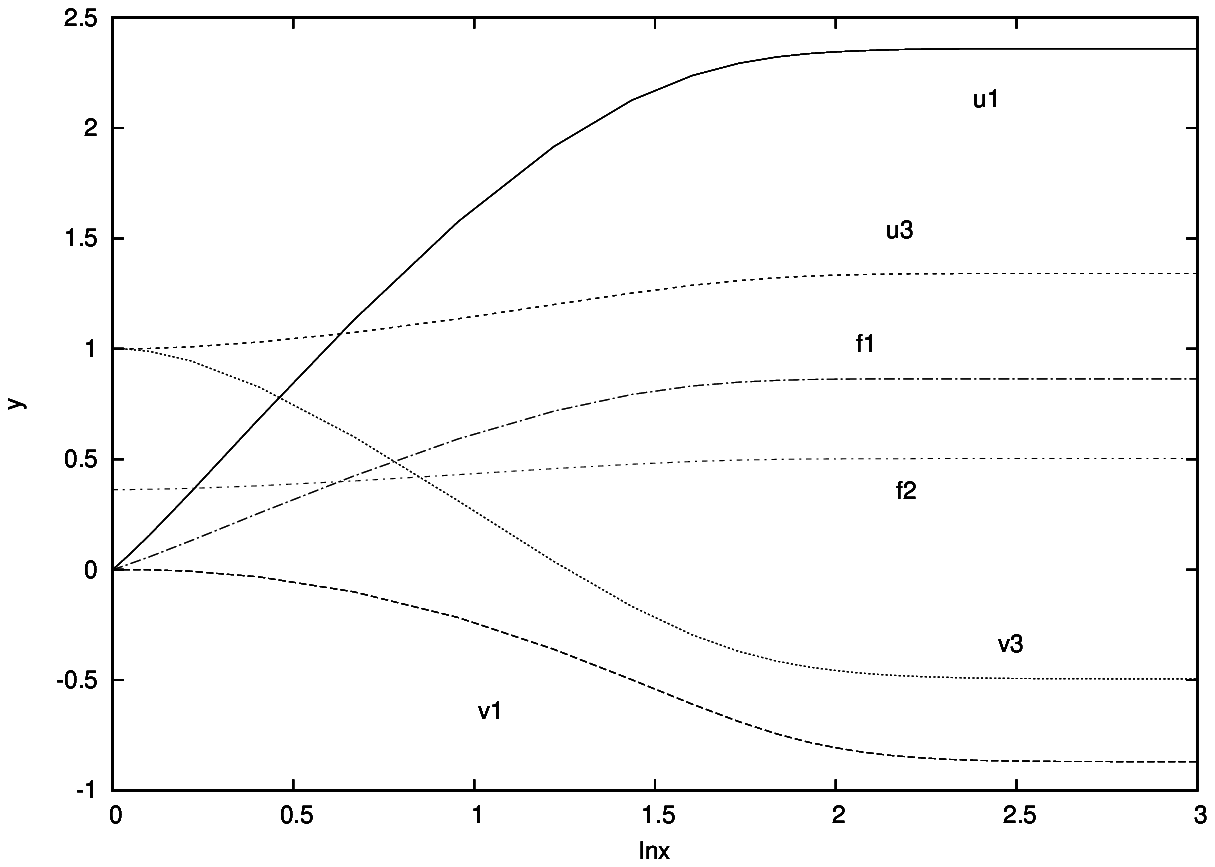}}
\hspace{5mm}%
\psfrag{x}{{${{\cal I}}$}}
\psfrag{y}{{$E/2\pi n$}}
\psfrag{title}{}
\psfrag{n1nu1}{{$n=\nu=1$}}
\psfrag{n2nu2}{{$n=\nu=2$}}
\psfrag{n2nu1}{{$n=2,\nu=1$}}
\psfrag{n2nu3}{{$n=2,\nu=3$}}
        \resizebox{8.5cm}{7cm}{\includegraphics{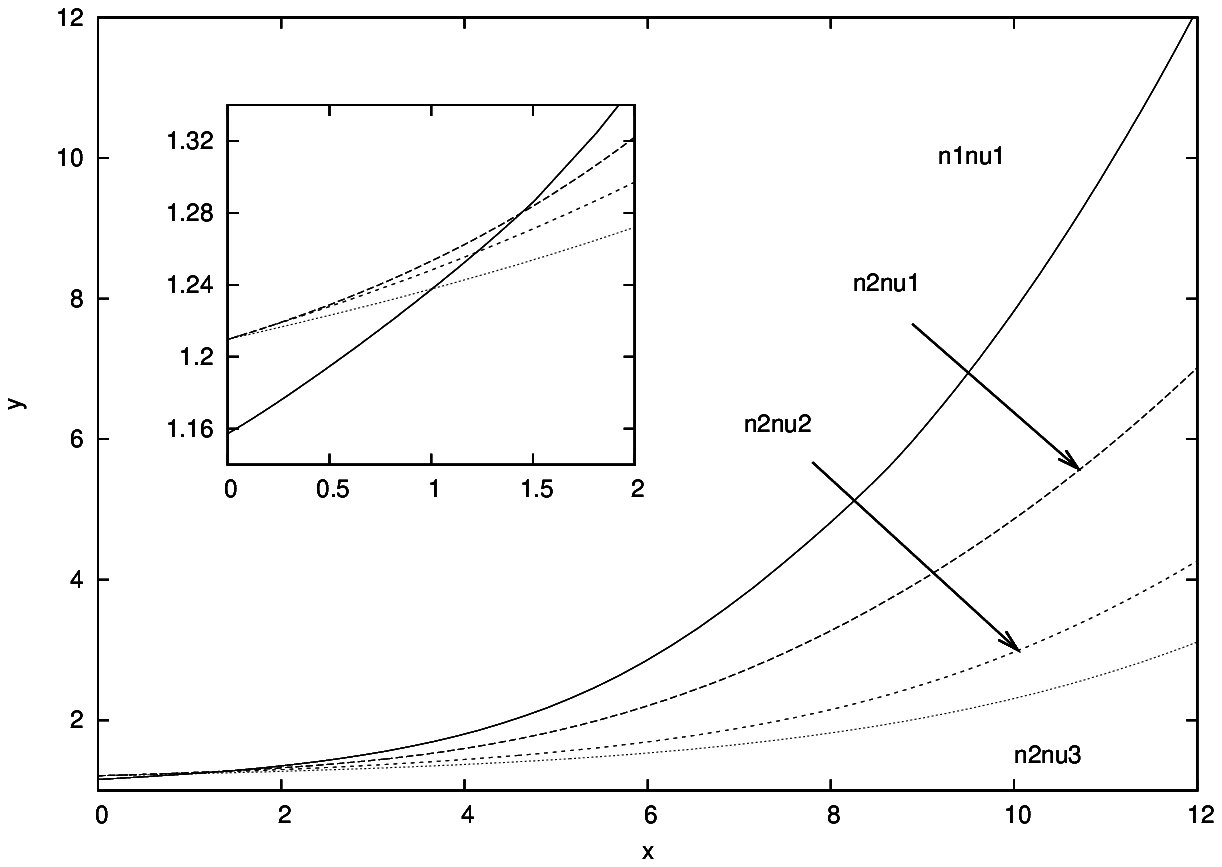}}	
\hss}

\caption{\small  Features of the $n=\nu=1$, 
$\beta=2$ solutions in the isospin limit, $g=1$. 
The profiles for  $\sigma=0.5$ 
(left) and the energy  
against the restframe current (right). The insertion shows 
that $E/n$ increases with ${\cal I}$ 
faster for $n=1$ than for $n=2$, 
so that the vortices repel 
each other for small currents and attract otherwise. 
}
\label{FIG_isospin}
\end{figure}

If $g^\prime=\sin\thetaw$ is small then  
the hypercharge field $\A_\mu$ scales as $g^{\prime 2}$ so that 
for $g^\prime\to 0$ the theory \eqref{L} reduces to
\be                             \label{LLL}
{\cal L}=
-\frac{1}{4}\,\WW^a_{\mu\nu}\WW^{a\mu\nu}
+(D_\mu\Phi)^\dagger D^\mu\Phi
-\frac{\beta}{8}\left(\Phi^\dagger\Phi-1\right)^2,
\ee
with $
D_\mu\Phi
=\left(\partial_\mu
-\frac{i}{2}\,\tau^a \WW^a_\mu\right)\Phi\,.
$
The SU(2) is now local while the U(1) is global. The perturbative mass spectrum contains 
three massive vector bosons with the mass $\mz$ 
and a Higgs boson
with the mass $\mh$. The field equations in the axially symmetric case
are obtained by setting in \eqref{ee3}--\eqref{CONS} 
$u(\rho)=const.$, $v(\rho)=2n-\nu$, which implies that the    
U(1) part of the field \eqref{003a} vanishes and that one has in 
Eqs.\eqref{orig},\eqref{inf},\eqref{rec} $a_1=c_1=u$, 
$c_2=2n-\nu$ and $Q=c_3=c_4=0$. 
The boundary conditions for the remaining field amplitudes  read 
\begin{align}               \label{rec1}
0\leftarrow\, & \Om_1 \to -u\sin\gamma\, \,, \nonumber \\ 
1\leftarrow\, & \Om_3 \to -u\cos\gamma\, \,, \nonumber \\  
0\leftarrow\, & \W_1 \to (\nu-2n)\sin\gamma\,,              \nonumber \\ 
\nu\leftarrow\, & \W_3 \to (\nu-2n)\cos\gamma\,,              \nonumber \\ 
a_3 \rho^{\n}\leftarrow\, & \f \to \cos\frac{\gamma}{2}\,,          \nonumber \\  
q\,\rho^{|\n-\nu|}\leftarrow\, & \p \to \sin\frac{\gamma}{2}\,,   
\end{align}
The simplest solution is obtained for $q=0$, this is the Z string \eqref{Zsol} 
restricted to $\thetaw=0$. 
%\be
%u_1=v_1=f_2=0,~~~u_3=1,~~~v_3=2v_{\mbox{\tiny ANO}}+\nu-2n,~~~~
%f_1=f_{\mbox{\tiny ANO}}
%\ee
For $q\neq 0$ one obtains  current-carrying solutions.
The typical solution is shown in Fig.\ref{FIG_isospin}. 
These solutions relate to the lower boundary of the chiral diagram in 
Fig.\ref{FIG_chiral}, which implies that they exist for any $\beta>0$ 
and for $\nu=1,\ldots 2n-1$. Their current is global, 
related to the global U(1) invariance of the theory \eqref{LLL},
it can be expressed in terms of $J^0_\mu$ in 
Eq.\eqref{P0},
\be                                 \label{IIso}
I^{\rm U(1)}_\alpha=\int J^0_\alpha\,d^2x=
\pi\sigma_\alpha\int_0^\infty\left\{(u_3+u)f_1^2
+2u_1f_1f_2-(u_3-u)f_2^2\right\}\rho d\rho.
\ee 
As for all solutions described above, there are no apparent 
restrictions on its values. 
The energy density is defined by Eqs.\eqref{EEE1},\eqref{EEE2} with
$u,v$  restricted to the values specified above.  
%one should omit terms with $u^{\prime 2}$ and 
%$v^{\prime 2}$ -- they vanish for $g^\prime\to 0$ since for small 
%$g^\prime$ one has $u^\prime \sim v^\prime \sim g^{\prime 2}$.  
The total energy  is finite, it 
always {\it increases} with current (see Fig.\ref{FIG_isospin}). 

%%%%%%%%%%%%%%%%%%%%%%%%%%%%%%%%%%%%%%%%%%%%%%%%%%%%%%%%%%%%%%ù
%%%%%%%%%%%%%%%%%%%%%%%%%%%%%%%%%%%%%%%%%%%%%%%%%%%%%%

These $\thetaw=0$ solutions 
can also be obtained from the generic electroweak vortices 
in the limit $\thetaw\to 0$.
Similarly  the $\thetaw=\pi/2$ solutions can be obtained in the limit $\thetaw\to \pi/2$.
The generic field configurations then approach the limiting ones pointwise,
although non-uniformly at large $\rho$, since they support 
a long-range
field $\sim Q\ln\rho$, while in the limit one has $Q=0$.   
%Solutions for $\thetaw=\pi/2$
%can similarly be obtained by the limiting procedure. 
Since the 
electromagnetic current vanishes in the limit, 
one can wonder how the global currents \eqref{Isemi} and \eqref{IIso}
can  be obtained via the limiting procedure.  
The answer is that they
 can be obtained as limits of the rescaled
 electromagnetic current $I_\alpha/(gg^\prime)$.
%where $e=gg^\prime$ is the dimensionless version of the electron charge.
The following relations take place,
\bea                                    \label{idd}
\frac{1}{gg^\prime}\,J_\alpha&=&
\frac{1}{gg^\prime}\,\partial^\mu F_{\mu\alpha}=
\frac{1}{g^{\prime 2}}\,\partial^\mu B_{\mu\alpha}
-\frac{1}{g^{2}}\,\partial^\mu (n^a\WW^a_{\mu\alpha})=  \notag \\
&=&
J^0_\alpha-\frac{1}{g^{2}}\,\partial^\mu (n^a\WW^a_{\mu\alpha})=  \notag \\
&=&\frac{1}{g^{\prime 2}}\,\partial^\mu B_{\mu\alpha}
-n^aJ^a_\alpha
-\frac{1}{g^{2}}\,(D^\mu n^a)W^a_{\mu\alpha}\,.
\eea
Here we have used Eqs.\eqref{P0},\eqref{P1} and the identity  
$\partial_\sigma (n^a\WW^a_{\mu\alpha})=(D_\sigma n^a)W^a_{\mu\alpha}+
n^aD_\sigma\WW^a_{\mu\alpha}$ where
$D_\sigma n^a=\partial_\sigma n^a+\epsilon_{abc}\WW^b_\sigma n^c$. 
Integrating the first line of these relations over the vortex cross section
gives $\frac{1}{gg^\prime}\,I_\alpha$. Integral of the second line 
reduces in the limit $g^\prime\to 0$ to the global current \eqref{IIso}, 
since the total derivative term in this line gives no contribution, because
the field $\WW^a_{\mu\nu}$ becomes short-ranged in this limit. 

Let us now notice that, since the whole expression in \eqref{idd} 
has the total derivative structure, its integral will not change 
upon replacing 
the vector $n^a=(\Phi^\dagger\tau^a\Phi)/(\Phi^\dagger\Phi)$ 
by any other unit vector with the same value at infinity. 
Let us therefore replace $n^a$ by $\tilde{n}^a=\delta^a_3$ in the third line
in \eqref{idd}. One will have $D_\mu \tilde{n}^a=\epsilon_{ab3}\WW^b_\mu$,
which scales as $g^2$ 
for small $g$, so that the last term in the third line 
vanishes in the limit $g\to 0$. The second term 
gives upon integration the global current \eqref{Isemi}
(up to the sign),
while the first term gives zero, since the field $B_\mu$ 
becomes massive
in the limit. We therefore conclude that 
\be
I^{\rm U(1)}_\alpha\leftarrow\frac{1}{gg^\prime}\,J_\alpha\to
-I^{\rm SU(2)}_\alpha
\ee
for $g^\prime\to 0$ and $g\to 0$, respectively, so that the global 
currents can indeed be obtained from the local 
electromagnetic current via the limiting
procedure.

\subsection{Special chiral solutions  \label{chirlim}}

Superconducting strings in the Witten model are usually constructed 
starting from the `dressed' $\sigma^2=0$  chiral/currentless string, 
viewed as  the `bare' ANO vortex stabilized 
by the condensate in its core.  
Increasing $\sigma^2$ produces the current, while 
the parameter $q$ decreases and for $q\to 0$ the solutions 
reduce to the `bare' ANO vortex  (see Appendix C). 

Superconducting vortices in the Weinberg-Salam theory are obtained 
in the opposite way, starting from the `bare' ANO vortex 
(Z string) and then decreasing the twist $\sigma$.  
One does not generically find chiral solutions in this case, 
since the limit $\sigma^2\to 0$ 
corresponds to  infinite and not zero restframe current.

However, chiral solutions can be obtained for special values of the 
parameters. As was discussed in Sec.\ref{weak}, for $q\ll 1$ they 
exist if $\beta,\thetaw,n,\nu$ belong to 
the chiral curves   shown in Fig.\ref{FIG_chiral}. 
 These curves are obtained by solving the 
linear eigenvalue problem \eqref{eqs1}. If $q$ is not small, 
then one should solve  the full system of equations \eqref{ee1}-\eqref{CONS} 
under the condition
\be                                 \label{chir123}
\sigma^2(\beta,\thetaw,n,\nu,q)=0\,.
\ee
This condition reduces by one the number of free
parameters in the local solutions \eqref{orig}, \eqref{inf}, so that 
in order to perform the matching one has to consider $\thetaw$ as one of the 
shooting parameters. For given $n,\nu,q$ this determines 
curves in the $(\beta,\thetaw)$ plane. 
For $\q\to 0$ these curves reduce to those shown in Fig.\ref{FIG_chiral},
while for $\q\neq 0$ they look qualitatively similar but shift  
{\it upwards}. 
As a result, given a point $\beta,\thetaw$ in an 
upper vicinity of a 
curve in Fig.\ref{FIG_chiral} one can adjust $\q$ such that the shifted 
curve will pass
through this point. This fine tuning determines 
the values of $\beta,\thetaw,\q$ for which there is  
a solution of Eqs.\eqref{ee1}--\eqref{ee8} with $\om^2=0$. 
The chiral solutions are therefore not generic but exist 
only for values of $\beta,\thetaw$ which are close to the chiral curves in 
Fig.\ref{FIG_chiral}. 

%%%%%%%%%%%%%%%%%%%%%%%%%%%%%%%%%%%%%%%%%%%%%%%%%%%%%%%%%%%%%%ù
\begin{figure}[ht]

\hbox to\linewidth{\hss%
\psfrag{lnx}{{$\ln(1+\rho)$}}
\psfrag{y}{{}}
\psfrag{title}{}
\psfrag{u1}{$u_1$}
\psfrag{u3}{$u_3$}
\psfrag{v1}{$v_1$}
\psfrag{v3}{$v_3$}
\psfrag{f1}{$f_1$}
\psfrag{f2}{$f_2$}
\psfrag{v3overnu}{{$v_3/\nu$}}
\psfrag{v}{$v$}
\psfrag{v2overnu}{$v_3/\nu$}
	\resizebox{8.5cm}{7cm}{\includegraphics{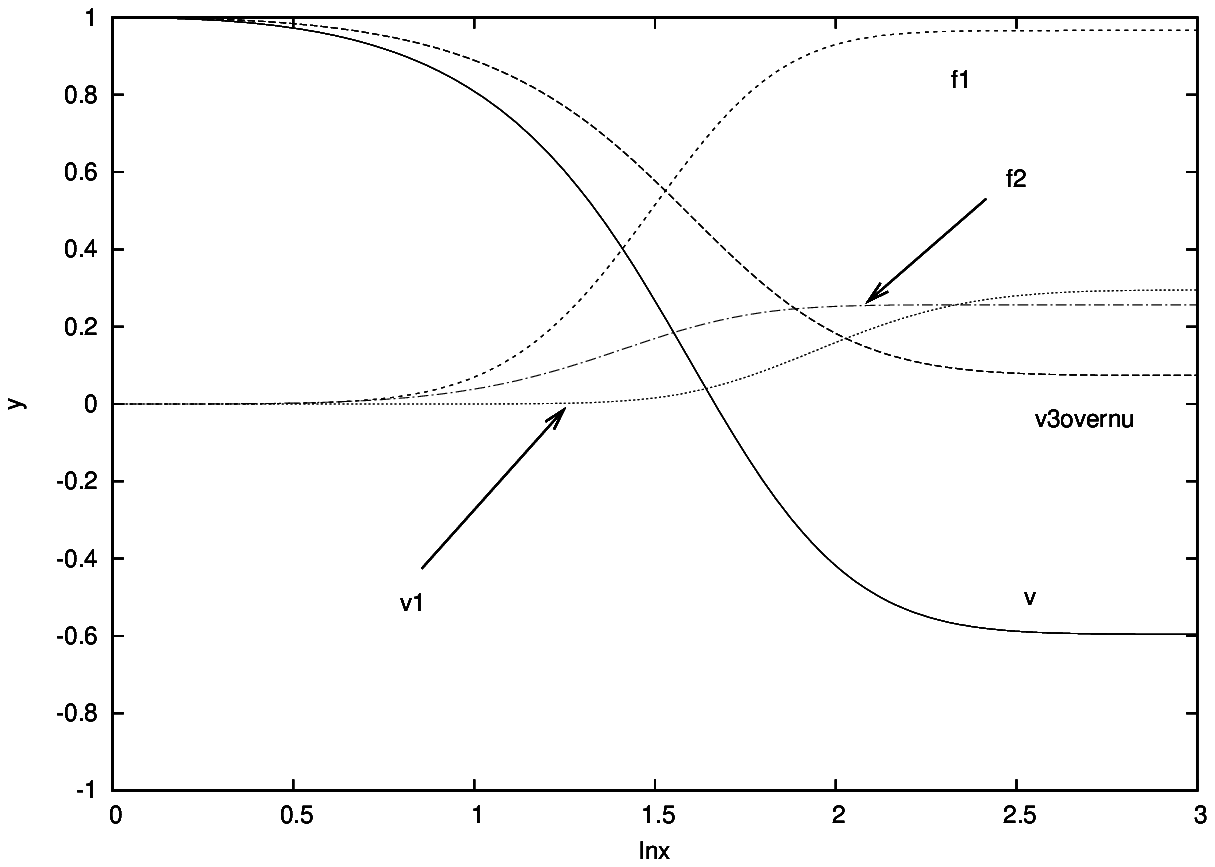}}
\hspace{1mm}%
\psfrag{x}{{${{\cal I}}$}}
\psfrag{title}{}
        \resizebox{8.5cm}{7cm}{\includegraphics{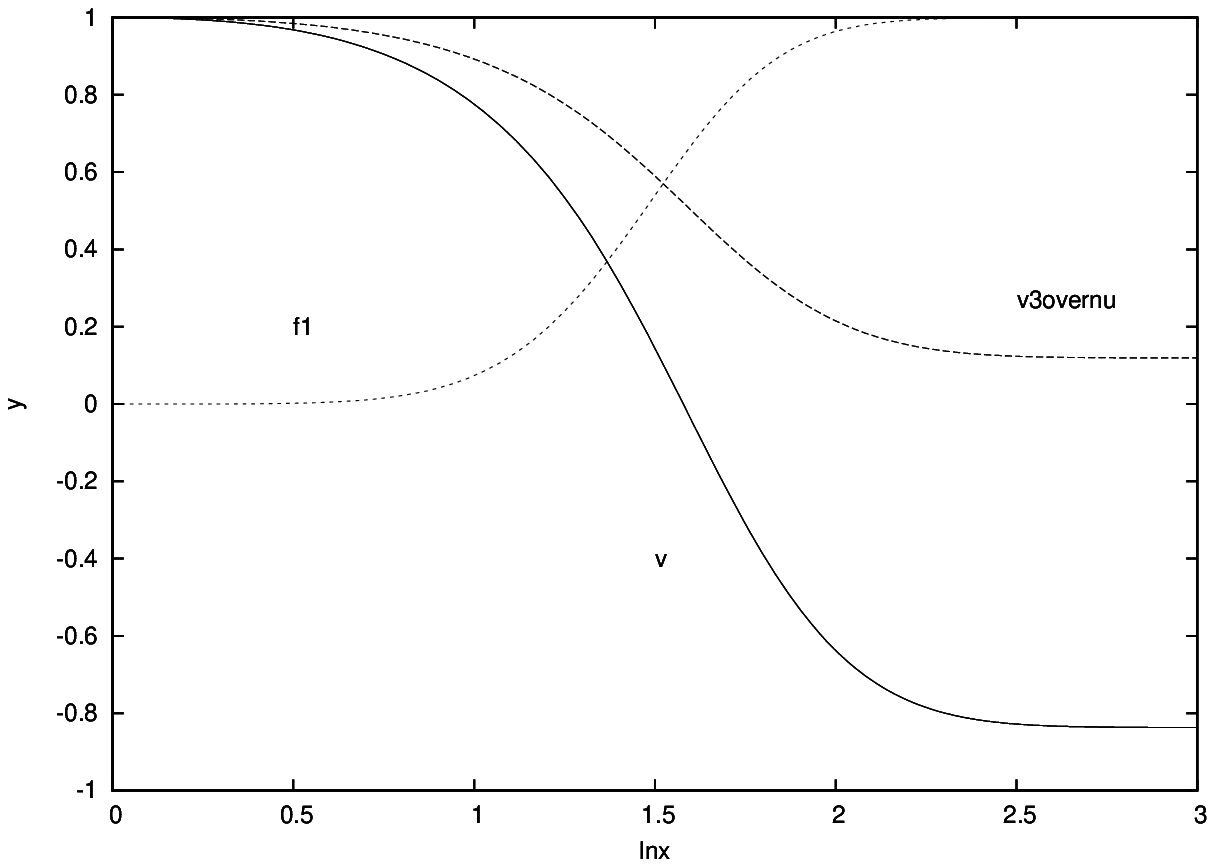}}	
\hss}

\caption{\small Profiles of the 
special chiral solution (left) and Z string (right)  for 
$n=4$, $\nu=7$, $\beta=2$, $g^{\prime 2}=0.23$. 
}
\label{FIG-chiral}
\end{figure}
%%%%%%%%%%%%%%%%%%%%%%%%%%%%%%%%%%%%%%%%%%%%%%%%%%%%%%

Reconstructing the fields \eqref{003a}, 
a given solution with $\sigma^2=\sigma_3^2-\sigma_0^2=0$ determines a family
of chiral vortices labeled by $\sigma_0=\pm\sigma_3$, different members of this family
being related by Lorentz boosts.  These vortices carry a chiral current and 
support a long-range field so that their energy 
is infinite. However, there is a distinguished solution with
$\sigma_0=\sigma_3=0$ for which the current is zero 
and the energy is finite. This special chiral solution
is not a Z string or its gauge copy, since it has  $\Psi_A\neq 0$
(see Fig.\ref{FIG-chiral}).

Such solutions remind of the `W-dressed Z strings' discussed some time ago 
\cite{per,Olesen}. By analogy with the `dressed' vortices in the Witten
model, these were supposed to be  Z strings stabilized by the W-condensate 
in the core. However, it seems that  this 
stabilization mechanism  
does not work in the electroweak case because  Z strings
are non-topological and can unwind into vacuum \cite{KO}. 
The numerical search for the `W-dressed Z strings' gives  
a negative result  \cite{perk} and so it seems they do not exist. 
Fortunately, as we have seen, the `W-dressed Z strings' are not necessary for 
constructing the superconducting vortices.

The special chiral solutions cannot be considered as the `dressed' Z strings,
first because 
they do not exist for generic values of the parameters, and secondly 
because 
they are not always less energetic than the 
Z string with the same $\beta,\thetaw,n$. 
Specifically, for $n=1$ they indeed 
have lower energy than the corresponding 
Z strings, but their parameters  $\beta,\thetaw$ lie in this case in the  
non-physical region (see Fig.\ref{FIG_chiral}), while
if they were the true `W-dressed Z strings' they would exist 
for any parameter values. 
The special chiral solutions can also be found   
in the physical region (see Fig.\ref{FIG-chiral}),  
 but only for higher values of 
$n,\nu$ starting from $n=4$, $\nu=7$ (see Fig.\ref{FIG_chiral}).
In this case  they turn out to be 
more energetic than the corresponding Z strings. 
%Presumably, these currentless solutions should rather be viewed as  
%superpositions of superconducting vortices with currents
%flowing in the opposite directions and canceling each other.

\begin{figure}[ht]

\hbox to\linewidth{\hss%
\psfrag{x}{{$q$}}
\psfrag{y}{}
\psfrag{I}{{$\mathcal{I}$}}
\psfrag{sigma}{{$\sigma$}}
       \resizebox{8.5cm}{7cm}{\includegraphics{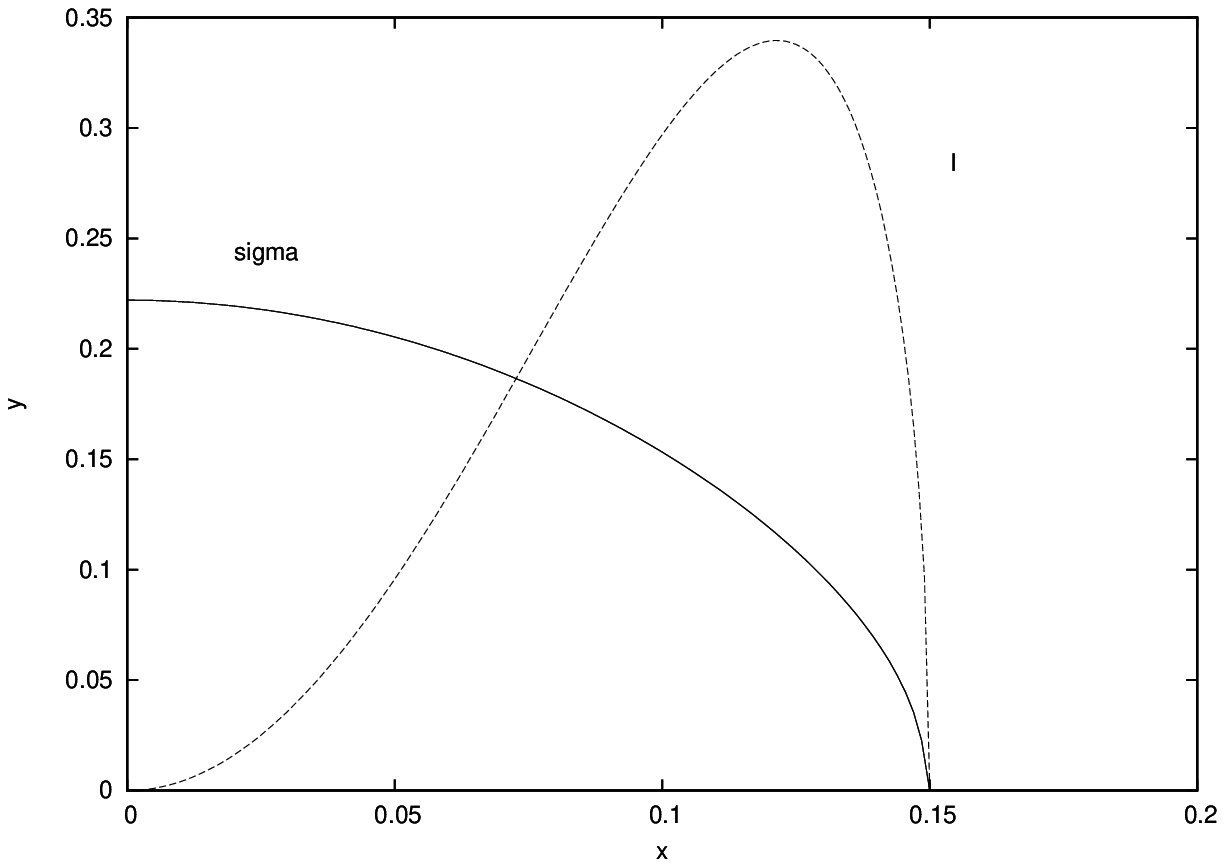}}
\hspace{5mm}%
\psfrag{x}{{$\mathcal{I}$}}
\psfrag{y}{{$\mathcal{E}$}}
\psfrag{q0}{{$q=0$}}
\psfrag{qstar}{{$q_\star$}}
        \resizebox{8.5cm}{7cm}{\includegraphics{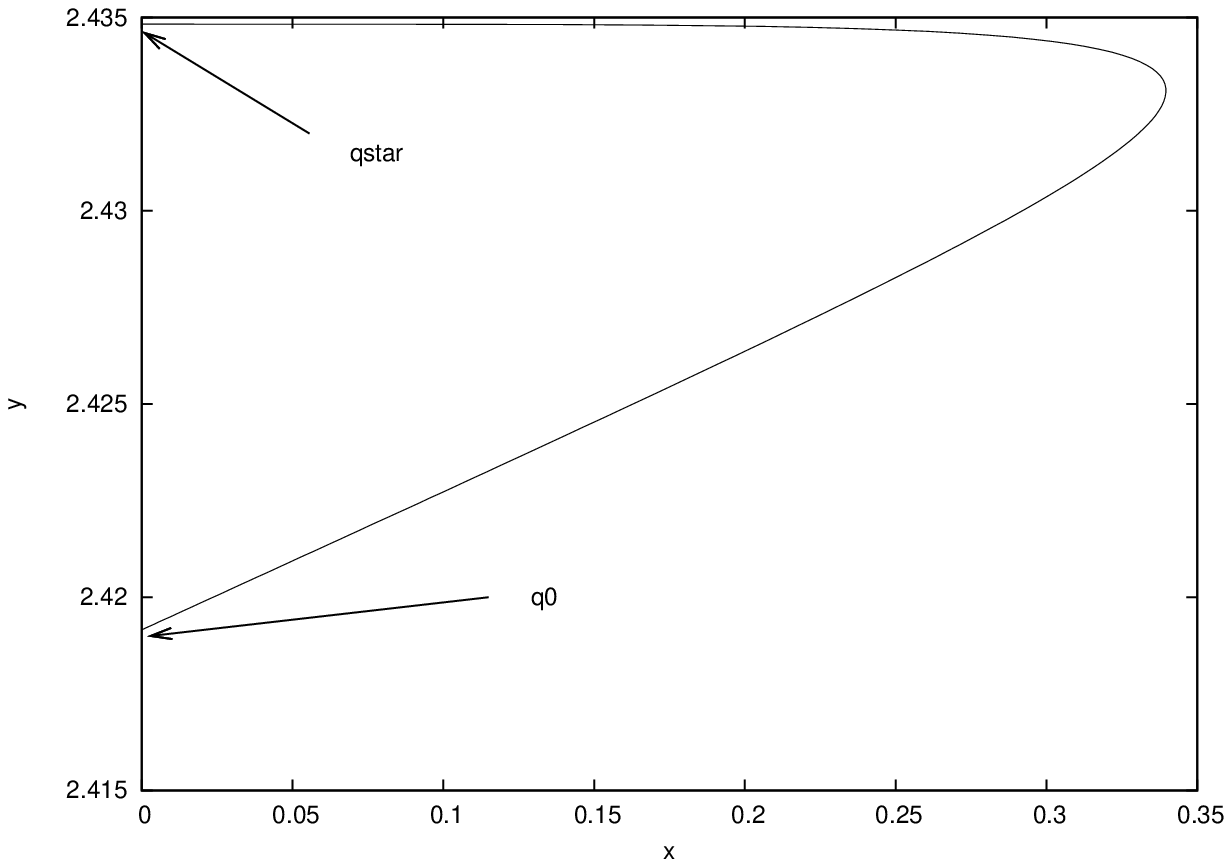}}
\hss}
\caption{\small  The current $\mathcal{I}$ and twist $\sigma$ against $q$ (left)
and the magnetic energy versus current (right) 
for the family of the $n=\nu=1$ electroweak vortices for $\beta=2$ and 
$\sin^2\thetaw=0.56$  that interpolates between the Z string for $q=0$
and the special chiral solution for $q=q_\star=0.15$. }
\label{FIG_QUENCH_param}
\end{figure}

Although the special chiral solutions are not `dressed' Z strings, one can 
relate them to  Z strings very much in the same way as one
interpolates between the `bare' and `dressed' vortices in the Witten model. 
Given a special chiral solution one can 
iteratively decrease $q$ relaxing at the same time the $\sigma^2=0$ condition. 
This gives solutions with  $\sigma^2>0$ which in the $q\to 0$ limit
reduce to the Z string. The current ${\cal I}(q)$ then shows 
(see Figs.\ref{FIG_QUENCH_param}) the quenched
behavior very similar for that for the Witten strings 
(see Fig.\ref{FIG-Wit}), but it should be stressed again that 
this type of behavior is not generic in the electroweak theory,
since it requires the fine-tuning of $\thetaw$.  
As is seen in Fig.\ref{FIG_QUENCH_param}, the energy 
of the special chiral solution can be higher than for the corresponding 
Z string.

\subsection{Infinite Higgs mass limit,  $\beta\to\infty$}

In this limit
the following constraint is enforced, 
\be	\label{BI_CONSTR}
\Phi^\dagger\Phi = f_1^2+f_2^2 = 1,
\ee
so that one can parametrize the Higgs field amplitudes as 
\be	\label{inf_beta}
f_1=\cos\Theta(\rho),~~~~~\ f_2=\sin\Theta(\rho).
\ee
The  equations for the gauge field amplitudes $u,u_1,u_3,v,v_1,v_3$
are then given by  \eqref{ee1},\eqref{ee2},\eqref{ee5}--\eqref{ee8} 
with $f_1,f_2$ defined by \eqref{inf_beta}, while the equation
for $\Theta$ reads
\begin{align}	\label{BI_MOTION}
% \frac{1}{\rho}\left(\rho u^\prime \right)^\prime &=
%	\frac{g^{\prime 2}}{2}\left( u+u_1\sin 2\Theta +u_3\cos2\Theta \right)  , \\
%\rho\left(\frac{v^\prime}{\rho}\right)^\prime &= 
%	\frac{g^{\prime 2}}{2}\left( v+v_1\sin 2\Theta +v_3\cos2\Theta \right)  , \\
%\frac{1}{\rho}\left(\rho u_1^\prime \right)^\prime &= 
%	-\frac{1}{\rho^2}\left(v_1u_3-v_3u_1 \right)v_3
%	+\frac{g^2}{2}\left( u_1+u\sin 2\Theta \right)  ,  \\
%\rho\left(\frac{v_1^\prime}{\rho}\right)^\prime &= 
%	\sigma^2\left(v_1u_3-v_3u_1 \right)u_3
%	+\frac{g^2}{2}\left( v_1+v\sin 2\Theta \right)  , \\
%\frac{1}{\rho}\left(\rho u_3^\prime \right)^\prime &= 
%	\frac{1}{\rho^2}\left(v_1u_3-v_3u_1 \right)v_1
%	+\frac{g^2}{2}\left( u_3+u\cos 2\Theta \right)  ,  \\
%\rho\left(\frac{v_3^\prime}{\rho}\right)^\prime &= 
%	-\sigma^2\left(v_1u_3-v_3u_1 \right)u_1
%	+\frac{g^2}{2}\left( v_3+v\cos 2\Theta \right)  ,  \\
\frac{1}{\rho}\left(\rho \Theta^\prime \right)^\prime &= 
	\frac{\sigma^2}{2}\left(u_1\cos2\Theta-u_3\sin2\Theta  \right)u
	+\frac{1}{2\rho^2}\left(v_1\cos2\Theta-v_3\sin2\Theta  \right)v  .
\end{align}
The constraint \eqref{CONS} becomes 
\be	\label{BI_CONSTR_EQ}
\sigma^2\left( u_1u_3^\prime-u_3u_1^\prime \right) 
+\frac{1}{\rho^2}\left( v_1v_3^\prime-v_3v_1^\prime \right) -g^2\Theta^\prime =0.
\ee

\begin{figure}[ht]
\hbox to\linewidth{\hss%
\psfrag{lnx}{{$\ln(1+\rho)$}}
\psfrag{y}{{}}
\psfrag{v}{$v$}
\psfrag{v1}{$v_1$}
\psfrag{v3}{$v_3$}
\psfrag{f1}{$f_1$}
\psfrag{f2}{$f_2$}
	\resizebox{8.5cm}{7cm}{\includegraphics{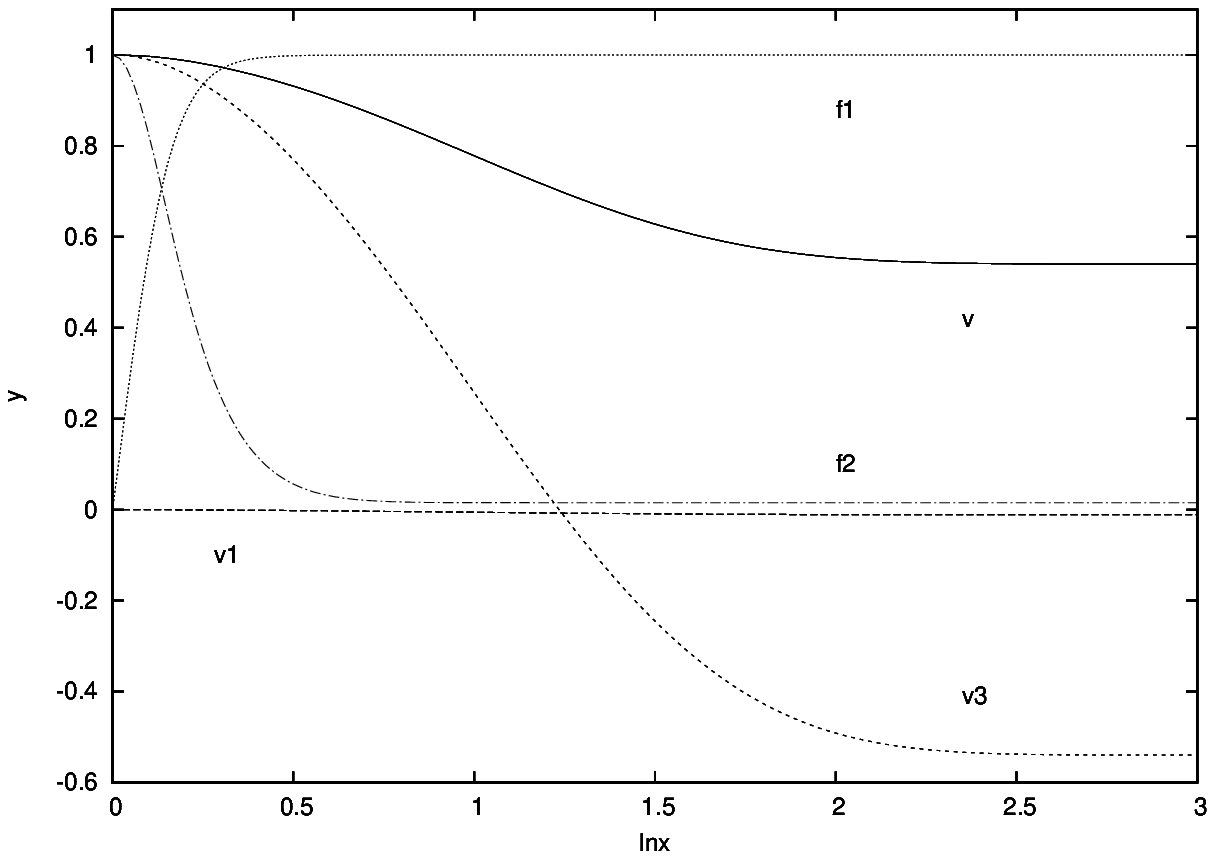}}
\hspace{5mm}%
\psfrag{lnx}{{$\ln(1+\rho)$}}
\psfrag{y}{{}}
\psfrag{v}{$v$}
\psfrag{v1}{$v_1$}
\psfrag{v3}{$v_3$}
\psfrag{f1}{$f_1$}
\psfrag{f2}{$f_2$}
        \resizebox{8.5cm}{7cm}{\includegraphics{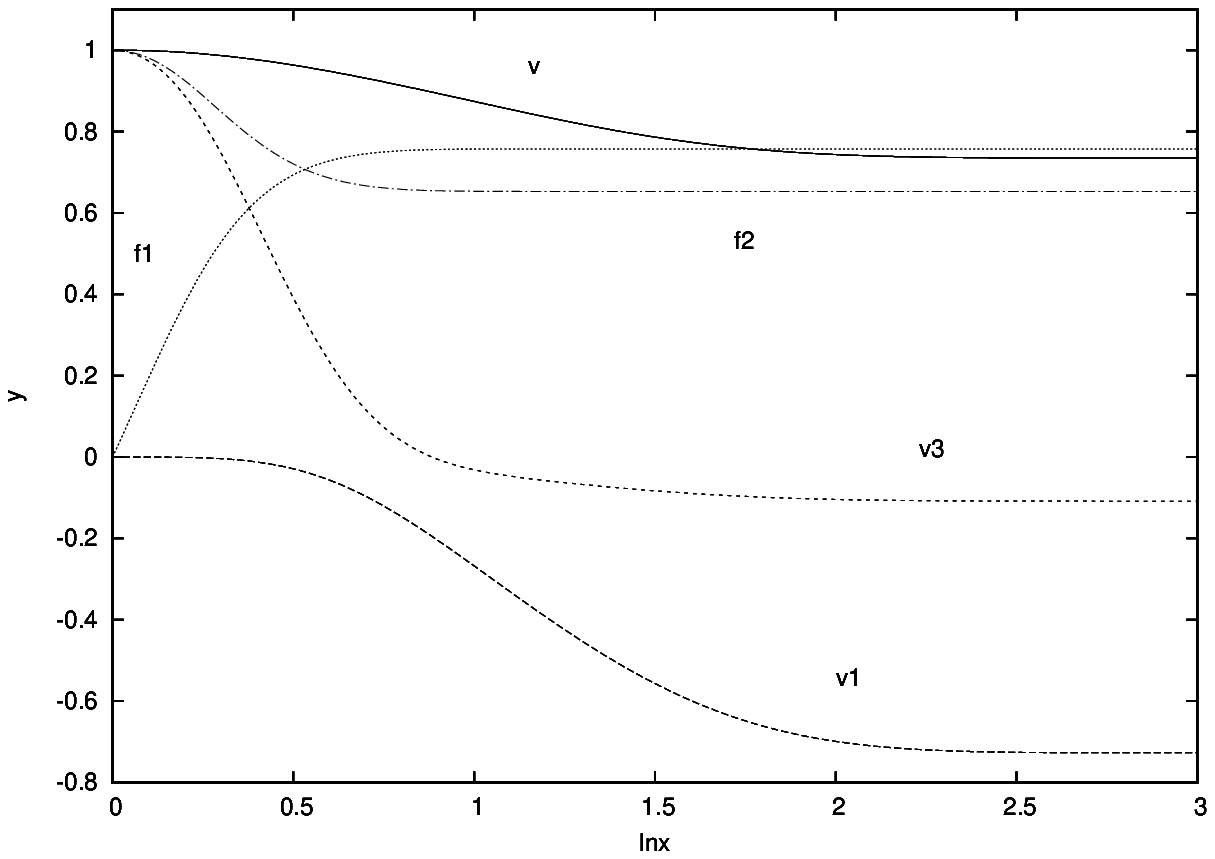}}
\hss}
\caption{\small  The $n=\nu=1$, $\beta=\infty$ solutions 
for $\sigma=5$ (left)
and for $\sigma=0.1$ (right). 
The electric amplitudes are not shown. 
}
\label{FIG_beta_inf}
\end{figure}

The boundary conditions for the function $\Theta$ for $\nu=n$ read
\be	\label{BI_LOCAL}
\frac{\pi}{2}+a_5\rho^n +\dots \leftarrow 
\Theta \rightarrow \frac{\gamma}{2}+\frac{c_5}{\sqrt{\rho}}\,
e^{-\int m_\sigma d\rho}+\dots ,
\ee
which implies that one always has $q=f_2(0)=1$. One cannot therefore use $q$
to label the solutions, but instead one can use $\sigma$.  
The boundary conditions for the gauge field amplitudes 
are given by Eqs.\eqref{orig},\eqref{inf}, provided that $\nu=n$. For $\nu\neq n$
the energy density diverges at the origin due to the term 
$(v-v_3)^2 f_2^2/\rho^2$ in \eqref{EEE2}. Therefore, solutions with $\nu\neq n$
become singular for $\beta\to\infty$, which suggests that 
 they can be viewed as excitations over the fundamental regular 
solution with $\nu=n$. 

The typical solutions are shown in Fig.\ref{FIG_beta_inf}.
As usual, the limit where $\sigma$ is small corresponds to large currents,
however, since $\Phi^\dagger\Phi=1$, the solutions do not show in this 
limit a region of vanishing Higgs field. This can be seen already 
from Eq.\eqref{rho-star}, since the size of the $\Phi=0$ region scales
as $\rho_\star\sim {\cal I}/\sqrt{\beta}$ and shrinks to zero  
when $\beta\to\infty$.

It seems that for $\beta=\infty$ there is no upper bound for $\sigma$.
Large $\sigma$'s correspond to small currents, the solutions then 
approaching Z strings. 
  The vacuum angle $\gamma$ then tends to zero, while 
the mass term \eqref{m-sigma} approaches infinity
(since $\sigma Q\sim{\cal I}\to 0$ and $m_\sigma\approx\sigma$)
in which case  
$\Theta$ changes very fast  at small $\rho$
to approach its asymptotic value.
For $\sigma=\infty$ one should set $\Theta(\rho)=0$, and then 
the solutions reduce to Z strings 
`in the London limit'.

\section{Summary and concluding remarks}

In summary, we have presented new solutions in the bosonic
sector of the Weinberg-Salam theory which describe straight vortices (strings) 
carrying a constant electric current. 
Such superconducting vortices contain a regular central core 
filled with the condensate of massive W bosons creating the electric current.
The current produces the long-range electromagnetic field. 
The solutions exist for any value of the Higgs boson mass
and for (almost) any weak mixing angle, in particular for the 
physical  values $\beta\in(1.5,3.5)$ and $\sin^2\thetaw=0.23$. 
They 
comprise a family that can be labeled by the four
parameters in the ansatz \eqref{003a}: $\sigma_\alpha$, 
$n$, $\nu$. 
These parameters determine the 
vortex electric charge density
$I_0\sim\sigma_0$, the electric current $I_3\sim\sigma_3$, 
the electromagnetic and Z fluxes $\Psi_F$ and $\Psi_Z$, as well as 
the vortex momentum $P\sim\sigma_0\sigma_3$ and angular momentum $M\sim\sigma_0$. 

The spacetime vector $I_\alpha=(I_0,I_3)\sim\sigma_\alpha$ is generically 
spacelike, so that its temporal component can be boosted away to give 
$I_\alpha=\delta^3_\alpha\,{\cal I}$
and $\sigma_\alpha=\delta^3_\alpha\,\sigma$. 
The restframe current ${\cal I}$ can assume any value. 
In the ${\cal I}\to 0$ limit the solutions reduce to Z strings, while the 
twist $\sigma$ reduces to the eigenvalue of the linear fluctuation operator 
 on the Z string background. 

For large currents
the solutions show a symmetric phase region of size $\sim{\cal I}$
where the magnetic field is so strong that it drives the Higgs field to zero. 
However, this does not destroy the vector boson superconductivity, 
since the scalar Higgs field is not the relevant order parameter in this case. 
The  current-carrying W-condensate is contained 
in the very centre of the symmetric phase, in a core of size $\sim 1/{\cal I}$, 
while the rest of this region 
is dominated by the massless electromagnetic and Z fields. 
The symmetric phase is surrounded by  the `crust' layer where the Higgs field 
relaxes to the broken phase while the Z field becomes massive and dies
away. Outside the crust there remains only the long-range  
Biot-Savart magnetic field, which produces a mild, 
logarithmic energy divergence 
at large distances for their core. 
However, finite vortex pieces, as for example vortex loops, will have finite energy. 

Straight, infinite vortices can have finite energy in special cases: 
either for 
$\thetaw=0,\pi/2$ 
when the massless fields decouple, or for $\sigma_\alpha=0$,
when the current vanishes. The latter case includes Z strings and also the 
currentless limit of the chiral solutions with 
$\sigma_3=\pm\sigma_0$.  However, the chiral solutions are not 
generic and exist only for special values of $\thetaw$.
% determined by 
%$\beta,n,\nu$ and $q$. 

 The W boson condensate producing the vortex current can be  
visualized as a superposition of two 
oppositely charged fluids made of $W^{+}$ and $W^{-}$, respectively, 
flowing in the opposite directions. 
In the vortex restframe the densities of both fluids are the same,
which is why the total momentum through the vortex cross section vanishes
while the current does not. Passing to a different Lorentz frame the 
fluid densities will no longer be equal, which will produce a net momentum. 
The vortex angular momentum can be explained in a similar way if one assumes
that the fluids perform the spiral motions in the opposite directions.

Stability of the new solutions is a very important issue. 
At first one may think that they should be unstable, 
since in the zero current limit they reduce to unstable Z strings.  
However, it seems that current can stabilize 
them. So far this has been checked 
only in the $\thetaw=\pi/2$ limit, where the complete
stability analysis has been carried out \cite{stab}. 
It turns out then 
that all negative modes of current-carrying vortices can be 
removed by imposing the periodic boundary conditions, 
while one cannot do the same in the zero current limit. 
It seems that the same conclusion applies also for $\thetaw\neq \pi/2$, 
although the detailed verification of this is still
in progress. However, the main argument is simple enough and  
goes as follows.  

As was mentioned above, the Z string perturbation eigenmodes  are proportional
to $e^{\pm i(\sigma_0 t+\sigma_3 z)}$ (see Eq.\eqref{003aZ1}) 
with  $\sigma_0=\sqrt{\sigma_3^2-\sigma^2}$ where $\sigma^2>0$ is the eigenvalue
of the spectral problem \eqref{eqs1}.  It follows that all modes with 
$\sigma_3<\sigma$ are unstable, since $\sigma_0$ is imaginary. 
Since 
\be
\lambda=\frac{2\pi}{\sigma_3}>\lambda_{\rm min}=\frac{2\pi}{\sigma},
\ee 
it follows 
 that all unstable modes are longer than $\lambda_{\rm min}$.
This suggests that one could eliminate them by imposing $z$-periodic
boundary conditions with a period $L\leq\lambda_{\rm min}$, which would leave  
`no room' for such modes to exist. However, this would not remove
one particular {\it homogeneous} mode with $\sigma_3=0$,
since it is independent of $z$ and so can be considered as periodic with any
period. 

Now, when passing to current-carrying solutions the situation will be essentially
the same.  It is clear on continuity grounds that,  at least for small currents, 
the solutions will still have negative modes with the wavelength 
bounded from below by a non-zero value $\lambda_{\rm min}$. However, 
the crucial point is that 
the homogeneous mode will disappear from the spectrum -- simply because 
the background solutions now have a non-trivial $z$-dependence, implying that 
all perturbation modes will depend on $z$ as well.  This was checked
in \cite{stab} for $\thetaw=\pi/2$ and it is very plausible that 
for generic $\thetaw$ the situation will be the same.
%since it is clear that adding more 
%$z$-dependent fields to the system will only make less likely the existence of 
%$z$-independent modes. 

Since the homogeneous mode is absent, imposing periodic
boundary  conditions with the period $L=2\pi/\sigma$ will remove 
all negative modes.  
This is somewhat similar to the hydrodynamical 
Plateau-Rayleigh instability of a water column \cite{Eggers} or to  
the gravitational Gregory-Laflamme instability of black strings
in the  theory of gravity in higher dimensions \cite{Gregory}, which 
manifest themselves only starting from a certain minimal length.  
Since $\sigma$ decreases with current,  
$L$ then {\it grows}.   
For large currents one has $\sigma\sim{\cal I}^{-3}$ (see the caption to Fig.\ref{fig3a})
so that the length of the stable vortex segment scales as ${\cal I}^3$ while its 
thickness grows as ${\cal I}$ (see Eq.\eqref{rho-star}). 

\begin{figure}[h]
\hbox to\linewidth{\hss%
% \resizebox{10cm}{5cm}{\includegraphics{segment.eps}}%
 \resizebox{5cm}{4cm}{\includegraphics{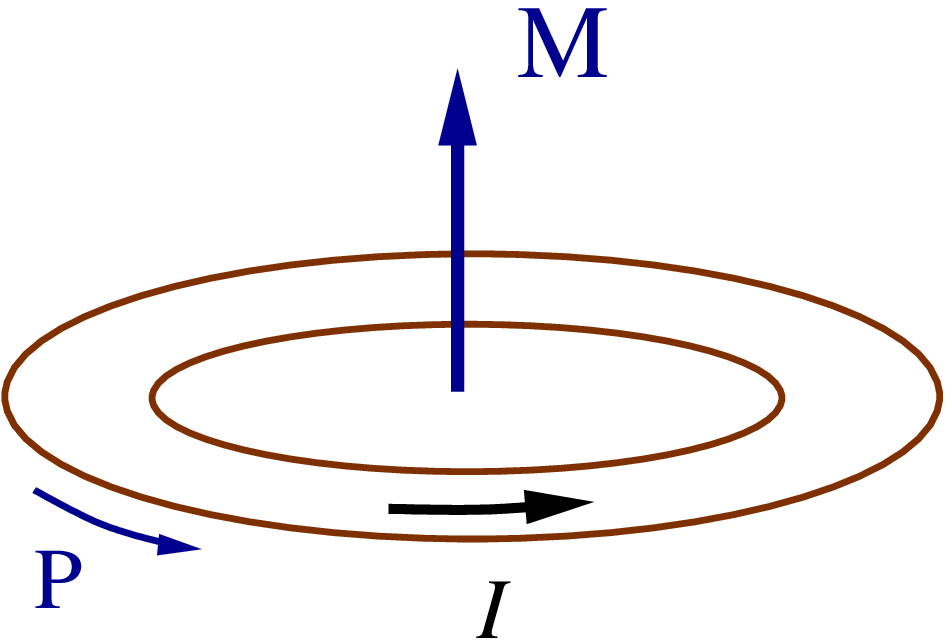}}%
\hspace{25mm}
%  \psfrag{y}{$\frac12(u-V_3)$} 
  \resizebox{7cm}{5cm}{\includegraphics{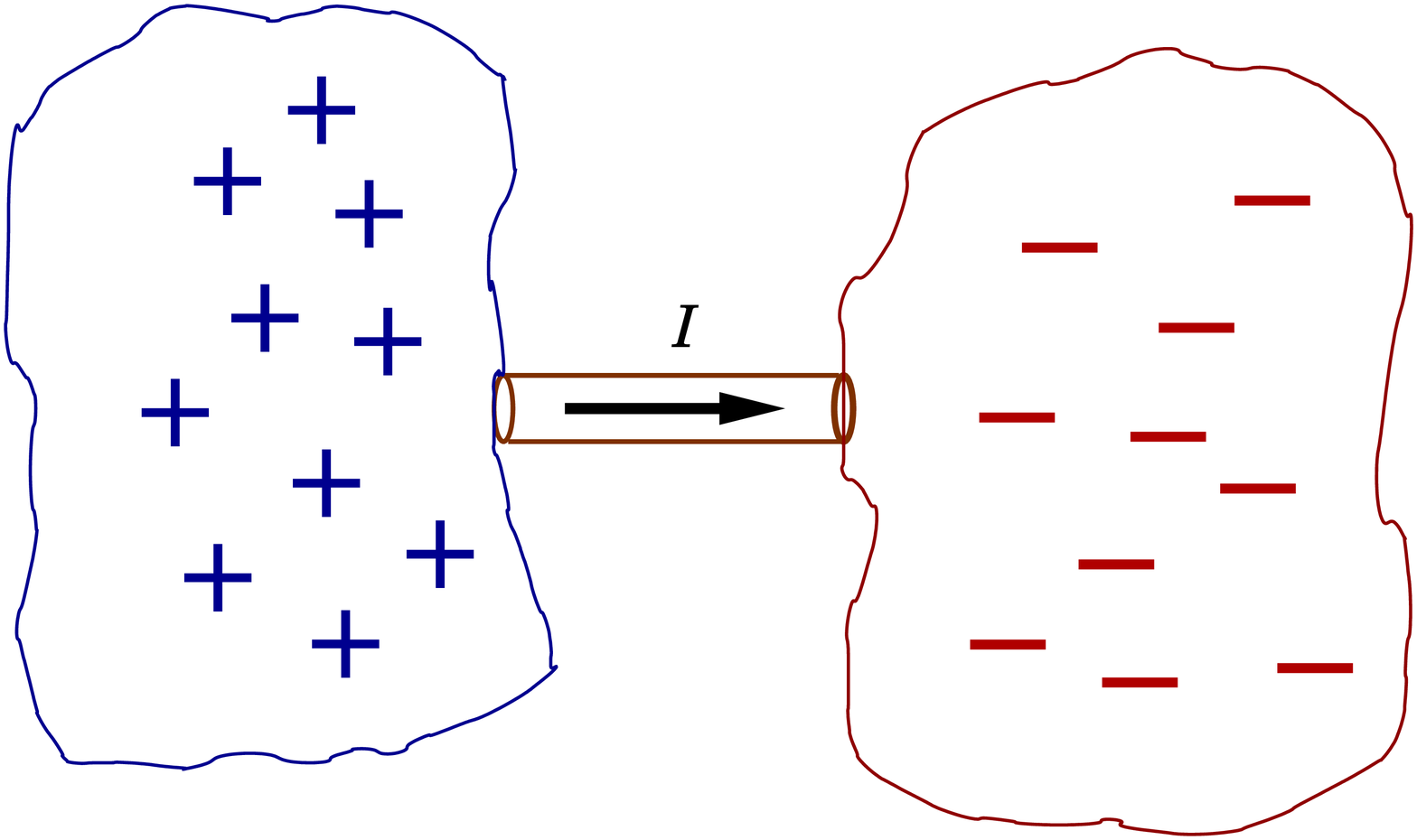}}%
\hss}
\caption{\small Vortex loops balanced by the centrifugal force (left) and 
vortex segments in a polarized medium (right).  
}
\label{FIG-loop}
\end{figure}

There could be different ways of imposing periodic boundary conditions on the 
vortex segment. One possibility is to bend it and identify the extremities to make a loop. 
If the vortex carries a momentum $P$ then  the loop will have an angular momentum $M$
which may balance it against contraction (see Fig.\ref{FIG-loop}).  
One may therefore conjecture that such electroweak analogues  of 
the `cosmic vortons' \cite{Davis} could exist and could perhaps be stable --
if they are made of stable vortex segments.    
Of course, verification of this conjecture requires serious efforts,
since so far vortons have been constructed explicitly  only 
in the global limit of the Witten model \cite{vortons}. 
However, even a remote possibility to have stable solitons  in the standard model
could be very important, since if 
electroweak vortons exist, they could be a dark matter candidate. 

%%%%%%%%%%%%%%%%%%%%%%%%%%%%%%%%%%%%%%%%%%%%%%%%%%%%%%%%%%%%%%%%%%%%%%%%%%
%%%%%%%%%%%%%%%%%%%%%%%%%%%%%%%%%%%%%%%%%%%%%%%%%%%%%%%%%%%%%%%%%%%%%%%%%%

Another possibility to impose periodic boundary conditions is to attach
the vortex ends to something. 
It is known that Z strings, since they are not 
topological, do not have to be always  infinitely long but can exist  
in the form of finite segments whose extremities look like a 
monopole-antimonopole pair \cite{Nambu},\cite{dumb}. 
This suggests that their current-carrying
generalizations could perhaps also exist in the form of finite segments joining 
oppositely polarized regions of space, 
similar to thunderbolts between clouds.
If some processes during the 
electroweak phase transition polarize the medium,  
then the `vortex thunderbolts' could provide a very efficient discharge mechanism -- 
in view of the very large currents  they can carry. They could be stable in this case
and exist as long as the `clouds' they join are not completely discharged.

There is however an important  difference with the ordinary  
atmospheric thunderbolts, which exist only in a polarized medium and emerge 
when the large electric field between the clouds 
creates the plasma channel and works against the resistance 
to drive the charges through. 
In superconducting vortices on the other hand the current flows without any resistance 
and no electric field is needed. In fact, 
inside the (restframe) vortex the electric field is zero. 
The vortices do not therefore need any polarization of the surrounding
medium to exist. 

%%%%%%%%%%%%%%%%%%%%%%%%%%%%%%%%%%%%%%%%%%%%%%%%%%%%%%%%%%%%%%%%%%%%%%%%%%
\begin{figure}[h]
\hbox to\linewidth{\hss%
% \resizebox{10cm}{5cm}{\includegraphics{segment.eps}}%
 \resizebox{6cm}{5cm}{\includegraphics{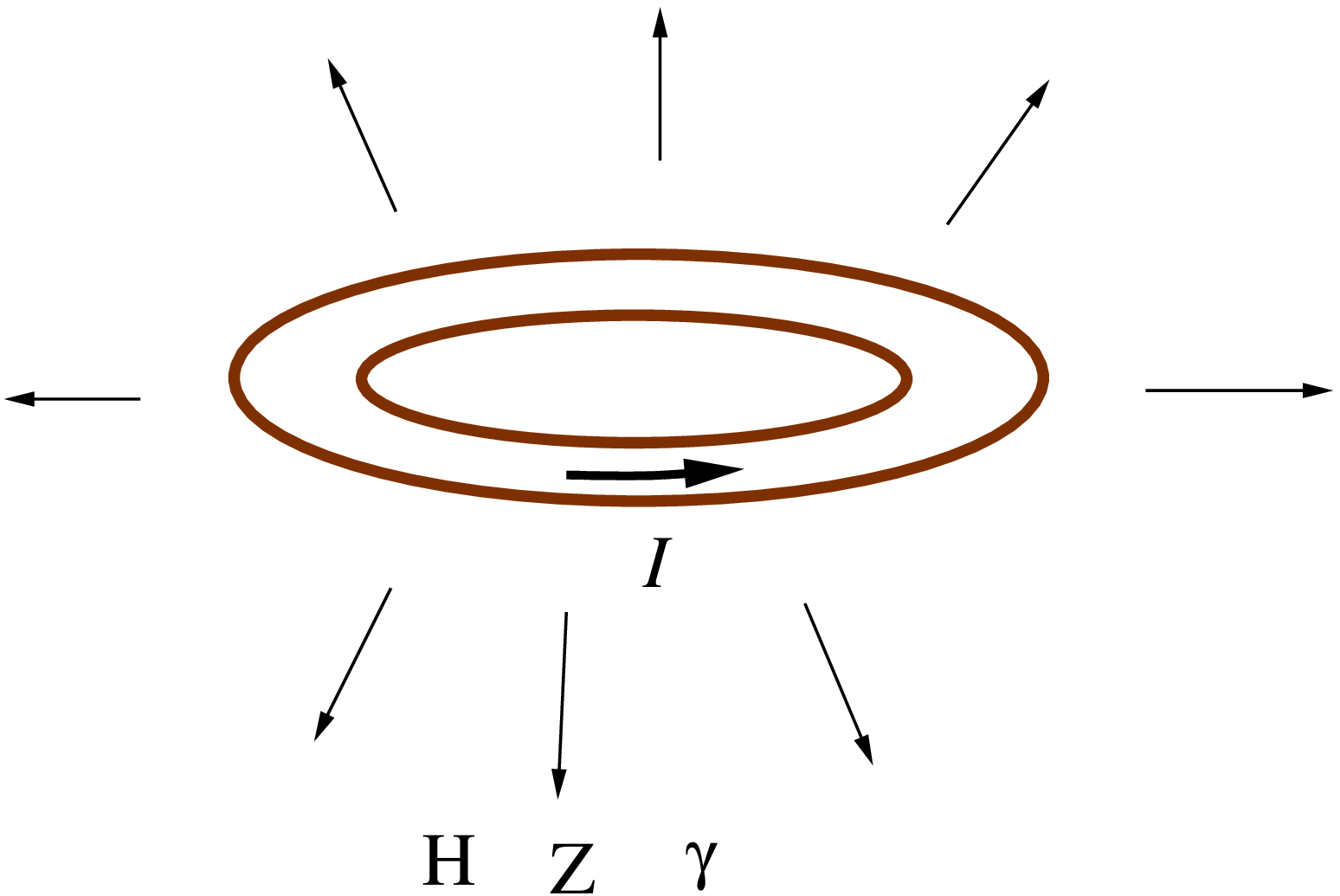}}%
\hspace{15mm}
%  \psfrag{y}{$\frac12(u-V_3)$} 
  \resizebox{8cm}{4.5cm}{\includegraphics{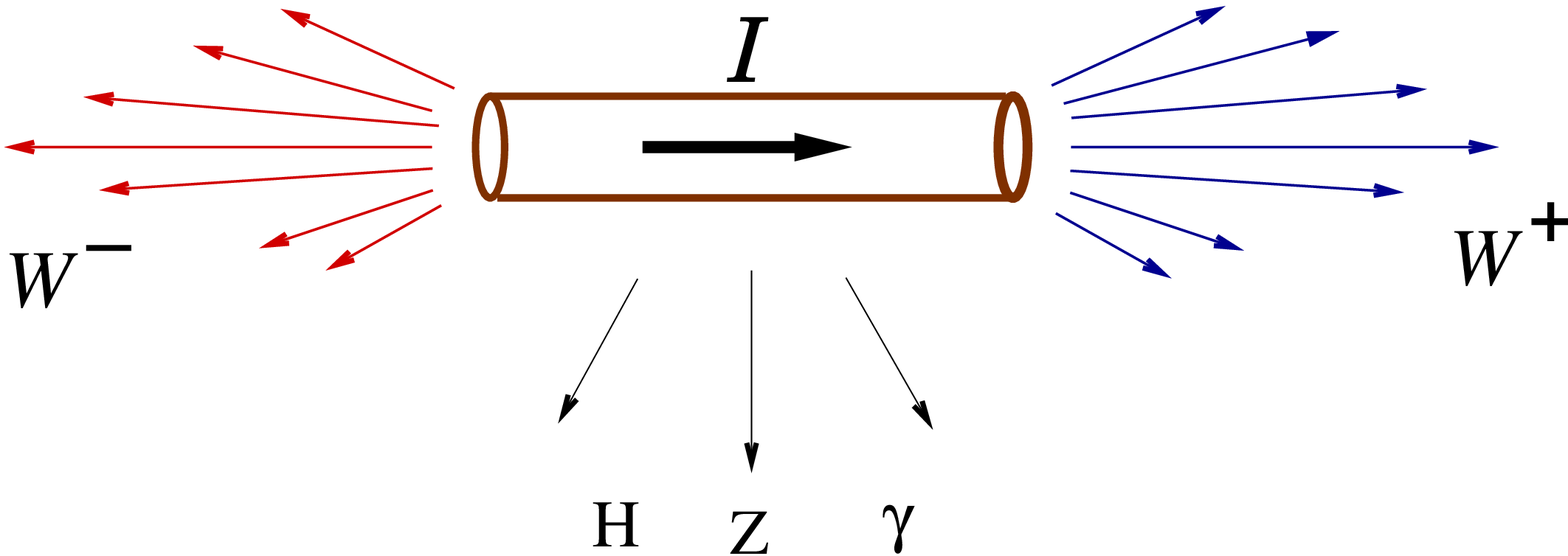}}%
\hss}
\caption{\small Disintegration of an unstable vortex loop (left)
and of an isolated vortex segment (right).
}
\label{FIG-seg}
\end{figure}
%%%%%%%%%%%%%%%%%%%%%%%%%%%%%%%%%%%%%%%%%%%%%%%%%%%%%%%%%%%%%%%%%%%%%%%%%%

Finite segments of superconducting vortices 
can probably be created at high temperatures or in high energy 
particle collisions. Once created, the current will start to leave the segment
through its extremities, so that the latter will  emit positive and negative charges 
(see Fig.\ref{FIG-seg}). Since the current is carried by  
W bosons, it follows that one vortex end will emit the $W^{+}$ while the other  
the $W^{-}$, until the segment radiates away all its energy.  
In addition, if the initial current is large enough, then the vortex will 
contain a zero Higgs field region, whose shrinking will be accompanied  
by a Higgs boson emission. The vortex segment will therefore end up 
in a blast of radiation, creating two jets of $W^{+}$ and $W^{-}$ 
and a shower of Higgs bosons. 
One can similarly argue that  
if the vortex segment is created in the form of an 
unstable loop (for example without angular momentum)
then it will  shrink emitting a shower of neutral bosons -- Higgs, Z
and photons.

The creation of superconducting vortex segments or loops with their subsequent 
disintegration can presumably be observed in the LHC experiments. 
It is also possible that processes of this type could be accompanied by a 
fermion number non-conservation, since already in the zero current limit the 
vortices (Z strings) admit the sphaleron interpretation \cite{KO}.   
However, a special analysis is needed to work out the details  of these 
processes and their experimental signatures. 
We leave this and other possible applications of the superconducting vortices  
for a separate study. 

\acknowledgments 
We thank Andrei Mironov for discussing possible applications of our solutions. 

%\appendix{A}

%\input{origin.tex}

\section*{Appendix A. Solutions near the symmetry axis}

\renewcommand{\theequation}{A.\arabic{equation}}
\setcounter{equation}{0}

The boundary conditions at the origin expressed by 
Eq.\eqref{ax} imply that at small $\rho$ one has 
\bea                 \label{dev0}
u&=&u(0)+\delta u,~~~
u_3=1+\delta u_3,~~~
u_1=\delta u_1,~~  \notag \\
v&=&2n-\nu+\delta v,~~~
v_3=\nu+\delta v_3,~~~
v_1=\delta v_1,~~~ \notag \\
f_1&=&\delta f_1,~~~
f_2=f_2(0)+\delta f_2,
\eea
where 
the deviations $\delta u,\ldots ,\delta f_2$ vanish for $\rho\to 0$
and 
$u(0)$ is a free parameter, while $f_2(0)$ is also a free parameter
if $\nu=n$, whereas $f_2(0)=0$ if $\nu\neq n$.  Inserting this to  
Eqs.(\ref{ee1})--(\ref{CONS})
and linearising with respect to the deviations the result is as follows.

Let us first consider the case where $\nu\neq n$ and $f_2(0)=0$. 
Then the equations read   
\bea
(\delta u)^{\prime\prime}+\frac{1}{\rho}\,(\delta u)^\prime&=&0,  \label{o1} \\  
(\delta u_3)^{\prime\prime}+\frac{1}{\rho}\,(\delta u_3)^\prime&=&0,        \label{o1a} \\
(\delta v)^{\prime\prime}-\frac{1}{\rho}\,(\delta v)^\prime&=&0,         \label{o2} \\ 
(\delta v_3)^{\prime\prime}-\frac{1}{\rho}\,(\delta v_3)^\prime&=&0,     \label{o2a}   \\
(\delta f_1)^{\prime\prime}
+\frac{1}{\rho}\,(\delta f_1)^\prime-\frac{n^2}{\rho^2}\,\delta f_1
&=&\frac14(\sigma^2(u(0)+1)^2-\beta)\delta f_1,                          \label{o3} \\              
(\delta f_2)^{\prime\prime}
+\frac{1}{\rho}\,(\delta f_2)^\prime-\frac{(n-\nu)^2}{\rho^2}\,\delta f_2
&=&\frac14(\sigma^2(u(0)-1)^2-\beta)\delta f_2,                         \label{o3a} 
\eea
plus a system of three coupled equations
\bea
(\delta u_1)^{\prime\prime}+\frac{1}{\rho}\,(\delta u_1)^\prime&=&
\frac{\nu}{\rho^2}\,(\nu\delta u_1-\delta v_1),       \label{o5}\\
(\delta v_1)^{\prime\prime}-\frac{1}{\rho}\,(\delta v_1)^\prime&
=&\sigma^2(\delta v_1-\nu\delta u_1),      \label{o6}\\
\sigma^2\,\delta u_1^\prime+\frac{\nu}{\rho^2}\,\delta v_1^\prime&=&0.   \label{o6a}
\eea                                  
  From Eqs.(\ref{o1})--(\ref{o2a}) one finds 
\bea                                           \label{o7}
\delta u&=&A_1+A_2\ln\rho,~~~~~\delta u_3=A_3+A_4\ln\rho,   \nonumber  \\
\delta v&=&A_5+A_6\,\rho^2,~~~~~\delta v_3=A_6+A_8\,\rho^2,
\eea
while solution of Eqs.\eqref{o3},\eqref{o3a} 
can be expressed in terms of the Bessel functions, such  that   
for small $\rho$ one has  
\bea                                 \label{o8}
\delta f_1&=&A_9\,\rho^n(1+\ldots)+ A_{10}\,\rho^{-n}(1+\ldots), \nonumber\\
\delta f_2&=&A_{11}\rho^{n-\nu} (1+\ldots)+ A_{12}\rho^{\nu-n}(1+\ldots).  
\eea
The three  coupled equations (\ref{o5}),(\ref{o6}),\eqref{o6a}  
can be reduced to one third order equation whose solution
can be obtained in terms of the hypergeometric functions. 
One then obtains for small $\rho$
\bea                             \label{o9}     
\delta u_1&=&
A_{13}\,\rho^\nu(1+\ldots)+
A_{14}\,\rho^{-\nu}(1+\ldots)+A_{15},~~~~\notag \\
\delta v_1&=&
-\frac{\sigma^2}{\nu+2}\,A_{13}\,\rho^{\nu+2}(1+\ldots)
+\frac{\sigma^2}{2-\nu}\,A_{14}\,\rho^{-\nu}(1+\ldots)+\nu A_{15}. 
\eea
Eqs.\eqref{o7}--\eqref{o9} give the most general solution of the equations,
since it contains the maximal number of the integration constants:
$A_1,\ldots,A_{15}$. Now, we have to suppress all the solutions that 
do not vanish for $\rho\to 0$, so that we set 
\be
A_1=A_2=A_3=A_4=A_5=A_7=A_{10}=A_{12}=A_{14}=A_{15}=0,
\ee
assuming that $n>\nu>0$, 
whereas  
$A_6$, $A_8$, $A_{9}$, $A_{11}$, $A_{13}$ 
remain free. 
This
gives the local solutions in the linear approximation
\bea                         \label{nearaxis}
u&=&a_1,                   \nonumber \\
u_1&=&a_2\,\rho^{\nu}+\ldots  ,\nonumber \\
u_3&=&1 ,                      \nonumber \\
v&=&2n-\nu+a_3\,\rho^2,                                      \nonumber \\
v_1&=&-\,\frac{\sigma^2 a_2}{\nu+2}\,\rho^{\nu+2}+\ldots,                \nonumber \\
v_3&=&\nu+a_4\,\rho^2,                                     \nonumber \\
\f&=&a_5\,\rho^{n}+\ldots ,                      \nonumber \\ 
\p&=&q\,\rho^{n-\nu}+\ldots                    
\eea
containing 6 free parameters: $a_1=u(0)$, $a_2=A_{13}$,
$a_3=A_6$, $a_4=A_8$, $a_5=A_9$, $q=A_{11}$. 

If $\nu=n$, $f_2(0)\neq 0$ then the linearized equations for 
$\delta u,\ldots \delta f_2$
are more complicated. They split into three coupled groups, whose solutions 
can be constructed in series.  
This gives essentially 
the same result as in \eqref{nearaxis}, 
\bea                         \label{nearaxis10}
u&=&a_1,                   \nonumber \\
u_1&=&a_2\,\rho^{n}+\ldots  \nonumber \\
u_3&=&1 ,                      \nonumber \\
v&=&n+a_3\,\rho^2+\ldots,                                      \nonumber \\
v_1&=&\frac{g^2q\,a_5-\sigma^2 a_2}{n+2}\,\rho^{n+2}+\ldots,                \nonumber \\
v_3&=&n+a_4\,\rho^2+\ldots,                                     \nonumber \\
\f&=&a_{5}\,\rho^{n}+\ldots ,                      \nonumber \\ 
\p&=&q,                     
\eea
which also contains 6 free parameters: 
$a_1,\ldots a_5$ and $q$. Eqs.\eqref{nearaxis} and \eqref{nearaxis10}
together imply Eq.\eqref{orig} in the main text.  

These local solutions are obtained in the linearized approximation. 
However, they can be used to take into account all the non-linear terms
in the equations when starting the numerical integration at $\rho=0$. 
This is achieved by rewriting the second order field equations Eqs.(\ref{ee1})--(\ref{ee8})
in the first order form,
\be                            \label{dynam}
y_r^\prime(\rho)={\cal F}_r(\rho, y_s),~~~~~r,s=1,\ldots 16
\ee
where 
\bea
y_{1}&=&u,~~~~~~~~~~~~y_{2}=\rho u^\prime\,,~~~~~~~~~~~~~~
y_{3}={u_1}/{\rho^{\nu}},~~~~~~~~~
~~~~~~y_{4}={u_1^\prime}/{\rho^{\nu-1}},\nonumber \\
y_{5}&=&u_3,~~~~~~~~~~~y_{6}=\rho u_3^\prime,~~~~~~~~~~~~~~
y_7={v},~~~~~~~~~~~~~~~~~~~~y_8={v^\prime}/{\rho}\,,\nonumber \\
y_9&=&v_1,~~~~~~~~~~~y_{10}=v_1^\prime\,,~~~~~~~~~~~~~~~
y_{11}={v_3},~~~~~~~~~~~~~~~~~y_{12}={v_3^\prime}/{\rho}\,,~~~~ \\
y_{13}&=&{\f}/{\rho^{n}},~~~~~~~y_{14}={\f^\prime}/{\rho^{n-1}}\,,~~~~~~~
y_{15}=\p/\rho^{|n-\nu|},~~~~~~~~~y_{16}=\p^\prime/\rho^{|n-\nu|-1} \notag 
\eea
with  the boundary conditions  
\bea                           \label{origin123}
y_1(0)&=&a_1,~~~~~~~~y_2(0)=0,~~~~~~~~~~~
y_3(0)=a_2,~~~~~~~~~y_4(0)=\nu a_2,\nonumber \\
y_5(0)&=&1,~~~~~~~~~
y_6(0)=0,~~~~~~~~~~~y_7(0)=2n-\nu,~~~~y_8(0)=2a_3,~~~~\nonumber \\
y_9(0)&=&0,~~~~~~~~~y_{10}(0)=0,~~~~~~~~~~
y_{11}(0)=\nu,~~~~~~~~~~y_{12}(0)=2a_4~~~~ \\
y_{13}(0)&=&a_5,~~~~~~~y_{14}(0)=na_5,~~~~~~~ \notag 
y_{15}(0)=q,~~~~~~~~~~y_{16}(0)=|n-\nu|q. \label{b0}
\eea
The right hand sides ${\cal F}_r(\rho, y_s)$
in Eqs.\eqref{dynam} are defined only for $\rho>0$. 
For example, one has 
$y_1^\prime={\cal F}_1=y_2/\rho$ which is undefined at $\rho=0$. 
However, using the field equations one can check that $y_r^\prime(0)=0$ 
for all $r$. One can therefore set {\it by definition} 
${\cal F}_r(\rho=0, y_s)=0
$, 
and then one can integrate Eqs.\eqref{dynam} 
with the boundary conditions \eqref{origin123} starting {\it exactly} at 
$\rho=0$, thus avoiding any  approximations.  
Such a procedure is sometimes called desingularization
of the equations at a singular point.

\section*{Appendix B. Solutions in the asymptotic region}

\renewcommand{\theequation}{B.\arabic{equation}}
\setcounter{equation}{0}

Solutions of Eqs.\eqref{ee1}--\eqref{ee8},\eqref{CONS1}
have to approach the Biot-Savart configuration \eqref{030} far away
from the vortex core. Therefore one has
\bea                 \label{dev}
u&=&\underline{u}+\delta u,~~~
u_3=-\underline{u}+\delta u_3,~~~
u_1=\delta u_1,~~  \notag \\
v&=&c_2+\delta v,~~~
v_3=-c_2+\delta v_3,~~~
v_1=\delta v_1,~~~ \notag \\
f_1&=&1+\delta f_1,~~~
f_2=\delta f_2,
\eea
where 
\be
\underline{u}=Q\ln\rho+c_1\,,
\ee
the deviations $\delta u,\ldots ,\delta f_2$ tend to zero as $r\to\infty$
and $c_1,c_2,Q$ are the same integration constants as in Eq.\eqref{030}. 
Inserting \eqref{dev} to Eqs.\eqref{ee1}--\eqref{CONS}
and linearising with respect to the deviations, the resulting 
linear system reduces to five independent  equations plus a subsystem of three
coupled equations. The first five equations are the following. 
The linear combinations 
\be
\delta u_Z=\delta u+\delta u_3,~~~~~~~~~~
\delta v_Z=\delta v+\delta v_3
\ee
fulfill two independent equations 
\bea
(\delta u_Z)^{\prime\prime}+\frac{1}{\rho}\,(\delta u_Z)^\prime &=&\frac12\,\delta u_Z, \notag \\
(\delta v_Z)^{\prime\prime}-\frac{1}{\rho}\,(\delta v_Z)^\prime &=&\frac12\,\delta v_Z. 
\eea
Their solutions vanishing for $r\to\infty$ are 
\be
\delta u_Z=\frac{c_3}{\sqrt{\rho}}\,e^{-\mz\rho}+\ldots,~~~~
\delta v_Z={c_4}{\sqrt{\rho}}\,e^{-\mz\rho}+\ldots,~~~
\ee
where $c_3,c_4$ are integration constants and the dots denote subleading terms.
This corresponds to the temporal and azimuthal components of the Z field with the   
mass  $\mz=1/\sqrt{2}$.  
The linear combinations 
\be
\delta u_A=g^2 \delta u-g^{\prime 2}\delta u_3,~~~~~~~~~~
\delta v_A=g^2 \delta v-g^{\prime 2}\delta v_3
\ee
correspond to the electromagnetic field and satisfy 
\bea
(\delta u_A)^{\prime\prime}+\frac{1}{\rho}\,(\delta u_A)^\prime &=&0, \notag \\
(\delta v_A)^{\prime\prime}-\frac{1}{\rho}\,(\delta v_A)^\prime &=&0. 
\eea
The only solution of these equations that vanishes at infinity is the trivial one, 
\be
\delta u_A=\delta v_A=0,
\ee
since the electromagnetic degrees of freedom are already incorporated into the 
background configuration \eqref{030}. One has also 
\be
(\delta f_1)^{\prime\prime}+\frac{1}{\rho}\,(\delta f_1)^\prime =\frac{\beta}{2}\,\delta f_1, 
\ee
whose solution for large $\rho$ is 
\be
\delta f_1=\frac{c_5}{\sqrt{\rho}}\,e^{-\mh\rho}+\ldots,~~~~
\ee
which corresponds to the Higgs boson with the mass $\mh=\sqrt{\beta/2}$.  

Let us now consider the equations for $\delta u_1,\delta v_1,\delta f_2$
obtained by linearising 
Eqs.\eqref{ee4},\eqref{ee5},\eqref{ee7}. The analysis turns out to be 
more involved in this case. 
It is
convenient to replace one of these three equations by the first integral \eqref{C},
which gives a completely equivalent system.  Linearising and defining 
\be                               \label{ms}
\chi=\frac{g^2}{2}+\sigma^2\underline{u}^2\equiv m_\sigma^2
\ee
also
%\be
%-\mu\,X=U\delta f_2+\frac12\,\delta u_1,~~~~
%\mu\,Y=g^2\delta f_2-\sigma^2 U \delta u_1,
%\ee
\be                                 \label{q0}
\delta f_2=\sigma^2\underline{u}\,X+\frac12\,Y,~~~~\delta u_1=g^2X-\underline{u}\,Y
\ee
the resulting equations read
\begin{subequations}                \label{qq}
\begin{align}
X^{\prime\prime}+(\frac{1}{\rho}+\frac{\chi^\prime}{\chi})\,X^\prime 
-(\frac{c_2^2}{\rho^2}+\chi)\,X
&=\frac{Q}{\rho \chi}\,Y^\prime, 			     \label{q1}      \\
(\delta v_1)^{\prime\prime}-\frac{1}{\rho}\,(\delta v_1)^\prime -\chi\,\delta v_1
&=c_2\chi Y, 			 				\label{q2}	   \\
\frac{c_2}{\rho^2}\,(\delta v_1)^\prime-\chi Y^\prime-\frac{2g^2\sigma^2Q}{\rho}\,X
&=\frac{C}{\rho},							\label{q3}
\end{align}
\end{subequations}
where $C$ is the same integration constant 
as in Eq.\eqref{C}. 
These three equations  are equivalent to one fourth 
order equation 
\be								\label{fourth}
Z^{\prime\prime\prime\prime}-2(\mu^2 Z^\prime)^\prime+
\left(\mu^4-3(\mu^2)^{\prime\prime}-\frac{6}{\rho}\,(\mu^2)^\prime
+\frac{4(c_2^2-1)}{\rho^4}\right)Z=C\,\frac{c_2(5+4\rho^2\mu^2)}{4\rho^{5/2}}\,
\ee
provided that  
\be                                         \label{q4}
\delta v_1=\int \frac{d\rho}{\sqrt{\rho}}\,Z,~~~~~~~~~
\mu^2=\chi+\frac{4c_2^2-5}{4\rho^2}.
\ee
Let us first set $C=0$. Then the asymptotic form of the two independent solutions 
that vanish at infinity is given by 
\be                                   \label{Zinf}
Z=\left(A_1 \mu^{-5/2}(1+\ldots)+A_2\,\rho\sqrt{\mu}\,(1+\frac{1}{2\rho\mu}
+\ldots)\right)\exp(-\int\mu\, d\rho).
\ee
The original field amplitudes can  be reconstructed with 
Eqs.\eqref{q2},\eqref{q3},\eqref{q4},\eqref{q0},
\be 						\label{sol5}
\delta f_2=\frac{c_6}{\sqrt{\rho}}\,e^{-\int\mu\, d\rho}+\ldots,~~~
\delta u_1=\frac{c_7}{\sqrt{\rho}}\,e^{-\int\mu\, d\rho}+\ldots,~~~
\delta v_1={c_8}{\sqrt{\rho}}\,e^{-\int\mu\, d\rho}+\ldots,~~~
\ee
where $c_6,c_7,c_8$ are expressed in terms of the two 
independent integration constants $A_1,A_2$, for example 
$
c_8=-{A_2}/{\sqrt{\mu}}.
$

Let us now consider the remaining four solutions of Eqs.\eqref{qq} -- there should be  
altogether  {\it six} independent solutions, since the equations contain five 
derivatives and one integration constant, $C$. Two additional solutions are obtained 
by choosing the plus sign in the exponent Eq.\eqref{Zinf}, but they are unbounded and 
should be excluded from the analysis. Yet another solution,
always for $C=0$, is given by $X=0$, $\delta v_1=-c_2 Y=const.$, which 
does not satisfy the asymptotic conditions and should be excluded as well. 

The last solution is obtained by setting $C\neq 0$, which gives rise to a solution 
that behaves as $Z=Cc_2/(\mu^2\sqrt{\rho})$ for large $\rho$, so that 
$\delta v_1\sim  C\ln\ln\rho$. 
Let us call this solution $C$-mode.
One can think at first that it should also be excluded.
However, the constraint equation \eqref{CONS} requires that $C=0$,
and so, even if the $C$-mode is included, it will be 
eventually removed by the constraint. The constraint is explicitly imposed 
on  local solutions { at small $\rho$},  which 
is sufficient to insure that it holds everywhere, so that it would be redundant 
to impose it  again. One can therefore add the $C$-mode to \eqref{sol5}, 
since at the end, when the matching is performed to construct 
the global solutions, it will be removed anyway.

Since this mode will finally disappear, 
one may wonder why one should consider it at all.  
However, including  it is essential, since it
increases the dimension of space of local solutions. 
From the technical viewpoint this mode 
is not very convenient, though, 
because it is not localized. 
However, since its main effect is to increase the dimension of function space,
any other mode with a non-zero overlap with it can actually be employed. 
Therefore, we simply use \eqref{sol5} with $c_6,c_7,c_8$ 
considered as {\it three independent parameters}, which 
leads to a good numerical convergence. In fact, these coefficients 
contain $\mu\sim\ln\rho$,  but since this is a slowly varying function,  
one can treat 
$c_6,c_7,c_8$ as if they were constants. 
In addition, without changing 
the leading asymptotic terms, one can replace 
$\int\mu\, d\rho$ in Eq.\eqref{sol5}  by 
$\int m_\sigma\, d\rho$. 

Summarizing, the large $\rho$ behaviour of the solutions is given by 
\bea                 \label{dev1}
u&=&Q\ln\rho+c_1+\frac{c_3 g^{\prime 2}}{\sqrt{\rho}}\,e^{-\mz\rho}+\ldots,~~~
u_3=-Q\ln\rho-c_1+\frac{c_3 g^{2}}{\sqrt{\rho}}\,e^{-\mz\rho}+\ldots,~~~\notag \\
v&=&c_2+{c_4 g^{\prime 2}}{\sqrt{\rho}}\,e^{-\mz\rho}+\ldots,~~~~~~~~~~~
v_3=-c_2+{c_4 g^{2}}{\sqrt{\rho}}\,e^{-\mz\rho}+\ldots,~~~\notag \\
f_1&=&1+\frac{c_5}{\sqrt{\rho}}\,e^{-\mh\rho}+\ldots,~~~ ~~~~~~~~~~~~
f_2=\frac{c_6}{\sqrt{\rho}}\,e^{-\int m_\sigma\, d\rho}+\ldots,~~~ \notag \\
u_1&=&\frac{c_7}{\sqrt{\rho}}\,e^{-\int m_\sigma\, d\rho}+\ldots,~~~~~~~~~~~~~~~~~~
v_1={c_8}{\sqrt{\rho}}\,e^{-\int m_\sigma\, d\rho}+\ldots,~~~
\eea
where $Q,c_1,\ldots, c_8$ are 9 independent parameters 
and $m_\sigma$ is defined in \eqref{ms}. 
This gives rise to Eq.\eqref{inf} in the main text.

\section*{Appendix C. Superconducting strings in the Witten model}
\renewcommand{\theequation}{C.\arabic{equation}}
\setcounter{equation}{0}

Numerical construction of Witten's 
superconducting strings was performed by many authors \cite{SS} 
(see \cite{Vilenkin-Shellard} for a review). 
In this Appendix we reproduce these solutions within the same approach
and using the same notation as in the main text. We tried to make this 
Appendix self-consistent in order to illustrate our procedure 
on a relatively simple example.

The U(1)$\times$U(1) Witten model is defined by the Lagrangian \cite{Witten}
\be  							\label{W00}
{\mathcal L}_{W}=-\frac14\sum_{a=1,2} F^{(a)}_{\mu\nu}F^{(a)\mu\nu}  
+\sum_{a=1,2}(D_\mu \phi_{a})^\ast D^\mu \phi_{a}  
 -{\rm U},
%{\mathcal L}_{W}=-\frac14\sum_{a=1,2}
%(\partial_\mu A^{(a)}_\nu-\partial_\nu A^{(a)}_\mu )^2
\ee 
with $F^{(a)}_{\mu\nu}=\partial_\mu A^{(a)}_\nu-\partial_\nu A^{(a)}_\mu$
and 
$D_\mu \phi_{a}=(\partial_\mu-ig_{a}A^{(a)}_\mu)\phi_{a}$ 
where $g_{a}$ are two gauge coupling constants. 
The scalar field potential 
\be 
{\rm U}=\frac{\lambda_1}{4}\left(|\phi_1|^2-\eta_1^2\right)^2
	+\frac{\lambda_2}{4}(|\phi_2|^2-\eta_2^2)^2
	+\gamma|\phi_1|^2|\phi_2|^2 -\frac{\lambda_2}{4}\,\eta_2^4
\ee
vanishes for $|\phi_1|=1$ and $\phi_2=0$, and this minimum is global if 
$4\gamma^2>\lambda_1\lambda_2$.
The perturbative mass spectrum of the field excitations around this vacuum
contains two scalar bosons with  masses $m_1^2={\lambda_1}\eta_1^2$ 
and $m_2^2=\gamma\eta_1^2-\frac12\lambda_2\eta_2^2$ 
as well as a vector boson with the mass 
$m_v^2={2}g_1^2\eta^2_1$
and  a massless vector boson. 
Pairs $(A^{(1)}_\mu,\phi_{1})$ and $(A^{(2)}_\mu,\phi_{2})$ 
are sometimes called vortex fields
and condensate  fields, respectively.

Passing to cylindrical coordinates, where $x+iy=\rho e^{i\varphi}$, and 
making  the stationary, axially symmetric ansatz 
\begin{subequations}             \label{w-field}
\begin{align} 
A^{(1)}_\mu dx^\mu&=\frac{1}{g_1}\,\left(n-v(\rho)\right)d\varphi ,~~~~~~~~~~~~~~~~~~~
\phi_1=f_1(\rho)e^{in\varphi},              \\
{A}^{(2)}_\mu dx^\mu&=\frac{1}{g_2}\,(\sigma_0 dt+\sigma_3 dz)\,(1-u(\rho)),~~~~
{\phi}_2={f}_2(\rho)e^{i\sigma_0 t+i\sigma_3 z},
\end{align}
\end{subequations}
with $n\in\mathbb{Z}$, 
the field equations 
\begin{align}
\partial^\mu F^{(a)}_{\mu\nu}&=2g_a\Re(i\phi^\ast_a D_\nu\phi_a), \\
D_\mu D^\mu\phi_a&=-\frac{\partial U}{\partial\phi_a^\ast},~
\end{align}
reduce to  
\begin{subequations}           \label{w-eq}
\begin{align} 
\frac{1}{\rho}\left(\rho f_1^\prime\right)^\prime &= 
\left(\frac{v^2}{\rho^2}+\frac{\lambda_1}{2}(f_1^2-\eta_1^2)+\gamma f_2^2 \right)f_1 , \\
\frac{1}{\rho}\left(\rho f_2^\prime\right)^\prime &= \left(\sigma^2u^2
+\frac{\lambda_2}{2}(f_2^2-\eta_2^2)+\gamma f_1^2 \right)f_2 , \\
\rho\left(\frac{v^\prime}{\rho}\right)^\prime &= 2g_{1}^2 f_1^2v, \\
\frac{1}{\rho}\left(\rho u^\prime\right)^\prime &= 2g_{2}^2 f_2^2 u,
\end{align}
\end{subequations}  
where $\sigma^2=\sigma_3^2-\sigma_0^2$. Depending on sign of $\sigma^2$,
solutions of these equations are called magnetic ($\sigma^2>0$),
electric ($\sigma^2<0$), or chiral ($\sigma^2=0$) \cite{Carter}. 

Equations \eqref{w-eq} have singular points for $\rho=0,\infty$. 
Constructing their local series solutions in the vicinity of these points
 gives the boundary 
conditions for $0\leftarrow\rho\to\infty$, 
%\begin{subequations}             
\begin{align} 
1+\dots \leftarrow &u\rightarrow  c_1 + Q\ln\rho+\dots, \notag \\
n+a_2\rho^2+\dots \leftarrow &v\rightarrow c_2\sqrt{\rho}e^{-m_v\rho}+\dots ,\notag \\
a_1\rho+\dots \leftarrow &f_1\rightarrow 1+\frac{c_3}{\sqrt{\rho}}e^{-m_1\rho}+\dots ,\notag \\
q+\dots \leftarrow &f_2\rightarrow\frac{c_4}{\sqrt{\rho}}e^{-\int m_\sigma d\rho} +\dots, 
\label{w-bound}
\end{align}
%\end{subequations}  
where $a_1,a_2,q,Q,c_1,\dots c_4$ are integration constants and 
\be                                       \label{ms-wit}
m^2_\sigma= m_2^2+\sigma^2(c_1+Q\ln\rho)^2.
\ee
These boundary conditions imply that the fields are regular at the symemtry axis,
while at infinity the scalars approach vacuum, the massive vector vanishes,
whereas the massless vector becomes the Biot-Savart field 
produced by the string current. 
The current density is 
\be                             \label{cur-Witten}
J_\nu=\partial^\mu F^{(2)}_{\mu\nu}=2g_2\Re(i\phi^\ast_2 D_\nu\phi_2),
\ee
integrating which over the $x,y$ plane gives 
\be                                  \label{w-cur}
{I}_\alpha = \int J_\alpha\, d^2x=-\frac{2\pi Q\sigma_\alpha}{g_2},
\ee
with $\alpha=0,3$. Here $I_0$ is the charge per unit vortex length
while $I_3$ is the total current through the vortex cross section. 
As Eq.\eqref{ms-wit} shows, the long-range Biot-Savart field affects 
the mass of the second   scalar. 
For $\sigma^2\geq 0$ 
this only improves the field localization, 
but for $\sigma^2<0$ the $m^2_\sigma$ term becomes negative at large 
distances, making the condensate field oscillate at infinity, which 
renders the energy strongly divergent. Such solutions 
are more difficult to handle, and so we shall concentrate on the magnetic
and chiral cases. 
Since 
the structure of the fields \eqref{w-field} is invariant with 
respect to Lorentz boosts along the z-axis, 
this can be used in the magnetic case 
to pass to the rest frame where   
${I}_\alpha = \delta_\alpha^3{\cal I}$
with 
$
{\cal I}=-{2\pi Q\sigma}/{g_2}
$ 
where the `twist' $\sigma=\sqrt{\sigma^2}$.

To solve equations \eqref{w-eq} the relaxation method is usually employed 
\cite{SS}, \cite{Vilenkin-Shellard}. Within this method  
it is very natural to find first of all the `dressed' currentless solution
with $u=\sigma^2=0$, $f_2\neq 0$. After this one can increase $\sigma^2$
to obtain current-carrying strings with $u\neq 0$. 

However, we use instead  the multiple shooting method 
in which the local solutions in \eqref{w-bound} are numerically 
extended to an intermediate region where the matching conditions
are  imposed. To fulfil the 8 matching conditions for the 4 functions
and their first derivatives  
we have on our disposal 9 free parameters: 8 integration constants in 
\eqref{w-bound} and also $\sigma$. This leaves  after
the matching one free parameter to label the global
solutions in the interval $\rho\in[0,\infty)$. 
We choose this parameter to be the value of the 
condensate scalar  at the origin, $q=f_2(0)$. Within our approach it is 
then natural to start not from the `dressed' solution that is apriori unknown, 
but from the known ANO vortex.  

We therefore set $q=0$ and then one has $f_2=0$, $u=1$ everywhere 
so that the condensate fields
vanish. The remaining 
equations reduce to the ANO system whose solution is the ANO vortex \cite{ANO}. 
For $q\ll 1$ one can expect the solution to be its small deformation that 
can be considered within the linear perturbation
theory. Linearising the equations with respect to the deformations the ANO
background, one can see that it is consistent to set the deformations of $u,v,f_1$
to zero, while the deformation of $f_2$ satisfies 
\be                                         \label{wl}
\frac{1}{\rho}\left(\rho \delta f_2^\prime\right)^\prime = \left(\sigma^2
-\frac{\lambda_2}{2}\,\eta_2^2+\gamma f_1^2 \right)\delta f_2 .
\ee
This can be viewed as the spectral problem 
with the eigenvalue $\sigma^2$.
This problems admits
a bound state with $\sigma^2>0$, which  
leads to the dispersion relation 
\be
\sigma_0=\sqrt{\sigma_3^2-\sigma^2}.
\ee 
This interpretation of this is two-fold. If the wave-number $\sigma_3$ is small,
$\sigma_3^2<\sigma^2$, then the frequency $\sigma_0$ is imaginary,
implying that the 
ANO vortex within Witten's model is unstable 
\cite{Witten}. On the other hand, for 
$\sigma_3^2\geq \sigma^2$ the frequency is real
 and the solution of \eqref{wl}
describes a small 
stationary deformation of the ANO vortex by the condensate.  

\begin{figure}[ht]
\hbox to\linewidth{\hss%
\psfrag{x}{{$\ln(1+\rho)$}}
\psfrag{y}{}
\psfrag{u}{$u$}
\psfrag{f1}{$f_1$}
\psfrag{f2}{$f_2$}
\psfrag{v}{$v$}
\psfrag{sigma}{$\sigma$}
\psfrag{I}{{$\mathcal{I}$}}
\psfrag{omega}{{$\sigma$}}
       \resizebox{8.5cm}{7cm}{\includegraphics{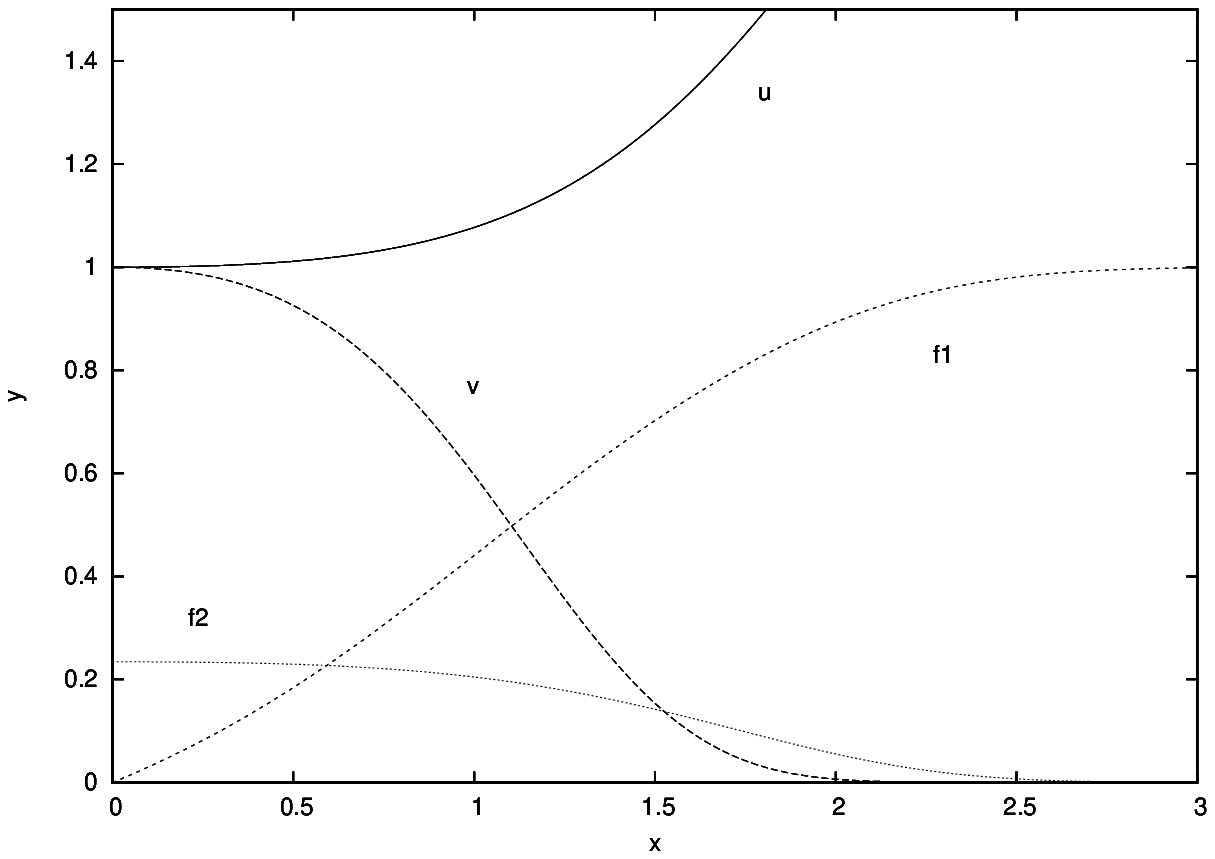}}
\hspace{5mm}%
\psfrag{x}{{$q$}}
        \resizebox{8.5cm}{7cm}{\includegraphics{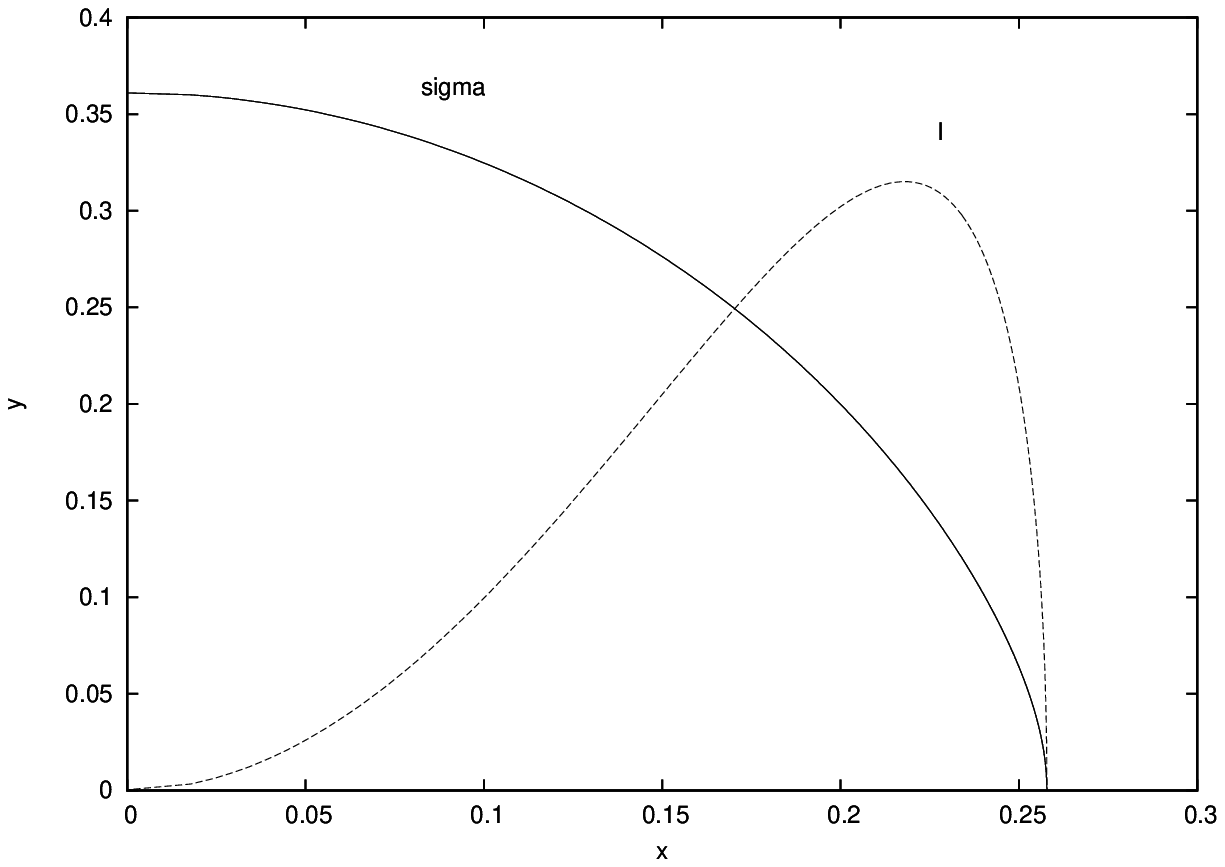}}
\hss}

\caption{\small  Left: profiles of a typical current-carrying solution of Eqs.\eqref{w-eq}
for 
$\lambda_1=0.1$, $\lambda_2=10$, $\eta_1=1$, $\eta_2=0.31$, $g_1=g_2=1$, 
$\gamma=0.6$ and  $\sigma=0.12$.
Right: the restframe current ${\cal I}$ and the twist $\sigma$ 
against the parameter $q$ for the same parameter values as in the 
left panel. 
}
\label{FIG-Wit}
\end{figure}

The perturbative bound state can be 
used as the starting configuration  
to iteratively construct the solution 
of the full non-linear system, still for small $q$. After this one can 
iteratively increase $q$, which gives the generic solutions. 
 The typical solution is shown in  Fig.\ref{FIG-Wit}. One finds then that 
the restframe current ${\cal I}$ first 
increases with $q$ but then starts to quench, and 
finally both ${\cal I}$ and $\sigma$ (but not $Q$)
vanish for some $q_\star$ as shown in  Fig.\ref{FIG-Wit}. 

It is worth reminding that in the Weinberg-Salam theory we had a completely 
different picture, as shown in Fig.\ref{FIG6} -- 
when $q$ approaches its maximal value $q_\star$ 
the current 
${\cal I}(q)$ does not quench but bends back to smaller values of $q$ and 
continues to grow.

Getting back to Witten's model, the solution for $q=q_\star$ 
(it looks qualitatively similar to other solutions) has $\sigma^2=0$
and so it is chiral, with values of $\sigma_0=\pm\sigma_3$ related to each other 
by Lorentz boosts. In this case 
one has ${\cal I}=0$ 
but the current $I_\alpha$ does not actually
vanish and becomes an isotropic vector.  The particular member of the chiral family 
is obtained for $\sigma_0=\sigma_3=0$, in which case one has $I_\alpha=0$
so that the solution is truly currentless, 
although it contains a non-trivial condensate. This is the `dressed' string.

\begin{figure}[ht]
\hbox to\linewidth{\hss%
\psfrag{x}{{$\sigma$}}
\psfrag{y}{}
\psfrag{q}{$q$}
\psfrag{q0}{$q=0$}
\psfrag{qstar}{$q_\star$}
\psfrag{v}{$v$}
\psfrag{I}{{$\mathcal{I}$}}
\psfrag{omega}{{$\sigma$}}
       \resizebox{8.5cm}{7cm}{\includegraphics{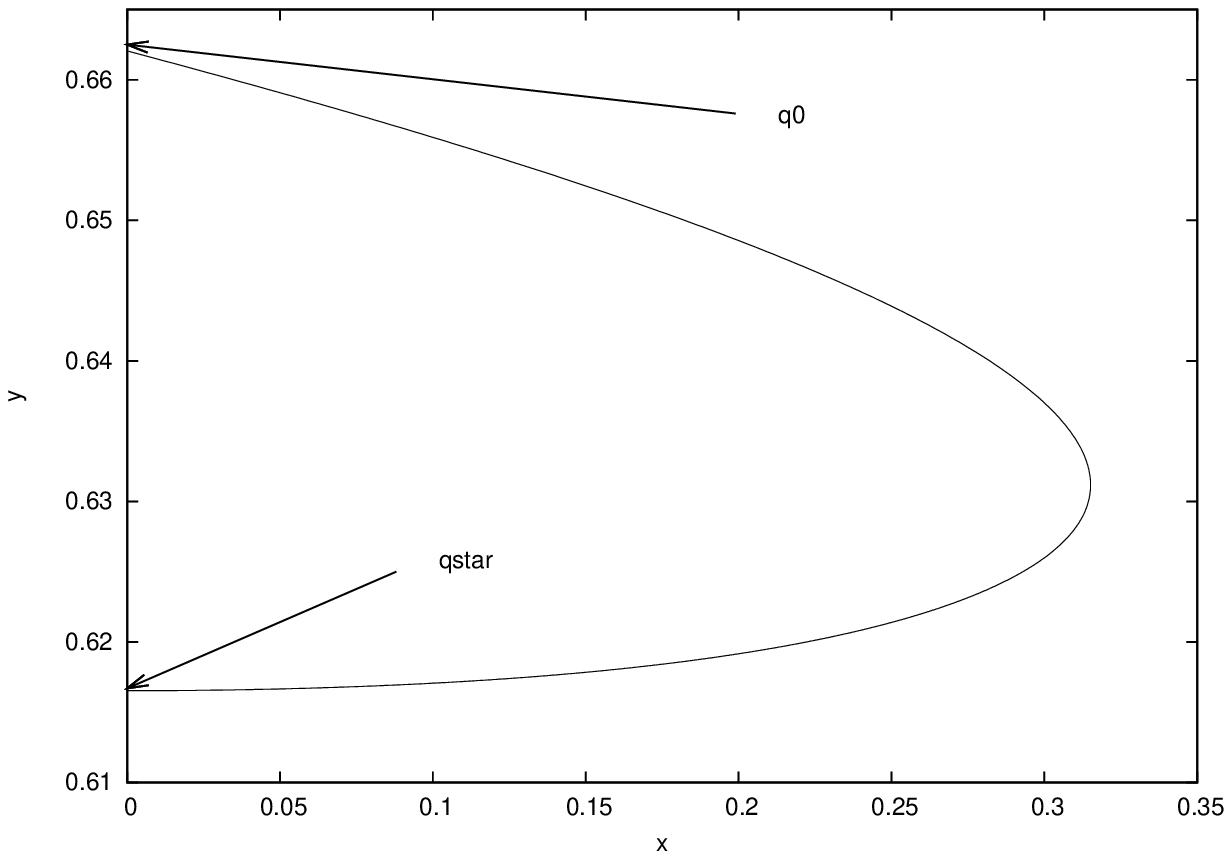}}
\hspace{5mm}%
\psfrag{x}{{$\sigma$}}
\psfrag{sigma}{$\sigma$}
        \resizebox{8.5cm}{7cm}{\includegraphics{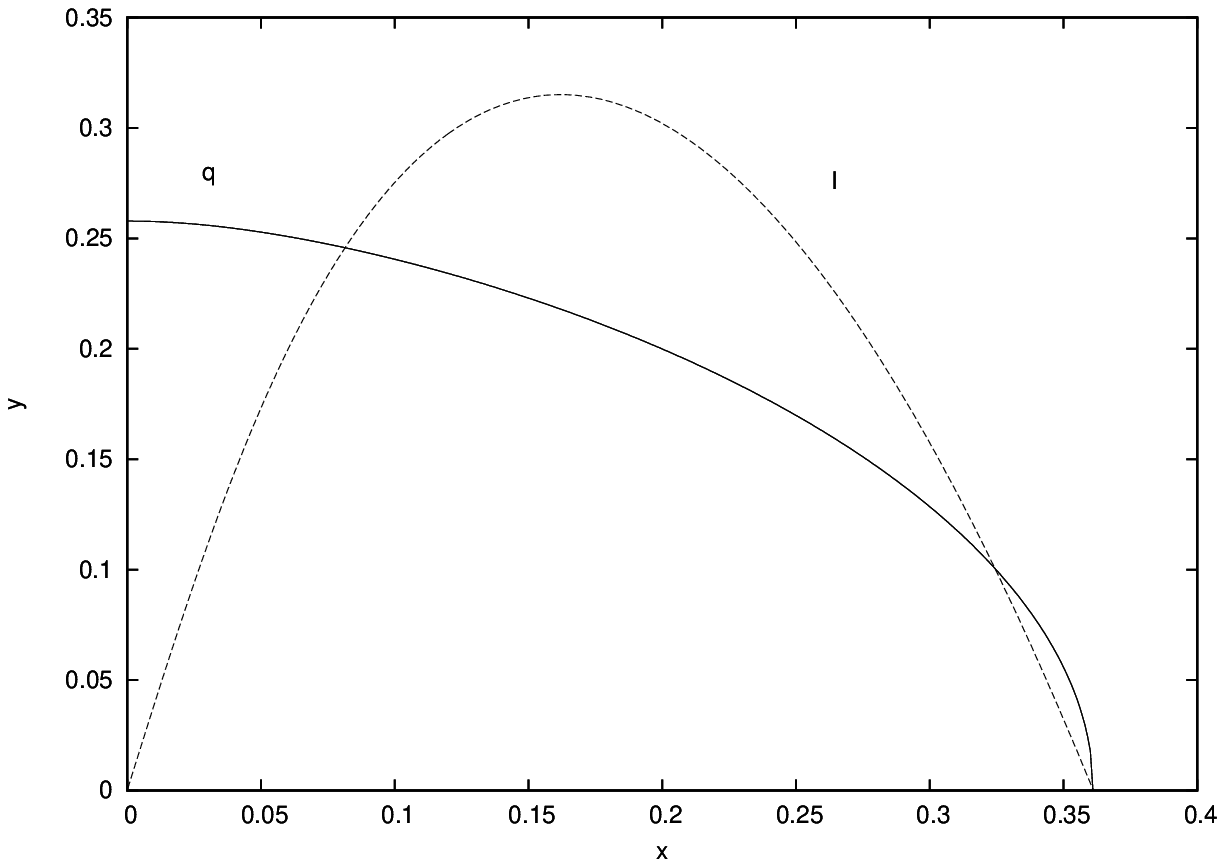}}
\hss}
\caption{\small  Magnetic energy (left) and 
the restframe current ${\cal I}$ and $q=f_2(0)$ (right)
against the twist $\sigma$  
for the same parameter values as in Fig.\ref{FIG-Wit}. 
}
\label{FIG-Wit1}
\end{figure}

Summarizing, when $q$ increases, the Witten strings 
interpolate between the `bare'  ANO vortex for $q=0$ and the `dressed'/chiral vortex for 
$q=q_\star$. Since the restframe current vanishes 
at the ends of the interval $[0,q_\star]$,
it passes through a maximum somewhere in between and so there is an upper
bound for it. The magnetic energy, which coincides with the total energy for the 
currentless strings,  decreases along the  interpolating 
sequence, as shown in Fig.\ref{FIG-Wit1}. The `dressed' vortex is therefore less
energetic than the `bare' one.

Within the standard approach based on the relaxation method the same 
solutions are obtained in the opposite order. 
First one finds the `dressed' solution with $q=q_\star$, $\sigma=0$, 
and then one increases the twist $\sigma$.
The functions $q(\sigma)$, ${\cal I}(\sigma)$ 
then exhibit the characteristic behaviour  shown in Fig.\ref{FIG-Wit1}
(which agrees with Fig.5.3 in Ref.\cite{Vilenkin-Shellard} where $\sigma$ is called 
`winding number density').  The condensate then always decreases, while   
the current passes through a maximum whose
value can be related to the parameters of the 
`dressed' solution \cite{Vilenkin-Shellard}. 
%This solution plays therefore the key role in this construction scheme. 

This scheme does not apply in the Weinberg-Salam theory, where
the `dressed' currentless 
solutions do not generically exist. We are therefore bound to start 
from the embedded ANO vortex with $q=0$. 
Increasing $q$ we find that its maximum value $q_\star$ exists,
although it corresponds not to the `dressed' currentless solution  
but to a turning point after which $q$ starts decreasing again, while the 
current continues to grow.

\end{document}